\documentclass[a4paper,12pt,oneside,headsepline]{scrbook}
\pdfoutput=1 
\usepackage{amsmath, amstext, amssymb}
\usepackage{setspace}
\usepackage{graphicx}
\usepackage{epsfig}
\usepackage[normalem]{ulem}
\usepackage{color}
\usepackage{multirow}
\usepackage{array}
\usepackage{units}
\usepackage{hyperref}
\begin{document}
\thispagestyle{empty}
\begin{Large}
\vspace*{-2cm}\hspace*{-2cm}Cover sheet for the eprint version:
\end{Large}
\vspace*{1cm}
\begin{center}
\begin{Large}Infrared Behaviour of Landau Gauge Yang-Mills Theory \\[0.2cm] 
with a Fundamentally Charged Scalar Field\end{Large}\\[.7cm]
diploma thesis, submitted May 2009 at the Karl-Franzens-Universit\"at Graz \\[1cm]
\end{center}
\begin{center}
Leonard Fister\\[0.3cm]
\begin{small}
{\it
Institut f\"ur Theoretische Physik, Ruprecht-Karls-Universit\"at Heidelberg, Philosophenweg 16, 69120 Heidelberg, Germany\\
ExtreMe Matter Institute EMMI, GSI Helmholtzzentrum f\"ur Schwerionenforschung, Planckstr. 1, 64291 Darmstadt, Germany\\
Institut f\"ur Physik, Karl-Franzens-Universit\"at Graz, Universit\"atsplatz
5, 8010 Graz, Austria}
\\[2cm]
\end{small}
\end{center}

\begin{center}\begin{Large}{\it Abstract} \end{Large} \end{center}
The infrared behaviour of the $n$-point functions of a Yang-Mills theory with a charged scalar field in the fundamental representation of {\it SU(N)} is studied in the formalism of Dyson-Schwinger equations. Assuming a stable skeleton expansion solutions in form of power laws for the Green functions are obtained. For a massless scalar field the uniform limit is sufficient to describe the infrared scaling behaviour of vertices. Not taking into account a possible Higgs-phase it turns out that kinematic singularities play an important role for the scaling solutions of massive scalars. On a qualitative level scalar Yang-Mills theory yields similar scaling solutions as recently obtained for QCD.\newpage
\begin{titlepage}
\begin{center}
\begin{Large}{\bf Leonard Fister} \end{Large}\\[2cm]
\begin{LARGE}{\bf Infrared Behaviour of Landau Gauge \\[0.1cm]
Yang-Mills Theory with a \\[0.2cm]
Fundamentally Charged Scalar Field}\end{LARGE}\\[9.5cm]
\begin{Large}{\bf Diploma Thesis} \end{Large} \\[0.4cm]
 Supervisor: {\bf Univ.-Prof. Dr. Reinhard Alkofer}\\[0.1cm]
 Co-Supervisor: {\bf Dr. Kai Schwenzer}\\[0.9cm]
\begin{Large} May 2009 \end{Large}\end{center}
\end{titlepage}
\newpage
\tableofcontents 
\thispagestyle{empty}
\newpage
\chapter{Introduction}

As we believe today there are four fundamental forces in nature, namely the gravitational, the electromagnetic, the weak and the strong force. One promising approach to unification of the different interactions is to describe the electromagnetic, weak and strong force by means of quantum field theory, which is usually called the standard model of elementary particle physics. Formulating gravity as a quantum field theory is not possible the usual way, but since the effects of gravity are much weaker on short distances compared to the ones of the other three forces the standard model provides a good description for high energy physics. \\
Within the standard model the gauge theory which specifies the strong interactions is called quantum chromodynamics (QCD), involving the dynamics of quarks and gluons. For high energies the theory is asymptotically free, as shown in the nobel prize work by Politzer, Gross and Wilczek \cite{Politzer:1973fx}. That means, that the coupling of the quarks and gluons vanishes for infinitely high momenta. Therefore perturbation theory, which assumes a free theory being disturbed by a small interaction, is justified for high energies. \\
As the QCD $\beta$-function shows, the strength of the coupling increases with decreasing momenta, and therefore perturbation theory cannot be applied anymore in this region. Thus there is a need for non-perturbative methods to find the origin of typical phenomena, as \textit{e.g.} dynamical chiral symmetry breaking or confinement, which means the absence of colour-charged particles in asymptotic states. Until today both features are not fully understood, neither their origin nor their possible relation to each other. For a review on the confinement problem and several possible explanations see \cite{Alkofer:2006fu}. \\

A promising approach to the whole energy regime is the formalism of Dyson-Schwinger equations (DSE), see \cite{Dyson:1949ha,Schwinger:1951ex,Schwinger:1951hq} for the original works or \cite{Roberts:1994dr,Alkofer:2000wg, Maris:2003vk,Fischer:2006ub} for reviews about the application of DSEs in hadron physics. Within this method the Landau gauge turned out to be the \textquotedblleft simplest\textquotedblright \ gauge, because of the transversality of the gluon propagator and the non-enhancement of the ghost-gluon vertex \cite{Taylor:1971ff,Marciano:1977su} in the infrared. The formalism of DSEs in Landau gauge has been very successful in the study of the infrared properties of both the Yang-Mills sector and QCD \cite{vonSmekal:1997vx,vonSmekal:1997is,Watson:2001yv,Lerche:2002ep,Fischer:2002hna,Fischer:2002eq,Alkofer:2003jj,Alkofer:2004it,Schleifenbaum:2004id,Fischer:2006vf,Alkofer:2008jy,Kellermann:2008iw,Huber:2007kc,Maas:2004se,Aguilar:2004sw,Aguilar:2007nf,Aguilar:2008xm,Binosi:2007pi,Boucaud:2006if,Boucaud:2008ji,Boucaud:2008ky,Bloch:2002eq,Bloch:2001wz,Zwanziger:2001kw,Fischer:2003rp,Alkofer:2006gz,Alkofer:2008tt,Fischer:2008yv,Schwenzer:2008vt}. \\
Also other tools prove useful in the analysis of Landau gauge QCD, as \textit{e.g.} functional renormalisation group equations (FRGs) \cite{Fischer:2006vf,Fischer:2008yv,Wetterich:1992yh,Litim:1998nf,Pawlowski:2005xe,Gies:2006wv,Ellwanger:1994iz,Bonini:1994kp,Ellwanger:1995qf,Ellwanger:1996wy,Ellwanger:1997tp,Bergerhoff:1997cv,Pawlowski:2003hq,Pawlowski:2004ip,Kato:2004ry,Fischer:2004uk,Braun:2007bx,Litim:2000ci,Litim:2001up,Litim:2001fd}, Gribov-Zwanziger action approaches \cite{Dudal:2005na,Dudal:2007cw,Capri:2007ix,Dudal:2008sp,Dudal:2008rm}, stochastic quantisation \cite{Zwanziger:2001kw,Zwanziger:2002ia,Zwanziger:2003cf} or lattice gauge theory \cite{Oliveira:2007dy,Bloch:2003sk,Sternbeck:2005tk,Sternbeck:2006cg,Cucchieri:2007md,Cucchieri:2007rg,Cucchieri:2008fc,Bogolubsky:2007ud,Sternbeck:2007ug,Bogolubsky:2005wf,Bogolubsky:2007bw,Bogolubsky:2009dc,Bornyakov:2008yx,Parrinello:1994wd,Ilgenfritz:2006he,Cucchieri:2008qm} (for an introduction see \cite{Lang:book,Rothe:2005nw} for textbooks or \cite{Greensite:2003bk} for a review about confinement in lattice gauge theory). All methods have their merits, and by comparison of different results one aims at getting a deeper understanding of the confinement mechanism.\\

The goal of this thesis is to make another little step further in the understanding of confinement. A possible mechanism for quark-confinement has recently been proposed in \cite{Alkofer:2008tt}, based on DSEs and a power law scaling behaviour of Green functions of QCD in the infrared. The authors showed, that this mechanism does not explicitly depend on the Dirac structure of the quark-gluon coupling, as Dirac scalar parts yield the same leading scaling behaviour as the Dirac vector parts. But the Dirac structure makes the calculation very complicated. Thus it would be desireable to have a simple model involving bosonic particles, that has the same qualitative behaviour in the infrared, \textit{i.e.} the same infrared scaling behaviour of the vertex function. The assumption that this is possible is based on the fact, that the string tension is similar for both types of particles in the fundamental representation \cite{Bock:1988kq,Bali:2000gf}.\\

In this thesis the system of scalars coupled to the Yang-Mills sector is investigated by means of DSEs with respect to its (non-)confining properties, \textit{i.e.} the infrared leading scaling behaviour is determined and compared with the scaling exponents in QCD. Herein the potential problem of finite tadpole contributions at zero temperature will be ignored. The big advantage of this model will turn out to be, that despite the fact that the infrared analysis, performed here, is complicated by bosonic self-interactions, it should simplify a numerical analysis of the mid-momentum region, which could not comprehensively be analysed for QCD yet. In addition it would simplify a comparison between the Dyson-Schwinger formalism and lattice gauge theory. The point is that on the lattice it is very intricate to simulate dynamical fermions, but for bosons this can be done more easily. As a result this model may be a suitable test for the mechanism of confinement, as it is proposed in \cite{Alkofer:2008tt,Schwenzer:2008vt,Schwenzer:2008prep}. \\

Fundamentally charged scalars coupled to Yang-Mills theory have not been dealt with by DSEs yet, but from lattice gauge theory some results for a fundamental Higgs-model in \textit{SU(2)} are known. From the Osterwalder-Seiler-Fradkin-Shenker theorem \cite{Osterwalder:1977pc,Fradkin:1978dv} one sees, that there is a free-charge phase, and additionally two other regions, the confinement and the Higgs region, which are analytically connected \cite{Kertesz:1989,Satz:2001zf}. Thus there is no thermodynamic phase transistion \cite{Fradkin:1978dv,Lang:1981qg,Evertz:1985fc}, and the distinction must be done by a non-local order parameter \cite{Langfeld:2002ic,Langfeld:2004vu,Caudy:2007sf}, which separates the two regions.\\

This thesis will be organised as follows: chapter 2 will give a short introduction into the underlying physical concepts, chapter 3 will deal with the specific methods that will be used, \textit{i.e.} DSEs and the infrared power counting analysis. The results of my work I will give in chapter 4. In chapter 5 the comparison of the scaling behaviour of scalar Yang-Mills theory with quenched QCD will be done, before in chapter 6 the conclusion and outlook to future work will be given. Conventions, details of the DSEs and parts of the calculation are put in various appendices.

\chapter{Yang-Mills Theory Including Fundamentally Charged Scalars}
In this chapter I will outline the basic concepts of gauge invariance and the consequences for the Lagrangian density, as the interaction terms and the charge of the scalar. I will start with a short introduction into group representations by comparing scalar particles in the adjoint representation to scalars in the fundamental representation of the group \emph{SU(N)}. The aspect of renormalisability will give further constraints for the possible interaction terms in the Lagrangian density. At the end of this chapter I will shortly discuss the model, which will be analysed in this thesis.

\section{Group Representations of {\it SU(N)}}
In the standard model of particle physics quantum chromodynamics (QCD) is a gauge theory, \textit{i.e.} a local quantum field theory which is symmetric under a special kind of transformations. For each local gauge symmetry the underlying group is a Lie group. For QCD the underlying symmetry group is \emph{SU(3)}. Thus the Lagrangian of the system must be symmetric under a so-called gauge transformation. \\
For one group \emph{SU(N)} the algebra  $su(N)$ can be specified by the commutation relation of the abstract generators $T^a$ of the group and the structure constants $f^{abc}$ of the group. For the generators of a Lie group the identity
\begin{equation}
 T^a T^b = \frac{1}{2N}\delta^{ab} 1_N + \frac{1}{2} \sum_{c=1}^{N^2-1} (if^{abc}+d^{abc})T^c
\end{equation}
holds, where the ($f^{abc}$) $d^{abc}$ are totally (anti-)symmetric. As the symmetric parts drop out in taking the commutator, it leaves the algebra defined by
\begin{equation}\label{comm}
  [T^a , T^b ] = i f^{abc} \ T^c.
\end{equation}
Fields can be described by an (irreducible) representation of \emph{SU(N)}. For \emph{SU(N)} the basic irreducible representation is the fundamental representation. It can always be constructed, and further that any other finite-dimensional irreducible representation can be generated from it, see \cite{Simon} for further details. In this representation a field is given by an $N$-dimensional vector and the unitary matrix representation $t^a$ of a generator $T^a$ is a $N\times N$ matrix, satisfying the commutation relation (\ref{comm}). For $N>2$ this representation is complex, \textit{i.e.} there is another inequivalent representation for a field $\bar{\psi}(x)$, sometimes called antifundamental representation. In order to find group invariants, one can always consider products of $\bar{\psi}(x) \psi(x)$.\\
A local gauge transformation can be written in the form
\begin{eqnarray} \label{trans}
 \psi(x) \rightarrow e^{i \alpha^{a}(x) t^a} \psi(x), \\[0.2cm]
 \big( \bar{\psi}(x) \rightarrow \bar{\psi}(x) e^{-i \alpha^{a}(x) t^a} \big) \nonumber
\end{eqnarray}
where $\psi(x)$ is an arbitrary matter field and $\alpha^a (x), \ a \in \{1,\ldots,N^2-1\}$ are continuous and differentiable functions, as \emph{SU(N)} is a compact Lie group. Note that $\alpha^a(x)$ is a space-time dependent quantity, \textit{i.e.} it parametrises a local symmetry.\\
Another important irreducible representation that exists for all \emph{SU(N)} is the adjoint representation. Given the structure constants $f^{abc}$, which contain the entire information about the algebra $su(N)$, one finds that the generators can be represented by $(N^2-1) \times (N^2-1)$ matrices given by
\begin{equation}
 (T^{a})^{bc} = -i f^{abc}.
\end{equation}

\section{Gauge Invariance}
Forming gauge invariants under an \emph{SU(N)} of the form $\bar{\psi}(x) \psi(x)$ is possible here, so there are no further problems in constructing mass terms in QCD. But a non-trivial theory must have a kinematic part, so necessarily there must be derivatives. The problem is, that ordinary derivatives are not gauge invariant by this construction, as they act not only on the field but also on the spacetime dependent group parameter. Thus they receive additional terms like in  
\begin{equation}\label{trans_deriv}
 \partial_{\mu} \psi(x) \rightarrow e^{i \alpha^{a}(x) t^a} \partial_{\mu} \psi(x) + i \left(\partial_{\mu} \alpha^{a}(x) \right) t^{a} e^{i \alpha^{a}(x) t^a}\psi(x).
\end{equation}
In order to get a derivative, which transforms as the field itself in eq. (\ref{trans}), one introduces a so-called covariant derivative, which has an additional space-time dependent term, that cancels the redundant term in eq. (\ref{trans_deriv}),
\begin{equation} \label{cov_deriv}
 \partial_{\mu} \rightarrow D_{\mu}  \ = \ \partial_{\mu} - i g A_{\mu}^{a} t^a. 
\end{equation}
In eq. (\ref{cov_deriv}) $D_\mu$ is the covariant derivative, $A_{\mu}^{a}$ is the gauge field, and $g$ is the coupling strength of the fields $\psi$ and $\bar{\psi}$ to the gauge field. \\
Since $D_{\mu}$ is covariant, also the commutator $\left[ D_{\mu},D_{\nu}\right]$ is covariant. Thus for non-Abelian gauge theories the field strength tensor is defined by, working in natural units and Euclidean metric, 
\begin{equation}
 F_{\mu \nu}^{a} = \frac{i}{g} \left[ D_{\mu}, D_{\nu} \right]= \partial_{\mu} A_{\nu}^{a} - \partial_{\nu} A_{\mu}^{a} - g f^{abc} A_{\mu}^{b}A_{\nu}^{c}.
\end{equation}
As a result the simplest unitary action, \textit{i.e.} an action with only two derivatives, is 
\begin{equation}
 S = \int d^4 x \left( -\frac{1}{4}F_{\mu \nu}^{a} F_{\mu \nu }^{a}\right).
\end{equation}
Due to gauge invariance we still have to choose a gauge, because the functional integral counts all configurations and thus overcounts the ones, that are in the same gauge orbit, \textit{i.e.} which are related by a simple gauge transformation. Faddeev and Popov showed a way how to insert this restriction in the funtional integral \cite{Faddeev:1967fc}, which I will shortly sketch.\\
The gauge fixing is done by a gauge condition $\delta(f^a(A))$, \textit{i.e.} considering only a hypersurface, such that each orbit subtends the hypersurface only once\footnote{In fact problems emerge, if one orbit intersects the hypersurface more than once. These \textquotedblleft copies\textquotedblright of the same orbit are called Gribov copies \cite{Gribov:1977wm}, thus Gribov's idea was to restrict the integration to the first Gribov region, where each orbit only intersects once. Later is was shown that still there is overcounting, thus one has to limit the functional integration to the fundamental modular region \cite{vanBaal:1991zw}, which causes severe troubles in practical calculations.}. In linear covariant gauges one can express this condition by $f^{a}(A)=\partial_{\mu} A_{\mu}^{a}$. But this insertion changes the measure of the integral. This can be compensated by the Faddeev-Popov determinant, which can be rewritten as another contribution in the Lagrangian\footnote{To be brief I only outline this topic, because it can be found in almost every text book on quantum field theory, as it is a standard technique. I refer the interested reader to the detailed description in \textit{e.g.}\cite{Ryder}, as there is also given an analogon to ordinary integration techniques, or other didactic presentations \cite{Kaku,Pokorski}.}. The corresponding particles are called ghosts, which are to be seen as scalar particles but underlying fermionic statistics. Note that ghosts are given in the adjoint representation, as Faddeev and Popov introduced them. With this technique of gauge fixing the action of the Yang-Mills theory reads
\begin{equation}\label{YMaction}
 S_{YM} = \int d^4 x \left( -\frac{1}{4}F_{\mu \nu}^{a} F_{\mu \nu }^{a} + \frac{1}{2 \zeta} (\partial _\mu A_{\mu}^{a})^2 + \bar{c}^a \partial_{\mu} D_{\mu}^{ab} c^b \right),
\end{equation}
where $\zeta$ is the gauge parameter, ($\bar{c^a}$) $c^a$ denotes the (anti)-ghost and $D_{\mu}^{ab}$ is the covariant derivative for ghost fields in the adjoint representation\footnote{Note that the $f^{abc}$ are the generators of the \emph{SU(N)} of the adjoint representation.}
\begin{equation}
 D_{\mu}^{ab} = \delta^{ab} \partial_{\mu} + g f^{abc} A_{\mu}^{c}. 
\end{equation}

\section{Lagrangian Density for Scalar Yang-Mills Theory}
Although pure gauge theory already bears enough subtleties, still one has to include matter in the Lagrangian to get a theory that is capable of predicting processes of the real world. In the standard model of particle physics quarks and leptons are the massive elementary particles, whose dynamics and interactions are to be described. Quarks are known to be fermions, \textit{i.e.} they obey the Fermi-Dirac spin-statistics. In search of a fundamental mechanism of confinement the Dirac-structure may not play the crucial role, but it definitely complexifies things, and may conceal the important aspects. Thus it would be desireable to have a model, that reproduces the qualitative properties of confinement but is free of the complications due to the Dirac structure.\\
In QCD the quark-gluon vertex is the link between matter and gauge sector. Parametrizing according to Ball and Chiu \cite{Ball:1980ay}, it leaves twelve tensor structures. Thus it is a system that is very hard to handle. The idea in this thesis is using fundamentally charged scalars instead of quarks. Scalar particles are bosons, in contrast to quarks, \textit{i.e.} scalars do not have a Dirac-structure. Looking at the vertex this means, that there are only two tensor components. So the vertex might be easier to handle - first of all in the Dyson-Schwinger formalism, as I will carry out in this thesis, but in lattice gauge theory as well, where dynamical fermions are still hard to implement, whereas bosons can be simulated efficiently. \\
So changing fermions to bosons yields a simpler model compared to QCD in the deep infrared region, which is also easier to handle in the mid-momentum region. But to be meaningful it furthermore must have the same qualitative scaling behaviour. The question is if such a severe modification yields the same qualitative behaviouro in the infrared region.\\
So the task is to include scalar particles into the Lagrangian. From the physical point of view some aspects must be considered:
\begin{itemize}
 \item{Scalars must be charged, in order to be able to define a conserved charge.}
 \item{Due to the gauge principle the Lagrangian has to be invariant.} 
 \item{In what representation does one include the scalars?}
 \item{The Lagrangian must be renormalisable.}
\end{itemize}
As quarks are in the fundamental representation, also for scalars the fundamental representation is chosen, and from now on I will write the scalar field as $\phi$ and its conjugate as $\phi^{*}$. As mentioned above these fields are two inequivalent irreducible representations of \emph{SU(N)}. \\
So as a result only terms with group invariants are permitted in the Lagrangian. Up to this point there is still an infinite number of possible interaction terms in the Lagrangian. But there is another restriction to the possible terms, \textit{i.e.} the theory has to be renormalisable. A quantum field theory naturally generates divergencies, as one sums over infinitely many internal modes in loops. But it turns out that although divergencies appear, the theory may bring a correct physical interpretation about\footnote{Consider the most intuitive example: the (divergent) self-energy of a point charge in classical electrodynamics. But dropping the divergent terms one finds a good description of the generated field of the point charge.}. Thus in order to yield a physically meaningful theory these divergencies must be cancelled such that one can predict observable (finite) quantities, at least, after adjusting parameters to the experiment. \\
So regularization and renormalisation is a method to render a theory finite and predictive. In my opinion the easiest way to understand the idea of renormalisation is by means of so-called counterterms. To get rid of divergencies by counterterms, one introduces additional terms into the Lagrangian\footnote{Note that in fact it is actually a dividing of terms, that are already contained in the Lagrangian, rather than adding further terms.}, which cancel the divergent terms. In order to cancel the divergencies up to all orders, the counterterms must have the same structure as the divergent functions. Another condition for renormalisability is, that there is only a finite amount of counterterms. An infinite amount of counterterms would correspond to infinitely many free parameters. Such a theory would not be predictive. Thus the renormalisation procedure is not valid any more.
\\
As a result of renormalisability only several interaction terms in the Lagrangian are possible for scalar fields, which can be shown by means of a dimensional analysis. Let's start from the fact, that the action 
\begin{equation}
 S = \int d^4 x \mathcal{L}
\end{equation}
must be dimensionless (in units of $\hbar$). The integration gives a length dimension of $4$, thus the Lagrangian must have length dimension of $-4$. Due to a Fourier transformation one sees that this is equivalent to a momentum dimension of $+4$. With respect to the kinetic terms the dimension of a scalar field must be $+1$, because of the derivative having dimension $+1$. In contrast to this fermions, as \textit{e.g.}quarks, have mass dimension $\frac{3}{2}$. So a renormalisable theory must only have a few distinct interaction terms, each of which have to be gauge invariant and renormalisable. These terms are
\begin{itemize}
 \item{the ones from the Yang-Mills Lagrangian}
 \item{a kinematic one for the fundamentally charged scalars. For gauge invariance this term must involve two covariant derivatives. These covariant derivatives imply a coupling of scalars to the gauge sector in terms of a (two-)scalar-gluon and a two-scalar-two-gluon vertex.}
 \item{a four-scalar coupling with a dimensionless coupling constant, differing from the gauge coupling.}
\end{itemize}
As a first result, the Lagrangian of the scalar Yang-Mills theory reads
\begin{equation}\label{Lagrangian}
 \mathcal{L}= \frac{1}{4} F_{\mu \nu}^{a} F_{\mu \nu}^{a} + \frac{1}{2 \zeta}  (\partial _\mu A_{\mu}^{a})^2 + \bar{c}^a \partial_{\mu} D_{\mu}^{ab} c^b + (D_{\mu,ij} \phi_j^{*}) (D_{\mu,ik} \phi_k) - m^2 \phi_{i}^{*} \phi_{i} - \frac{\lambda}{4!} (\phi_{i}^{*} \phi_{i})^2,
\end{equation}
with the covariant derivatives and the field tensor
\begin{align*}
D^{ab}_{\mu} & = \delta^{ab} \partial_{\mu} + g f^{abc}A_{\mu}^{c} \\
D_{\mu,ij} & = \delta_{ij} \partial_{\mu} - i g (\frac{t^{a}}{2})_{ij} A_{\mu}^{a} \\
F_{\mu \nu}^{a} & = \partial_{\mu} A_{\nu}^{a} - \partial_{\nu} A_{\mu}^{a} - g f^{abc} A_{\mu}^{b} A_{\nu}^{c}.
\end{align*}
For the convenience of the reader I summarise the nomenclature in this work:
\begin{itemize}
 \item{Greek indices are Lorentz indices, Roman superscripts are color indices. Einstein sum convention over identical indices is assumed.}
 \item{$F_{\mu \nu}^{a} $ is the field strength tensor.}
 \item{$\zeta$ is the gauge fixing parameter - in this thesis Landau gauge is used, \textit{i.e.} $\zeta \rightarrow 0$.} 
 \item{$(\phi_{i}^{*}) \phi_{i} $ is the complex (anti-) scalar field in the fundamental representation. Here $i \in \{1, \ldots, N\}$ for \emph{SU(N)}.}
 \item{$A_{\mu}^{a}$ is the gluon field.}
 \item{($\bar{c^a})$ $c^a$ is the Faddeev-Popov (anti-)ghost.}
 \item{$D_{\mu}^{ab}$, $D_{ij,\mu}$ are the covariant derivatives for the adjoint and fundamental representation, respectively, where $f^{abc}$ and $t^a$ are the various generators.}
 \item{$m$ is the mass of the scalar field, and $\lambda$ the coupling of the self-interaction of the scalars.}
\end{itemize}
In the following chapters the generating functional of Green functions will be very important for the derivation of Dyson-Schwinger equations. For the full Green functions it reads
\begin{equation}\label{gen_func}
 Z[J,J^{*}, \eta, \bar{\eta}, Q]= e^{-S+\int d^4 x \left( J \phi^{*}+J^{*}\phi+\bar{\eta} c + \bar{c} \eta + Q^a A^a \right)},
\end{equation}
where ($J$) $J^{*}$ are the sources of the (anti-)scalars, ($\eta$) $\bar{\eta}$ are the sources of the (anti-)ghosts and $Q^a$ the sources of the gluons.

\chapter{Infrared Behaviour of Scalar Yang-Mills Theory}\label{chapter2}
In this chapter the derivation of Dyson-Schwinger equations is sketched. The idea of a skeleton expansion is introduced, which will prove useful in justifying the methods, that will be used in later chapters. The relevant DSEs for scalar Yang-Mills theory are presented. For later application the power counting analysis is defined at the end of this chapter.

\hspace{1cm}
\section{Derivation of Dyson-Schwinger Equations}
In quantum field theories the quantities of interest are the Green functions, also called correlation functions or $n$-point functions. Dyson-Schwinger equations (DSEs) are the quantum equations of motion for Green functions and as they are non-perturbative they serve as a functional method to investigate the whole momentum range of QCD. Thus also the infrared region of QCD can be analysed, \textit{i.e.} where the related momenta are far below the momentum scale of QCD ($\Lambda_{QCD}$).\\

F. J. Dyson and J. S. Schwinger separately presented the formalism of DSEs, see \cite{Dyson:1949ha,Schwinger:1951ex,Schwinger:1951hq} for the original works or \cite{vonSmekal:1997vx,vonSmekal:1997is,Watson:2001yv,Lerche:2002ep,Alkofer:2003jj,Alkofer:2004it, Fischer:2006vf,Alkofer:2008jy,Bloch:2001wz,Fischer:2003rp,Alkofer:2006gz,Alkofer:2008tt} for reviews and recent applications in QCD. An alternative method, based on equal time commutation relations and Heisenberg's equations of motion, is presented in \cite{Rivers}.
The underlying idea of Dyson-Schwinger equations is that, due to translational invariance, the integral of a total derivative of the generating functional vanishes (as boundary terms vanish by definition)
\begin{equation} \label{DSEs_deriv}
 \frac{d}{d \varphi_i}  Z[J]= \int \mathcal{D} [\varphi] \frac{d}{d \varphi_i} e^{-S[\varphi]+\varphi_i J_i}= 0.
\end{equation}
Note that $J_i$ is the source of a field $\varphi_i$, wherein $\varphi$ is a reducible representation of the irreducible fields and in the index $i$ all color, space-time and representation indices are absorbed. $S[\varphi]$ is the action of the theory containing $\varphi$.
From eq. (\ref{DSEs_deriv}) one can derive a similar expression for the generating functional for connected ($W[J]$) or 1-particle-irreducible ($\Gamma[\varphi]$) Green functions. The modification is based on the relation of the generating functional for full Green functions to the generating functional for connected Green functions
\begin{equation}
 Z[J] = e^{W[J]}
\end{equation}
and a Legendre transformation
\begin{equation}
 \Gamma[\Phi] = - W[J] + \Phi_i J_i ,
\end{equation}
wherein $\Phi_i$ are the averaged fields in the presence of the sources
\begin{equation}
 \Phi_i \equiv \langle \varphi_i \rangle_J.
\end{equation}
The resulting equation 
\begin{equation}\label{DSE_1point}
 - \left. \frac{\delta S}{\delta \varphi_i} \right|_{\varphi_i = \Phi_i + \Delta_{ij}^{J} \frac{\delta}{\delta \Phi_i}} + \frac{\partial \Gamma}{\partial \Phi_i} = 0
\end{equation}
(where $\Delta_{ij}^{J}$ is a general propagator of the generic field $\varphi$), 
serves as a starting point for the derivation of the DSEs. Note that mixed propagators are contained until the external sources are set to zero.\\
Taking the functional derivative with respect to a scalar, gluon or ghost in eq. (\ref{DSE_1point}) leaves the so-called generating equations for the particles. By taking further functional derivatives one obtains the equations of motion for arbitrary higher n-point functions. The generating equation must be obtained by doing the analytic calculation by hand, but proceeding with this method the derivation for higher derivatives becomes a very tedious task. Fortunately there is also a graphical method, which I will sketch shortly in the following, for a detailed explanation for an algorithm of deriving DSEs I refer the interested reader to \cite{Alkofer:2008nt}, where also a Mathematica package \emph{DoDSE} is presented, which derives DSEs by the given algorithm.\\
\begin{table}
\centering
\begin{tabular}{|p{4cm}|p{4cm}|}
\hline
 \parbox[0cm][1cm][c]{4cm}{\centering \includegraphics[width=3cm]{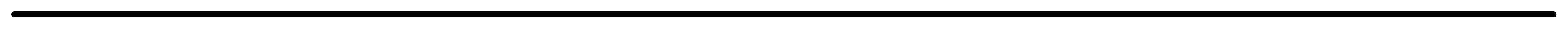} } & scalar\\
\hline
 \parbox[0cm][1cm][c]{4cm}{\centering \includegraphics[width=3cm]{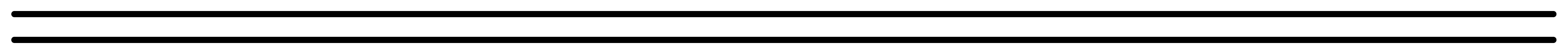} } & generic field\\
\hline 
\parbox[0cm][1cm][c]{4cm}{\centering \includegraphics[width=3cm]{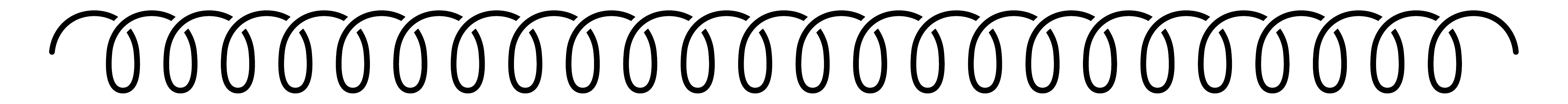} } & gluon\\
\hline
  \parbox[0cm][1cm][c]{4cm}{\centering \includegraphics[width=3cm]{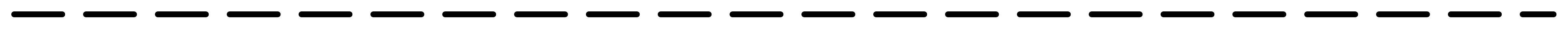} } & ghost\\
\hline
 \parbox[0cm][1cm][c]{4cm}{\centering \includegraphics[width=0.5cm]{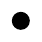} } & bare vertex\\
\hline
 \parbox[0cm][1cm][c]{4cm}{\centering \includegraphics[width=0.5cm]{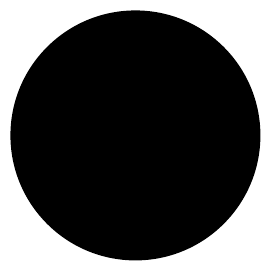} } & dressed vertex\\
\hline
 \parbox[0cm][1cm][c]{4cm}{\centering \includegraphics[width=0.5cm]{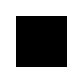} } & mixed propagator\\
\hline
 \parbox[0cm][1cm][c]{4cm}{\centering \includegraphics[width=0.5cm]{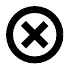} } & external field\\
\hline
\end{tabular}\caption{\small Graphical representations for emerging quantities in the derivation of DSEs.}\label{table:graphics}
 \end{table}

\begin{figure}
\centering
\caption{\small Replacement rules for the graphical derivation of DSEs. The crossed dot represents an external field, which is removed by derivation, a mixed propagator, marked as a box, yields a 3-field vertex with two attached mixed propagators, and an arbitrary $n$-point function receives an additional leg. \vspace{0.5cm}}
\includegraphics[width=13cm]{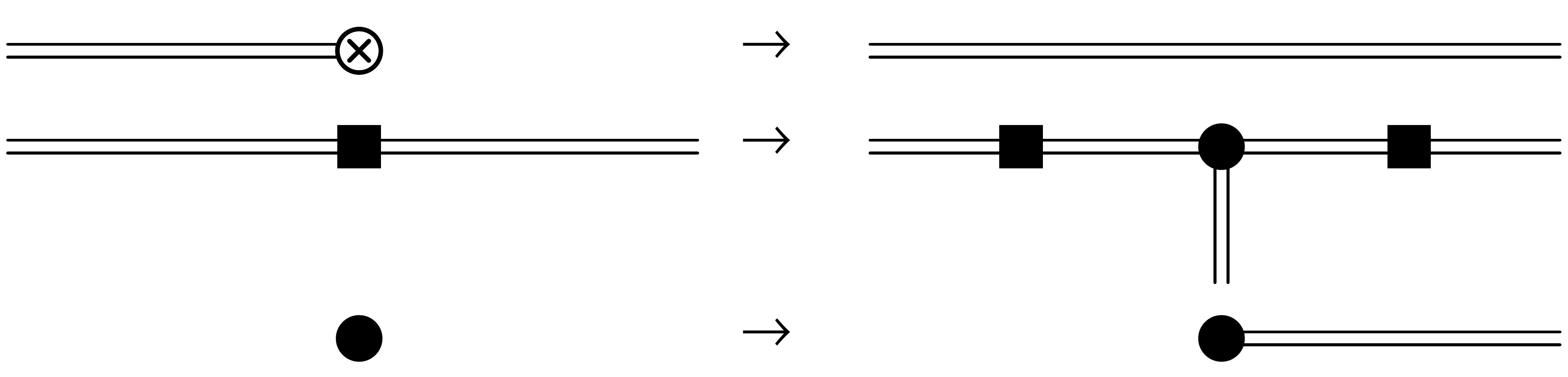}
\label{fig:graph_deriv}
\end{figure}

From now on in all diagrams in this thesis a straight line will represent a propagating scalar particle, a spring line a gluon and a dashed straight line a ghost. Double straight lines stand for the generic field, dots denote bare vertices, blobs dressed vertices, crossed circles external fields and squares mixed propagators, see table \ref{table:graphics}.\\
Functional derivatives of the various quantities give
\begin{eqnarray}
&& \frac{\delta}{\delta \Phi_j} \Phi_j = \delta_{ij}, \\
&& \frac{\delta}{\delta \Phi_j} \Delta_{ik}^{J} = \Delta_{im}^{J} \Gamma_{mjn}^{J} \Delta_{nk}^{J}, \\
&& \frac{\delta}{\delta \Phi_j} \Gamma_{i_1 \ldots i_n}^{J} = \Gamma_{ji_1 \ldots i_n}^{J}.
\end{eqnarray}
These structures can easily by described by graphical means, as can be seen in fig. \ref{fig:graph_deriv}, where I omitted all indices as the figure serves for pure illustrational purposes. The arrows restate the process of taking the functional derivative. The last blob stands for an n-point function which receives a further leg. \\
After having replaced the structures consecutively one sets the external sources equal to zero, \textit{i.e.} all generic field propagators become gluon, scalar or ghost propagators.\\

\hspace{1cm}
\section{Skeleton Expansion}
DSEs are by construction an infinite set of coupled non-linear integral equations, as every equation involves also higher n-point functions, which can be seen in the diagrams of the following sections. To handle such a system a classification is needed, which can be done by means of a skeleton expansion.\\
The skeleton expansion is based on an idea of the skeleton of an arbitrary diagram. Every diagram can be decomposed into its skeleton graph by contracting all self-energy and primitively divergent vertex contributions to a point. Therefore each diagram has a uniquely determined skeleton. But following these thoughts one can construct every diagram that can emerge, if one has all different skeletons attached with suitable insertions, which are given in fig. \ref{fig:skel_exp} for Yang-Mills theory with scalars in the defining representation. \\
Applying this method to the DSEs leaves a system of equations, that only contains different skeletons, each containing one bare and further dressed primitively divergent vertices. Arbitrary high loop-orders can be obtained by subsequent insertions of the allowed structures.\\
The validity of the skeleton expansion is based on some assumptions. First of all higher loop-orders must not increase the order of singularity compared to the skeleton diagrams. Further the prefactors of the diagrams must \textquotedblleft behave well\textquotedblright, \textit{i.e.} they do neither vanish nor combine in such a way that different diagrams cancel against each other. Under these assumptions it suffices to analyse the first order system via a power counting analysis to gain a self-consistent solution about the scaling behaviour of all vertex functions in the deep infrared. \\
For illustrative purposes I give a standard example for the skeleton expansion of a diagram in the three-gluon vertex, which contains a two-ghost-two-gluon interaction, which is not a primitively divergent vertex. Thus it has to be expanded, and it can be seen in fig. \ref{fig:sg_skel_exp} that all higher orders can be generated by proper insertions.\\
This section shall only give an idea of the skeleton expansion, the explicit expansion and further subleties will be defered to the following chapters.


\hspace{1cm}
\section{Dyson-Schwinger Equations of Fundamentally Charged Scalars in the IR-Limit}
\begin{figure}[!htb]
 \centering
 \includegraphics[width=13.8cm]{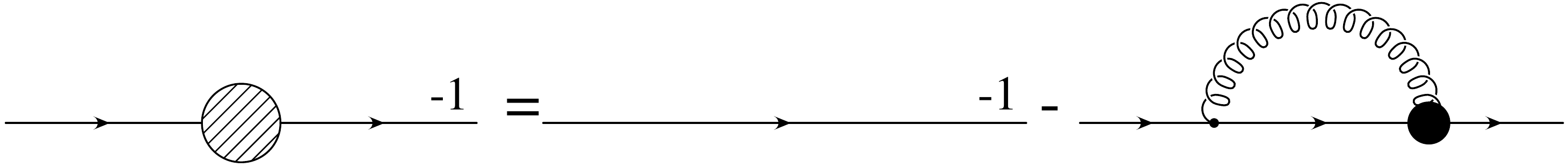}
 \caption{\small{Dyson-Schwinger equation for the quark propagator in QCD.}}
 \label{fig:quark_prop}
\end{figure}

\begin{figure}[!htb]
 \centering
 \includegraphics[width=13.8cm]{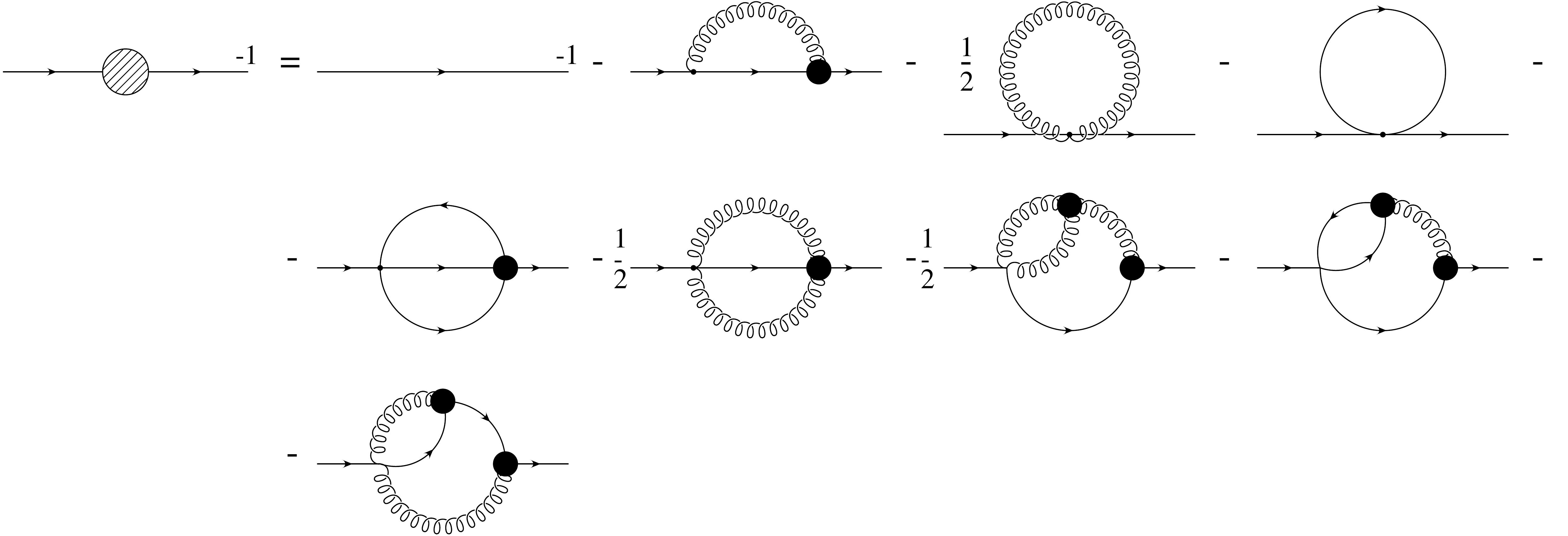}
 \caption{\small{Dyson-Schwinger equation for the scalar propagator.}}
 \label{fig:scal_prop}
\end{figure}

\begin{figure}
 \centering
 \includegraphics[width=13.8cm]{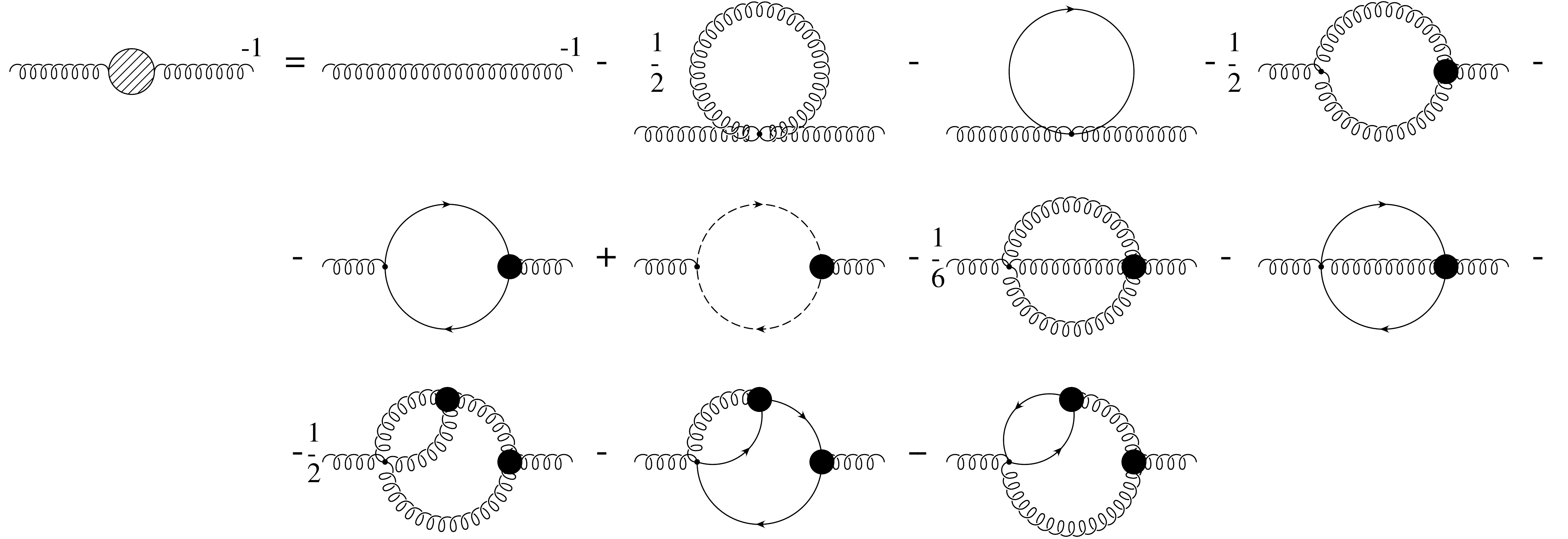}
 \caption{\small{Dyson-Schwinger equation for the gluon propagator.}}
 \label{fig:g_prop}
\end{figure}

\begin{figure}
 \centering
 \includegraphics[width=9.5cm]{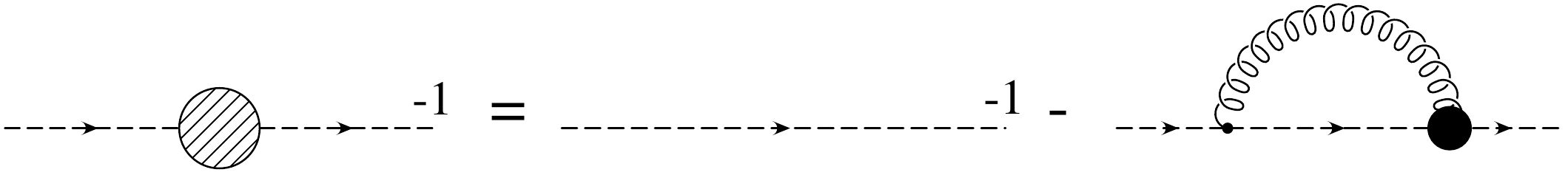}
 \caption{\small{Dyson-Schwinger equation for the ghost propagator in QCD as well as for a gauge theory including fundamentally charged scalars.}}
 \label{fig:ghost_prop_text}
\end{figure}

\begin{figure}[!htb]
 \centering
 \includegraphics[width=13.8cm]{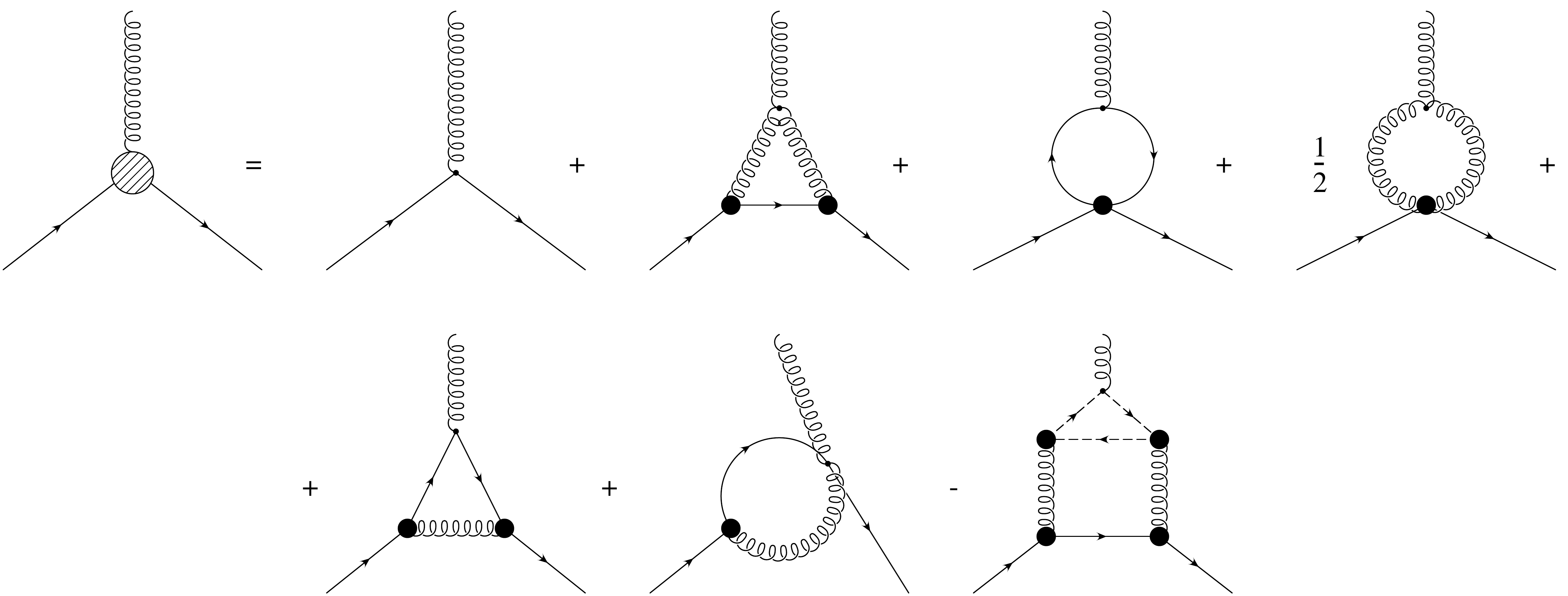}
 \caption{\small{Dyson-Schwinger equation for the scalar-gluon vertex.}}
 \label{fig:scalar_gluon_vertex_DSE}
\end{figure}

As the Yang-Mills system including fundamentally charged scalars shall be used as a model for QCD, I want to point out the differences in the basic equations already in this section. \\
The most obvious difference between quarks and scalars is the interaction of these types of particles. Quarks are fermions, whereas scalars obey bosonic statistics. Thus the interaction terms in the Lagrangian must be different, as explained above, which will influence the DSEs.\\
For QCD there are seven primitively divergent vertices, five stemming from the pure gauge sector, the quark propagator and the quark-gluon vertex as the link of the quark to the Yang-Mills sector. Including scalars instead of quarks, two additional vertices come into play, namely the four-scalar and the two-scalar-two-gluon vertex, which cause major modifications in the DSEs. All nine primitively divergent vertices are listed in table \ref{tab_primdiv}.
\begin{table}[!htb]
\caption{\small{Diagrammatic representation for the nine primitively divergent vertex functions of fundamentally charged scalar particles. Coupling scalars to the Yang-Mills sector yields 4 additional primitively divergent vertex functions.}} \centering
\begin{tabular}{p{4cm} p{2cm} p{4cm} p{2cm}}
\hline
\vspace{0.2cm} gluon propagator  & \vspace{0.2cm} \includegraphics[width=2cm]{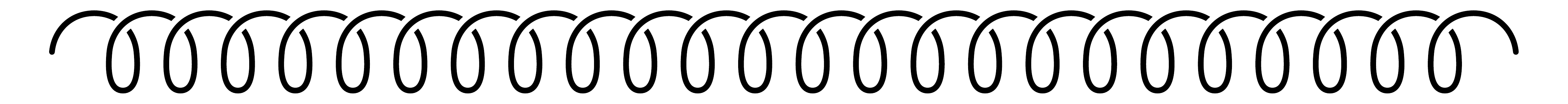} & \vspace{0.2cm} ghost propagator &  \vspace{0.2cm} \includegraphics[width=2cm]{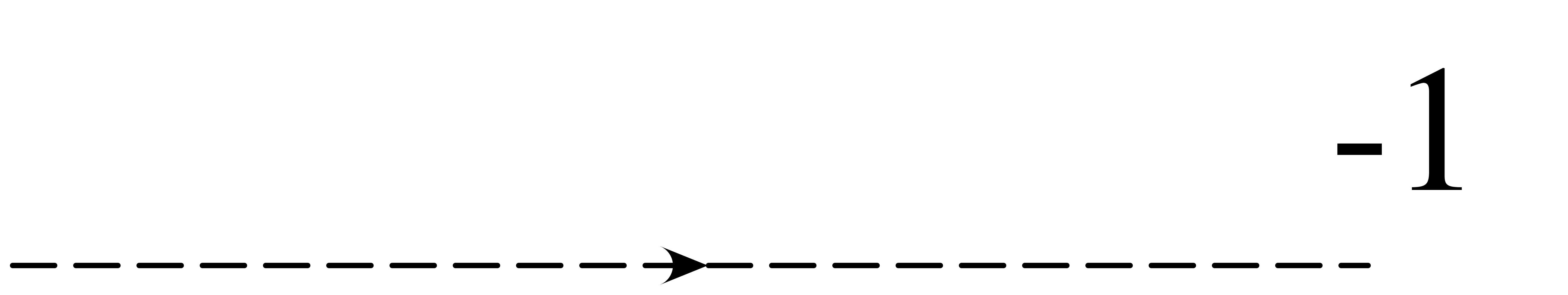} \vspace{1.2cm} \\
 scalar propagator &\includegraphics[width=2cm]{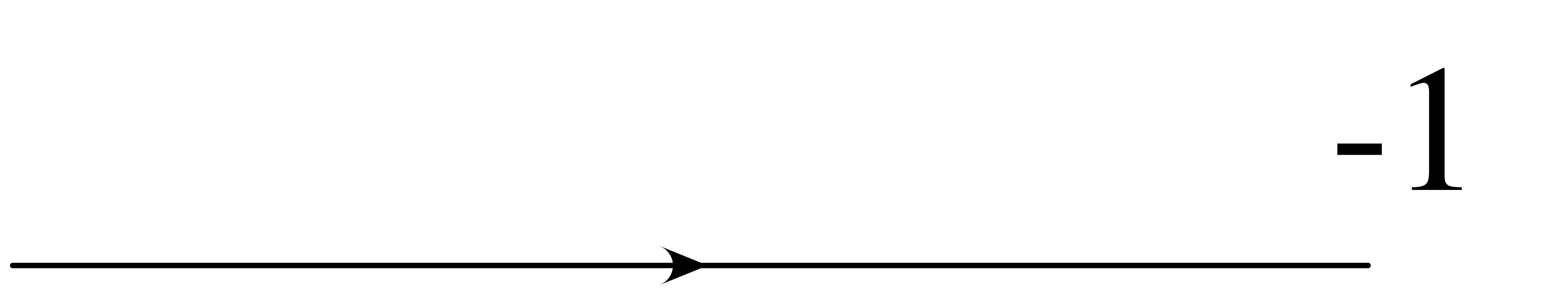}  & ghost-gluon vertex &   \vspace{-1.2cm} \includegraphics[width=2cm]{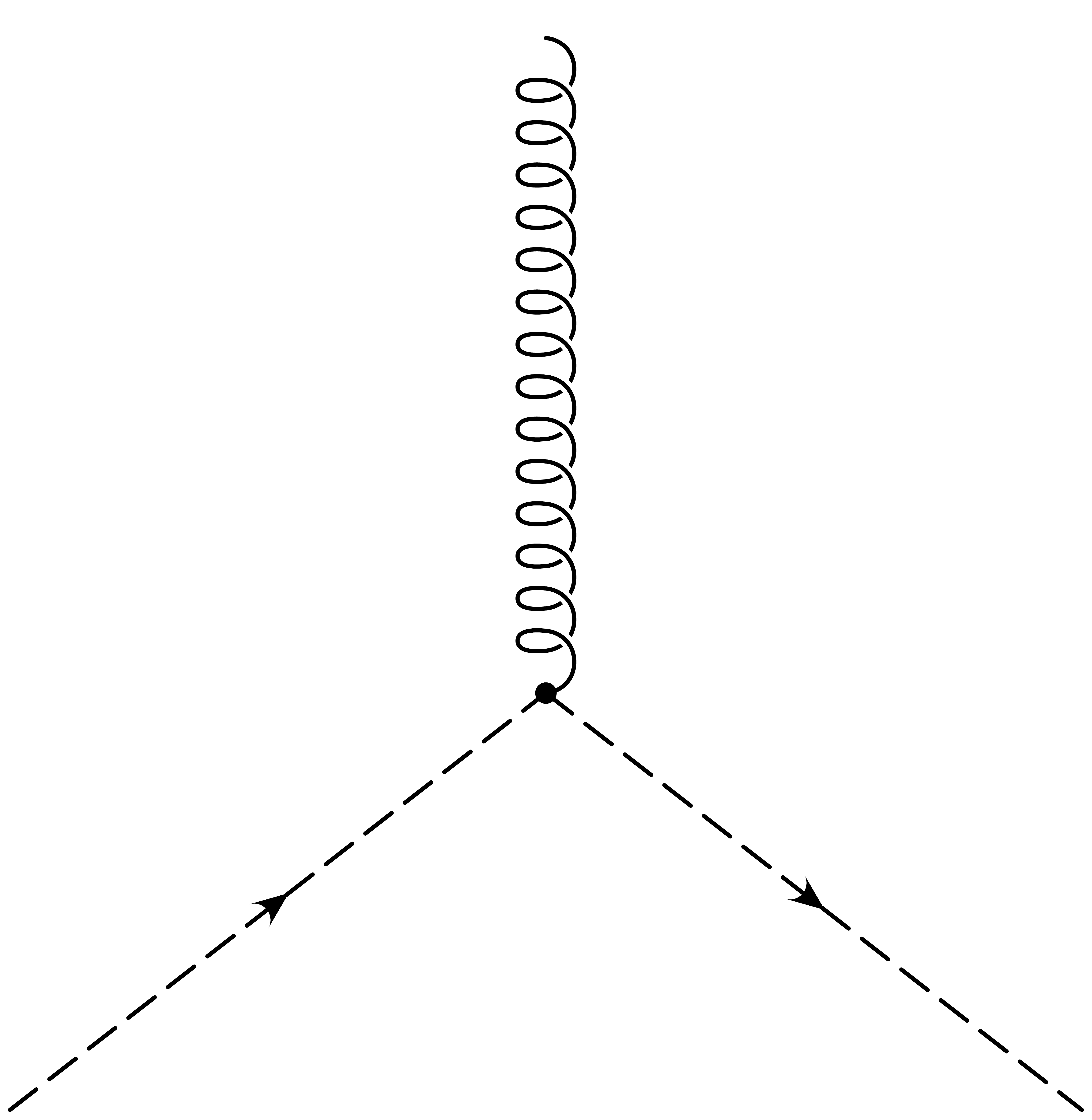} \vspace{1.2cm} \\
 scalar-gluon vertex & \vspace{-1.2cm} \includegraphics[width=2cm]{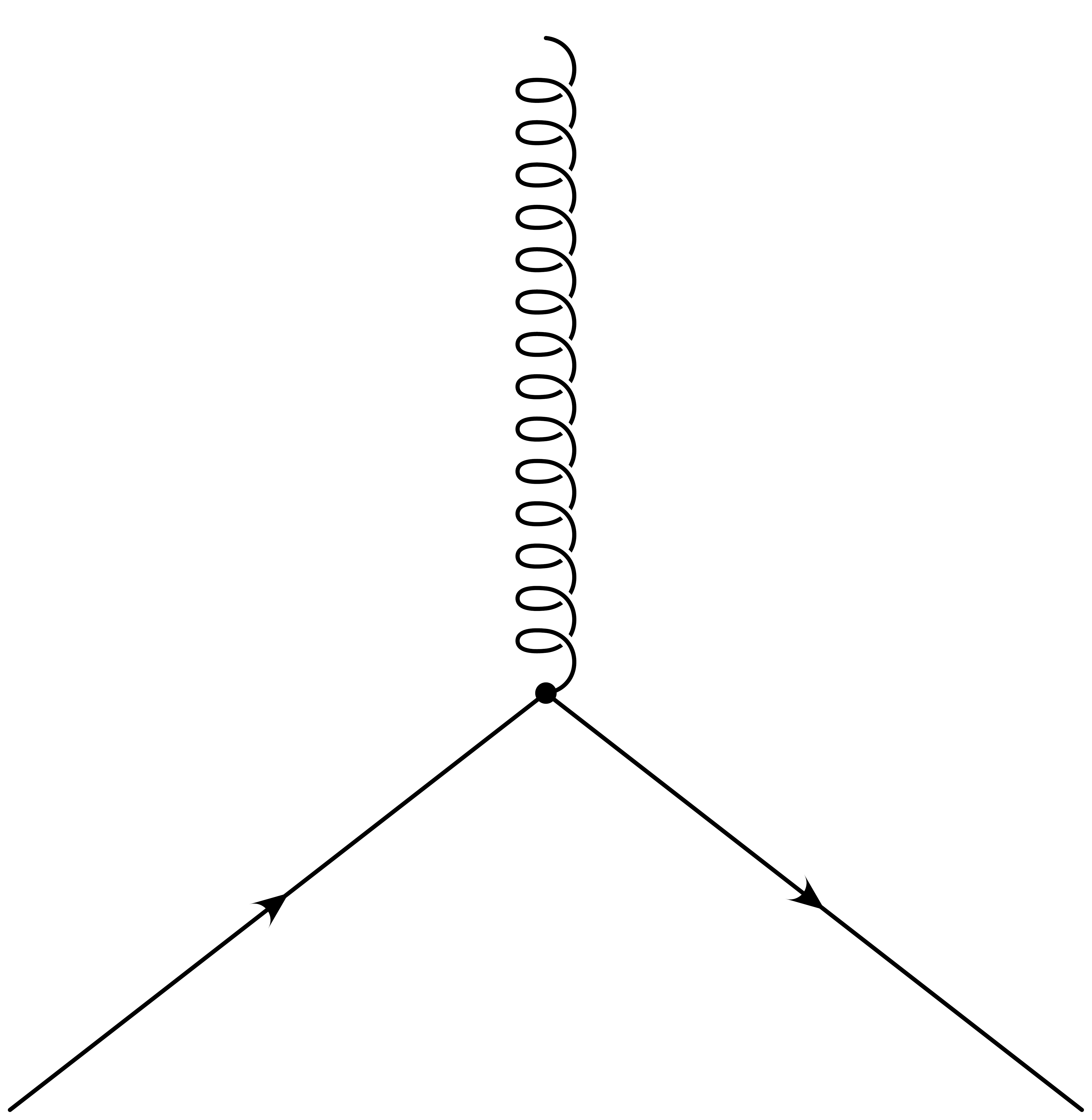} & three-gluon vertex &  \vspace{-1.2cm} \includegraphics[width=2cm]{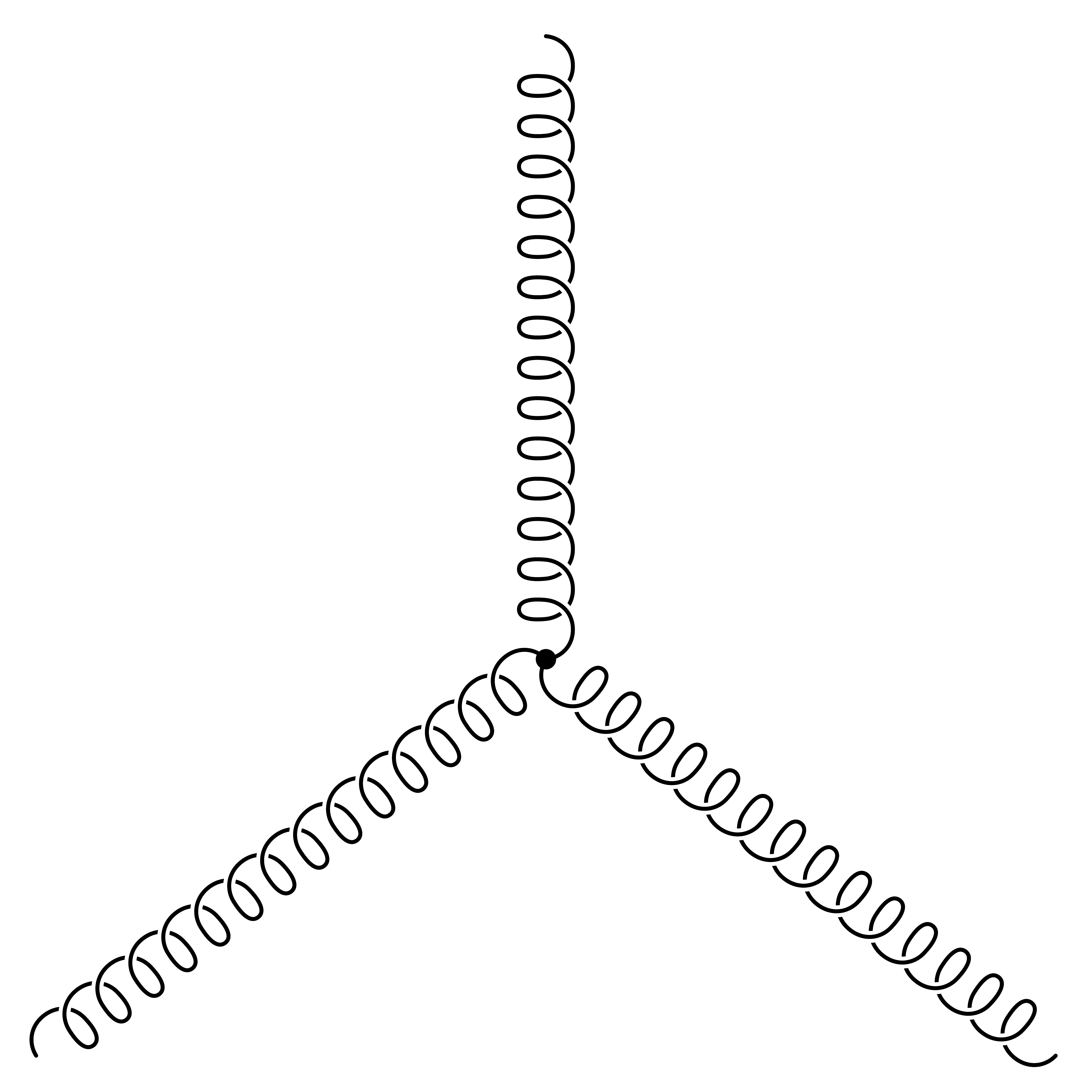} \vspace{1.2cm} \\
 four-gluon vertex & \vspace{-1.2cm} \includegraphics[width=2cm]{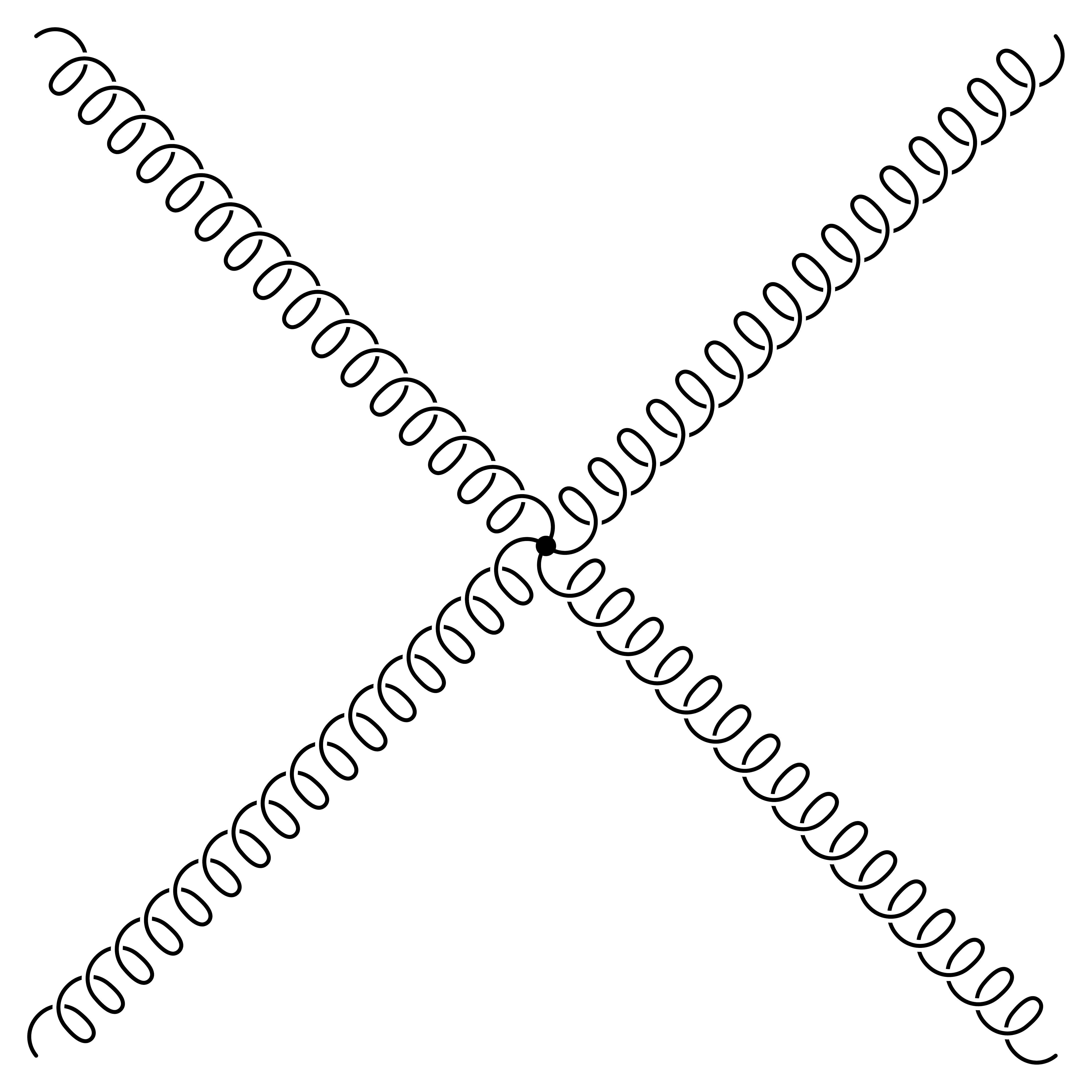} & four-scalar vertex & \vspace{-1.2cm} \includegraphics[width=2cm]{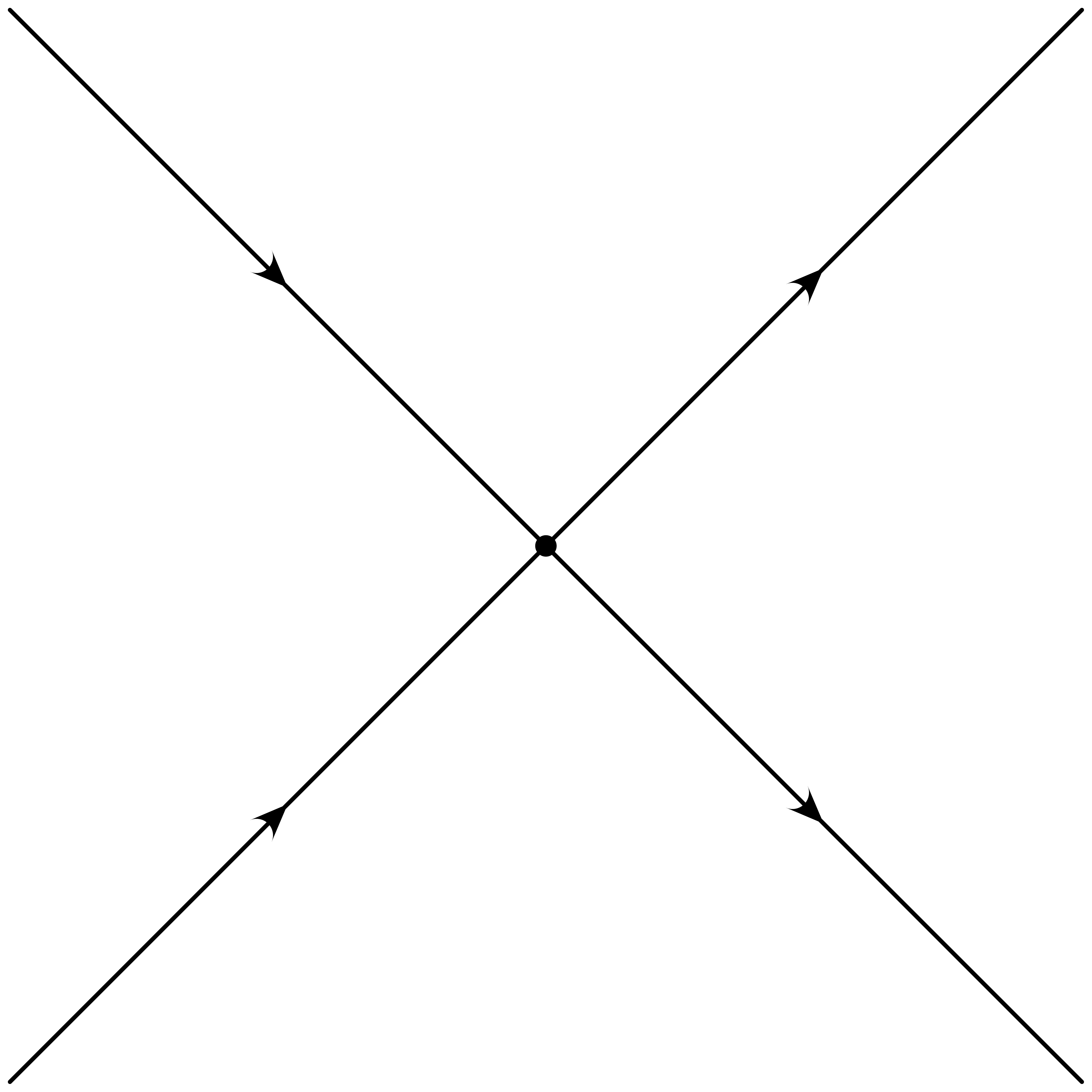} \vspace{1.2cm} \\
 two-scalar-two-gluon vertex  \vspace{0.4cm} & \vspace{-1.2cm} \includegraphics[width=2cm]{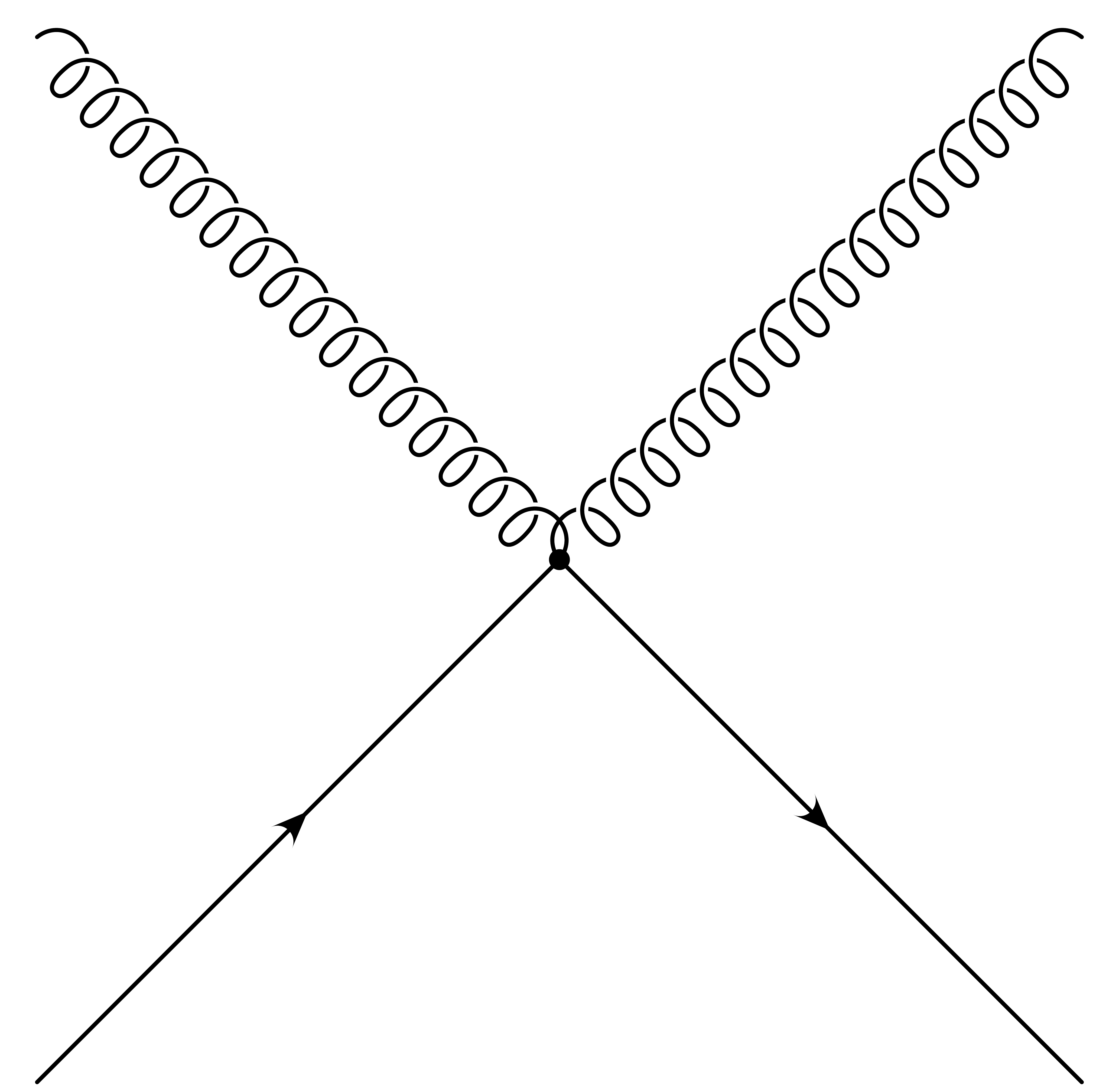}  \vspace{0.2cm}& \vspace{0.2cm} & \vspace{0.2cm} \\
\hline
\end{tabular}\label{tab_primdiv}
\end{table}
The DSE for the quark propagator\footnote{See \textit{e.g.} \cite{Alkofer:2000wg} for the derivation of the quark propagator DSE.} can be represented graphically, as given in fig. \ref{fig:quark_prop}. 
The analogous equation for the scalar propagator is complicated by the additional vertices. The scalar propagator\footnote{For the DSEs in this thesis I derived the various generating equations analytically and proceeded with the graphical method as explained above. After the publication of \cite{Alkofer:2008nt} the DSEs were rechecked with the \emph{Mathematica}-package \emph{DoDSE}.} DSE in fig. \ref{fig:scal_prop} has eight additional structures, each containing a bare two-scalar-two-gluon or four-scalar vertex.
Also in the gluon DSE additional vertices emerge, as can be seen in fig. \ref{fig:g_prop}. Nevertheless the ghost DSE\footnote{Note that in QCD the equations for the quark and the ghost are similar, which naturally does not hold for the scalar DSE.} is not affected by the additional interactions up to the graphic representation, see fig. \ref{fig:ghost_prop_text}.\\
The coupling of the scalars to the pure gauge sector is realised by the scalar-gluon and the two-scalar-two-gluon vertex. The corresponding DSE for the scalar-gluon vertex is given in fig. \ref{fig:scalar_gluon_vertex_DSE}, the DSE for the latter Green function can be found in App. \ref{DSEs}. \\
Note that only the full Green functions for the propagators are given, all the others are truncated in the following way. Due to the complexity of the DSEs stemming from the primitively divergent 4-point functions, only the one-loop diagrams are considered. The remaining system is skeleton expanded, thus there is another diagram in the DSE for the scalar-gluon vertex, involving the ghost-triangle and a gluon-\textquotedblleft box\textquotedblright. So this diagram is of two-loop order. It is crucial to take this graph into account, because there is no direct interaction between scalars and ghosts in a primitively divergent vertex. Thus there is no one-loop diagram that accounts for this interaction type, but in two-loop order, such a diagram can occur by interchanging gluons. Thus, leaving this two-loop diagram aside would restrict the possible solutions. Also in the two-scalar-two-gluon equation this aspect has to be considered. Further details to this will be given in chapter \ref{results}.\\
Already at this point I want to emphasise that this is a suitable truncation. The justification for this will be given below.

\hspace{1cm}
\section{Power Law Behaviour}\label{section:powerlawbehaviour}
\begin{figure}[!htb]
 \centering
 \includegraphics[width=5cm]{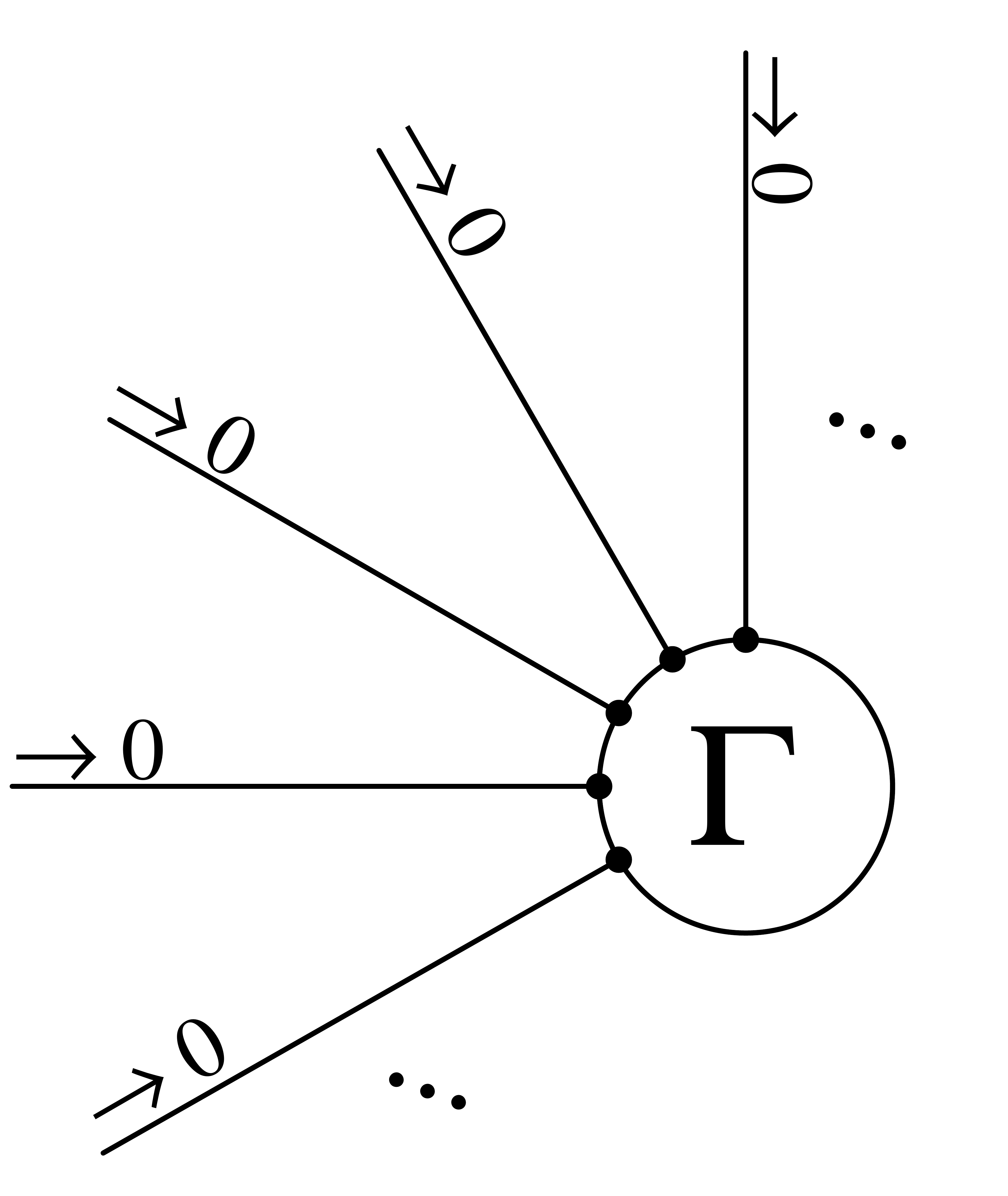}
 \caption{\small{Uniform limit for an arbitrary vertex function. Plain lines represent arbitrary in- or outgoing particles, the blob stands for an arbitrary vertex. There is only one vanishing scale in the infrared if all external momenta scale to zero.}}
 \label{fig:arbitrary}
\end{figure}

Considering the preceding Dyson-Schwinger equations it is obvious that they are part of an infinite tower of coupled integral equations. To find a solution for the vertices it seems as if one has to solve this coupled system. The authors of \cite{Alkofer:2004it} showed another possibility to give a statement about the qualitative infrared behaviour of the vertices. In their argumentation they pursue the following idea. On a classical level Yang-Mills theory is a scale-invariant theory, \textit{i.e.} the action is invariant under a dilatation of the coordinate system and a simultaneous transformation of the fields. For a dilatation parameter $\lambda$ the coordinate system is rescaled by $x\rightarrow \lambda x$, and the fields transform according to $A_{\mu}(x) \rightarrow \lambda A_{\mu}(\lambda x)$ for bosonic, and $\psi (x) \rightarrow \lambda^{\frac{3}{2}} \psi (\lambda x)$ for fermionic fields due to dimensional reasons. Therefore the action is invariant
$$ S = \int d^4 x \mathcal{L}_{YM} \rightarrow 	\int d^4 x \lambda^{4} \mathcal{L}_{YM}(\lambda x) = \int d^4 x' \mathcal{L}_{YM}(x') = S.$$
The classical scale invariance does not survive the process of quantization, \textit{i.e.} a dimensional parameter is generated, usually called $\Lambda_{QCD}$, which has the dimension of a momentum. The infrared region is the region where the relevant momenta are far below the momentum scale $\Lambda_{QCD}$. As distances are large it is a fair assumption due to renormalisation group arguments, that all vertex functions should be described as a power law. \\
This can be demonstrated by means of a general vertex function $\Gamma_{\mu \nu \rho \ldots}$, as given in eq. (\ref{eqVertexFunction}), given by a sum over the different tensor structures $T^{i}$ with the proper dressing functions $G_{i}$. The $p$'s denote the incoming or outgoing momenta in this vertex.
\begin{equation}\label{eqVertexFunction}
\Gamma_{\mu \nu \rho \ldots}(p_1,\ldots,p_n) \ = \ \sum_{i} G_i(p_1,\ldots,p_n) T_{\mu \nu \rho \ldots}^{i}(p_1,\ldots,p_n)
\end{equation}
For the dressing functions $G_{i}$ power law ans\"{a}tze are made as, cf \cite{Alkofer:2008jy},
$$G_i(p_1,\ldots,p_n) = \sum_{j} c_{i,j}\left( \frac{p_{1}^{2}}{q_{j}^{2}},\ldots,\frac{p_{n}^{2}}{q_{j}^{2}} \right) \ \big(q_{j}^{2}(p_1^2,\ldots,p_n^2)\big)^{\delta_{i,j}} .$$
The dressing functions thus scale as a power $\delta_i$ of a scaling variable $q_{j}$, which is a function of all external momenta, whereas the index $j$ denotes possibly different scaling limits. The prefactors $c_{i,j}$ must be constructed in such a way, that all vertex functions can be described by the same scaling variable. Note that all internal momenta in the integrals must transform into external momenta due to dimensional reasons, which can be seen in analytic solutions for two- and three-point functions \cite{Davydychev:1991va,Anastasiou:1999ui}. \\
The most obvious scaling behaviour is the uniform limit. Uniform scaling describes the fact that for a given Green function there is only one scaling variable that vanishes if and only if all external momenta scale to zero uniformly, as illustrated in fig. \ref{fig:arbitrary}. This limit is therefore defined as
$$ q^{2}(p_1^2,\ldots,p_n^2) \rightarrow 0 \Leftrightarrow p_1,\ldots,p_n \rightarrow 0 \ \wedge \ \frac{p_1^2}{q^2},\ldots ,\frac{p_n^2}{q^2} \ \textnormal{constant}.$$
\\
In principle the singularities that occur when all external momenta vanish are not the only ones that can occur. There can also be divergencies, if only a subset of the external momenta scales to zero (so-called soft momenta), while others stay finite (referred to as hard momenta). Note that hard momenta are not necessarily smaller than $\Lambda_{QCD}$ to ensure the power law behaviour. 
\\

In the Dyson-Schwinger equations these ans\"{a}tze make it possible to give a statement about the qualitative infrared behaviour of the theory. By inserting the dressing functions and the tensor structures in the DSEs for the nine primitively divergent vertex functions a system of equations for the infrared exponents can be extracted. Although an arbitrary DSE contains many different tensor structures, it is not the goal of the power counting analysis to find the scaling behaviour of the various tensors, but rather to give a statement about the scaling of the whole vertex. Thus one aims at finding the infrared exponent of the IR dominating tensor, although this analysis does not determine which tensor is the leading one. Solving the system of coupled equations for the infrared exponents yields the infrared scaling behaviour of all vertex functions.\\
The parametrization of the dressing functions can be done arbitrarily, because the scaling exponents are independent of the parametrization. The parametrization in this work is done with respect to the bare vertices, which can be found in App. \ref{app_bare}. The dressed vertex is parametrised by a product of the bare quantity with a dressing function. Thus the system of equations involves canonical dimensions, \textit{i.e.} momentum powers of the bare vertices, and anomalous dimensions, that describe the scaling of the dressing function.\\
To give some examples for the parametrization, the propagators for the scalar particle $S_{ij}$, gluon $D_{\mu \nu}$ and ghost $D^{G}$, with the corresponding dressing functions $B_s$, $Z$ and $G$, working in Euclidean momentum space and Landau gauge, \textit{i.e.} taking the limit for the Feynman parameter $\zeta \rightarrow 0$ in the Lagrangian, are shown in eq. (\ref{props}). Similar parametrizations are made for the other primitively divergent vertex functions.
\begin{equation} \label{props}
S_{ij}(p^2) = - \delta_{ij} \frac{B_s(p^2)}{p^2}  \ , \ D_{\mu \nu}(p^2)= \left( \delta_{\mu \nu} -\frac{p_{\mu} p_{\nu}}{p^2} \right)  \frac{Z(p^2)}{p^2}  \ , \ D^{G}(p^2)=-\frac{G(p^2)}{p^2} .
\end{equation}
\\
The abbrevations for the anomalous dimensions in the uniform limit are chosen as given in table \ref{dimensions}, motivated by the nomenclature in \cite{Alkofer:2008jy}. Further kinematic cases, where only a subset of external momenta vanishes are denoted by additional superscripts, that symbolise those external momenta that tend to zero, whereas the remaining ones stay finite. 
\begin{table}[!htb]
\caption{\small{Infrared exponents in the uniform limit of the primitively divergent $n$-point functions.}}
\begin{center}
\begin{tabular}[t]{|l|p{2cm}||l|p{2cm}|} 
\hline
\small{$n$-point function} & \small{anomalous dimension} & \small{$n$-point function }& \small{anomalous dimension} \\
\hline
\small{scalar propagator} & \small{$\delta_s $} & \small{gluon propagator}& \small{$\delta_g $}\\
\small{ghost propagator} & \small{$\delta_{gh} $ }& \small{scalar-gluon vertex} & \small{$\delta_{sg}$}\\
\small{ghost-gluon vertex} &\small{$\delta_{ggh} $} & \small{three-gluon vertex} & \small{$\delta_{3g} $}\\
\small{four-gluon vertex} & \small{$\delta_{4g} $ }& \small{four-scalar vertex} & \small{$ \delta_{4s} $}\\
\small{two-scalar-two-gluon vertex} & \small{$\delta_{ssgg}$} & & \\
\hline
\end{tabular} \label{dimensions}
\end{center}
\end{table}

\chapter{Results}\label{results}
In this chapter an infrared power counting analysis for the previously derived Dyson-Schwinger equations is done for a massless as well as a massive fundamentally charged scalar particle in Yang-Mills theory. The uniform limit is calculated for the full system without any truncation. In the case of a massive scalar the importance of kinematic divergencies becomes obvious, whose consequences for the scaling behaviour of the system are presented in the last section of this chapter. This analysis yields self-consistent solutions for the exponents of the leading tensor structures in each Dyson-Schwinger equation.

\section{Uniform Scaling}
\begin{figure}[ht]
\centering
\includegraphics[width=5cm]{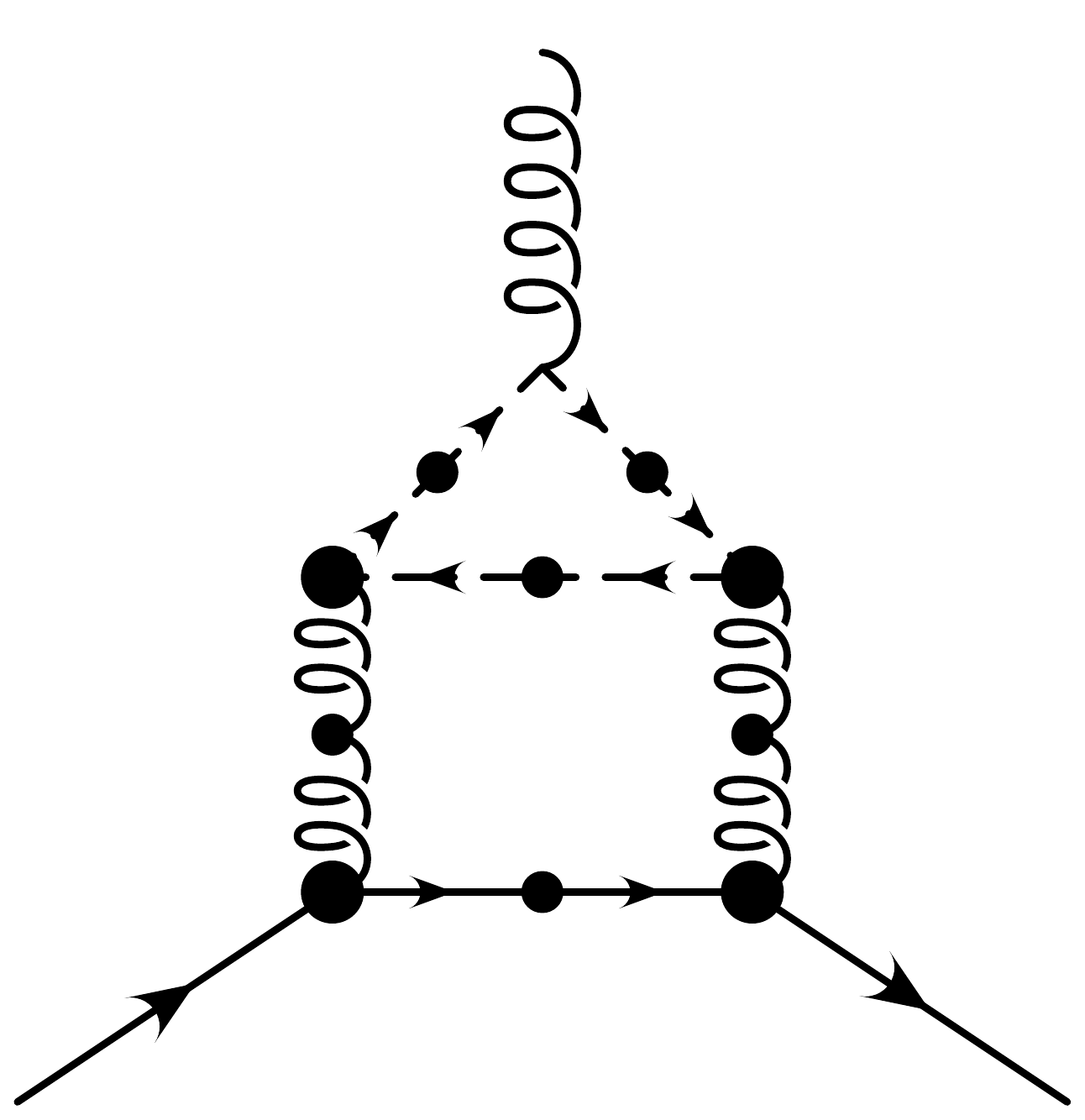}
\caption{\small{Ghost contribution to the scalar-gluon vertex.}}
\label{fig:bsp}
\end{figure}

In the Dyson-Schwinger equations for the $n$-point functions the powers of the particular diagrams are made up by their various constituents, whose powers are counted according to the ans\"{a}tze above. The sum of all contributing exponents of the $n$-point functions and integrals is therefore the infrared exponent of the diagram. The easiest way to explain this procedure may be by the help of an example, given in figure \ref{fig:bsp}, which is a diagram in the skeleton expansion of the scalar-gluon vertex DSE. In figure \ref{fig:bsp} the blobs of the internal dressed propagators are drawn explicitly to remind of their contribution in the vertex.
\\
The power counting proceeds as follows: Within the diagram in fig. \ref{fig:bsp} there are two loops. This means, that in the integral equation there are two momentum integrals for the exchanged momentum within the loop. In four dimensions one hence gets an additional momentum power of $4$ from each loop-integral, \textit{i.e.} an additional exponent of $4$. The next contributing parts are the propagators, and due to their parametrization, see eq. (\ref{props}), they account for $(p^2)^{-1+\delta_{s}}, \ (p^2)^{-1+\delta_{g}}$ and $ \ (p^2)^{-1+\delta_{gh}}$. The vertex powers depend on whether the vertex is dressed or bare. By construction of the DSEs there is exactly one bare vertex in each term of the equation, and consequently in each diagram too. The bare vertex has no anomalous dimension but contributes with its canonical dimension. With regard to App. \ref{app_bare} the parametrization of the exponents is chosen such, that the canonical dimension for the scalar-gluon-, ghost-gluon- and three-gluon vertex equals $\frac{1}{2}$. In the diagram in fig. \ref{fig:bsp} there are three ghost-, two gluon- and one scalar-propagator, two scalar-gluon and two ghost-gluon vertices. Counting the appropriate parts yields
\begin{small}
\begin{eqnarray*}
\Big(\! \big(p^2\big)^{2}\! \Big)^{2}
\big(p^2\big)^{-1+\delta_{s}}
\Big(\! \big(p^2\big)^{-1+\delta_{g}} \!\Big)^{2}
\Big(\! \big(p^2\big)^{-1+\delta_{gh}} \!\Big)^{3}
\big(p^2\big)^{\tfrac{1}{2}}
\Big(\! \big(p^2\big)^{\tfrac{1}{2}+\delta_{ggh}}\!\Big)^{2}
\Big(\! \big(p^2\big)^{\tfrac{1}{2}+\delta_{sg}}\!\Big)^{2}.
\end{eqnarray*}
\end{small}\\
Regarding only the exponents of the constituents of fig. \ref{fig:bsp} leaves a sum of momentum powers
\begin{eqnarray*}
\underbrace{4}_{\parbox{0.7cm}{\tiny loop-\\integ.}}\underbrace{-\!1\!+\!\delta_s}_{\parbox{1.3cm}{\tiny scalar\\propagator}}\!
+\! \underbrace{2(\!-\!1\!+\!\delta_{g}\!)}_{\parbox{1.7cm}{\tiny gluon \\propagators}}\!
+\! \underbrace{ 3(\!-\!1\!+\!\delta_{gh}\!)}_{\parbox{1.7cm}{\tiny ghost\\propagators}}
 +\underbrace{\tfrac{1}{2}}_{\parbox{1.5cm}{\tiny bare ghost-\\gluon vertex}}
+ \underbrace{\!2(\tfrac{1}{2}\!+\!\delta_{ggh}\!)}_{\parbox{1.8cm}{\tiny dressed ghost-\\gluon vertices}}
 + \underbrace{\!2(\tfrac{1}{2}\!+\!\delta_{sg}\!)}_{\parbox{1.9cm}{\tiny scalar-gluon \\vertices}}\!
=\\= \tfrac{1}{2}+\delta_{s}+2\delta_{g}+3\delta_{gh}+2\delta_{ggh}+2\delta_{sg}. 
\end{eqnarray*}
\\
In the same way the diagrams in the DSEs in Appendix \ref{DSEs} are counted. \\
To be the most singular and thus leading vertex function in the infrared requires that the order of the singularity is the highest of all, and thus the exponent of interest has to be the smallest of all tensor structures in the DSE. 
\\
Proceeding with the analysis for the theory including fundamental scalar charges, already in the power counting analysis three qualitatively different cases have to be distinguished \footnote{It is not fair to speak about real \textquotedblleft phases\textquotedblleft \  of the system, because according to the Fradkin-Shenker-Osterwalder-Seiler theorem \cite{Osterwalder:1977pc,Fradkin:1978dv} there is no thermodynamic phase transition between the confinement and Higgs-phases. Thus the word \textquotedblleft phases\textquotedblleft \ rather means regions in this context.}: the first one is an idealised case that the scalar field has no mass initially and no mass is generated either. \\
The second possibility is that the scalar particle carries a mass (which in general depends on the momentum). Here another distinction has to be done, where after a gauge fixing the residual global symmetry is considered. In the Higgs phase the remnant symmetry is spontaneously broken. For a detailed investigation of this remnant symmetry see  \cite{Greensite:2004ke}. A suitable quantity for the distinction of the two phases, wherein the gauge has been fixed to Landau gauge \cite{Langfeld:2002ic, Langfeld:2004vu} is the vacuum expectation value of
\begin{equation}\label{order_parameter}
 Q=\left( \int d^4 x \phi \right) \left( \int d^4 x \phi^{\dagger} \right).
\end{equation}
In the confinement, \textit{i.e.} unbroken phase $\langle Q \rangle$ vanishes, whereas it approaches a finite value in the Higgs, \textit{i.e.} broken phase. \\
All three phases have different properties, thus each case has to be analysed separately. They can be specified by the parameters:
\begin{itemize}
 \item{massless scalar particle $m=0$ (and  $\langle Q\rangle=0$)}
 \item{massive scalar particle $m\neq0$, but unbroken phase $\langle Q\rangle=0$}
 \item{massive scalar particle $m\neq0$, spontaneously broken phase $\langle Q\rangle\neq0 $ (Higgs phase)}
\end{itemize}
In this work the massless and the massive but unbroken cases are studied. For these two cases the DSEs are topologically equal, so the diagrammatic representations in the Appendix \ref{DSEs} are valid for both cases. This is no longer true, if there is a scalar condensate, because then new primitely divergent vertices can emerge.\\

\begin{figure}
 \centering
 \includegraphics[width=13.8cm]{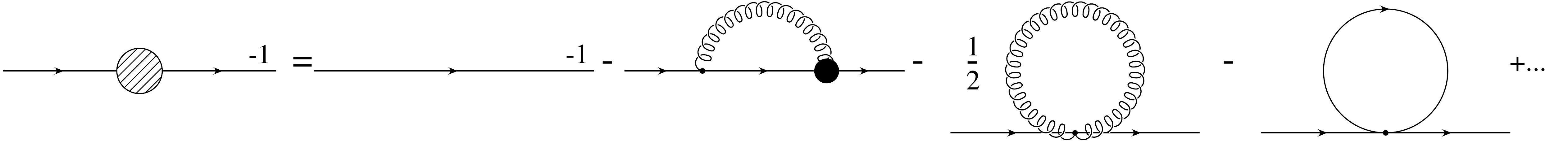}
\caption{\small{Truncated DSE for the scalar propagator.}}
 \label{fig:scalar_prop_oneloop}
\end{figure}

The calculation for the systems of equations for the massless and the massive scalar are similar for the determination of some infrared exponents. I will determine the various exponents for both cases simultaneously and then analyse the differences of the two systems in the next subsections separately. \\
The diagrammatic form of the DSEs is the same in both cases. Due to a similar parametrization, see eq. (\ref{props}), of the dressed scalar propagators, only the term for the bare propagator is counted differently. Applying the truncation from above, the DSE for the scalar propagator simplifies to the terms given in fig. \ref{fig:scalar_prop_oneloop}.

\begin{figure}[!htb]
 \centering
 \includegraphics[width=9cm]{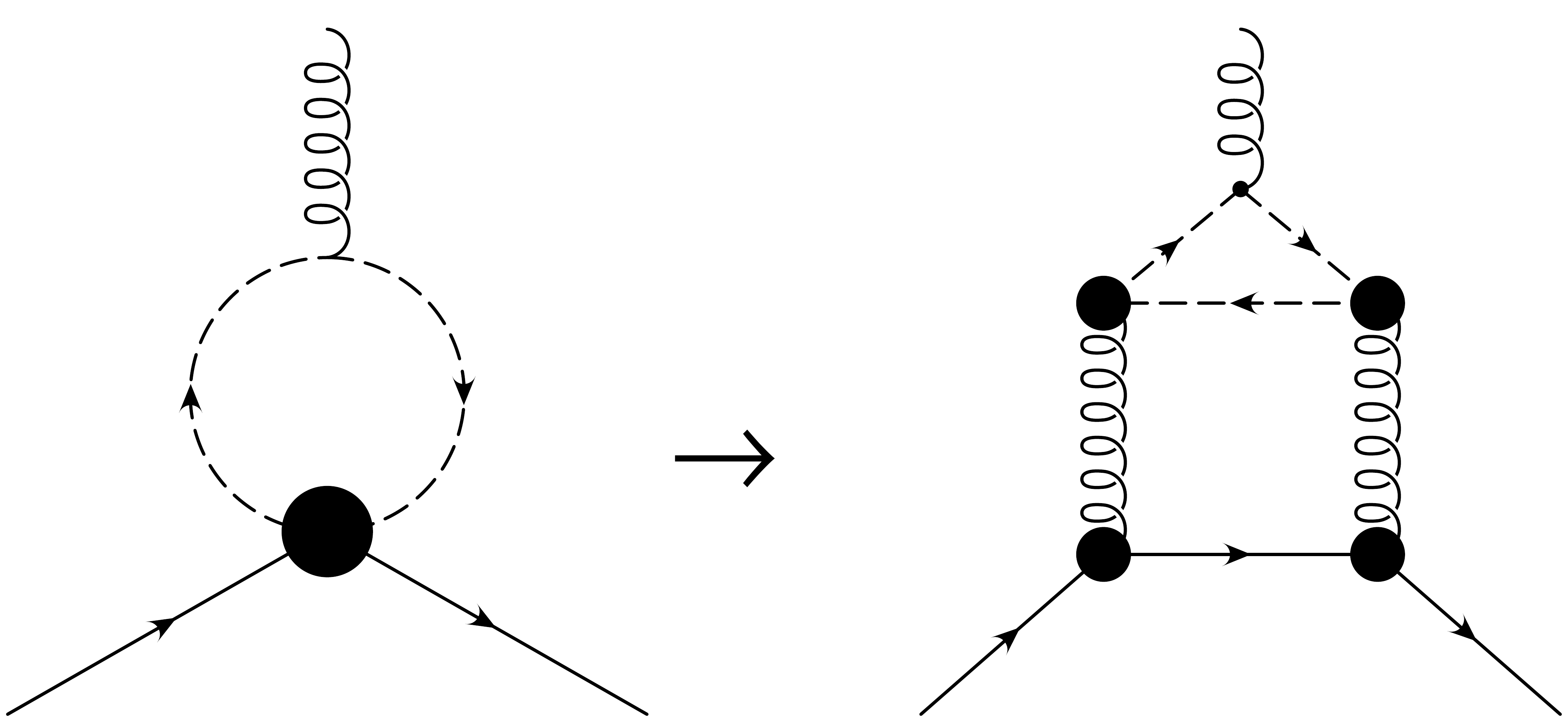}
 \caption{\small{The leading ghost contribution in the DSE for the scalar-gluon vertex. This two-loop diagram has to be included into the one-loop truncation, to give rise to a possible ghost-dominance.}}
 \label{gh_box}
\end{figure}

\begin{figure}[!htb]
 \centering
 \includegraphics[width=13.8cm]{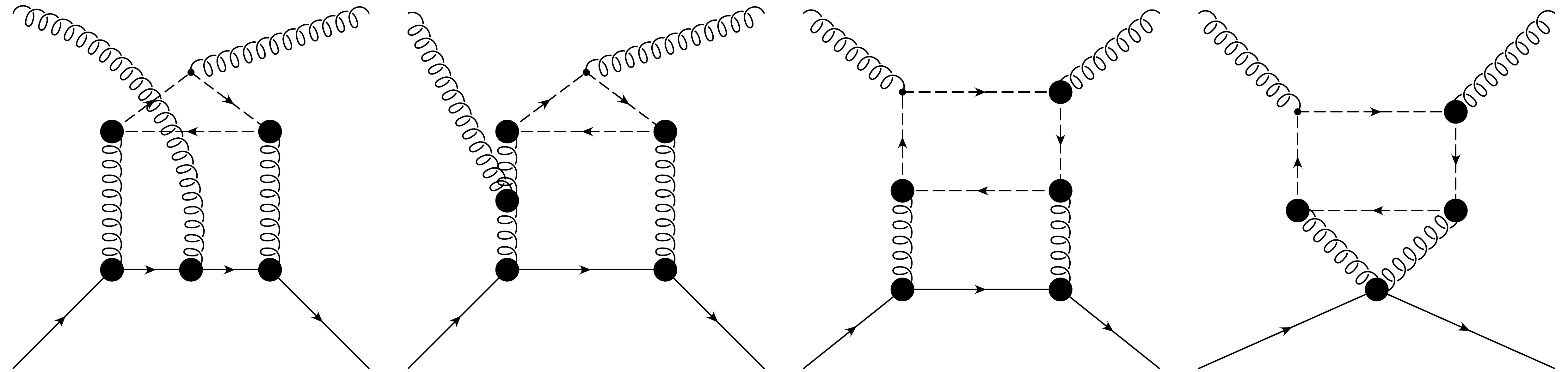}
 \caption{\small{Leading graphs of the expansion of the two-scalar-two-ghost interaction in the 2-scalar-2-gluon vertex DSE.}}
 \label{fig:gh_2s2g}
\end{figure}

\begin{figure}[ht]
 \centering
 \includegraphics[width = 13.8cm]{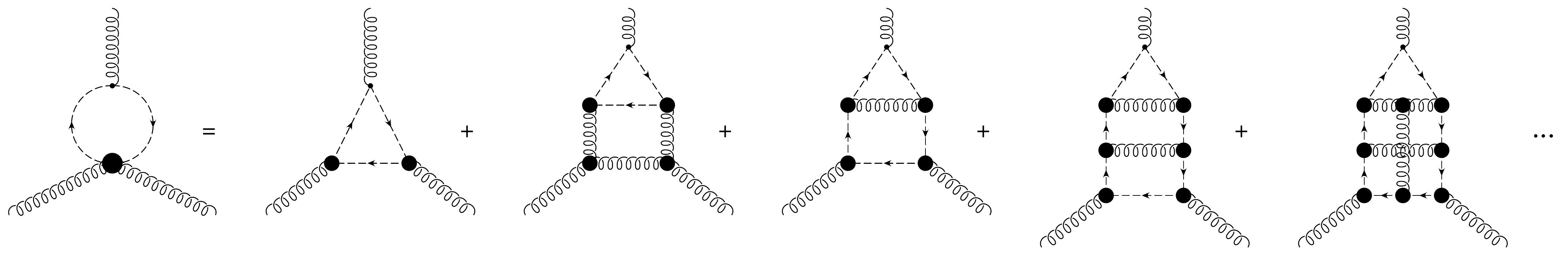}
 \caption{\small{Skeleton expansion of the 3-gluon vertex DSE.}}
 \label{fig:sg_skel_exp}
\end{figure}

\begin{figure}[ht]
 \centering
 \includegraphics[width=13.8cm]{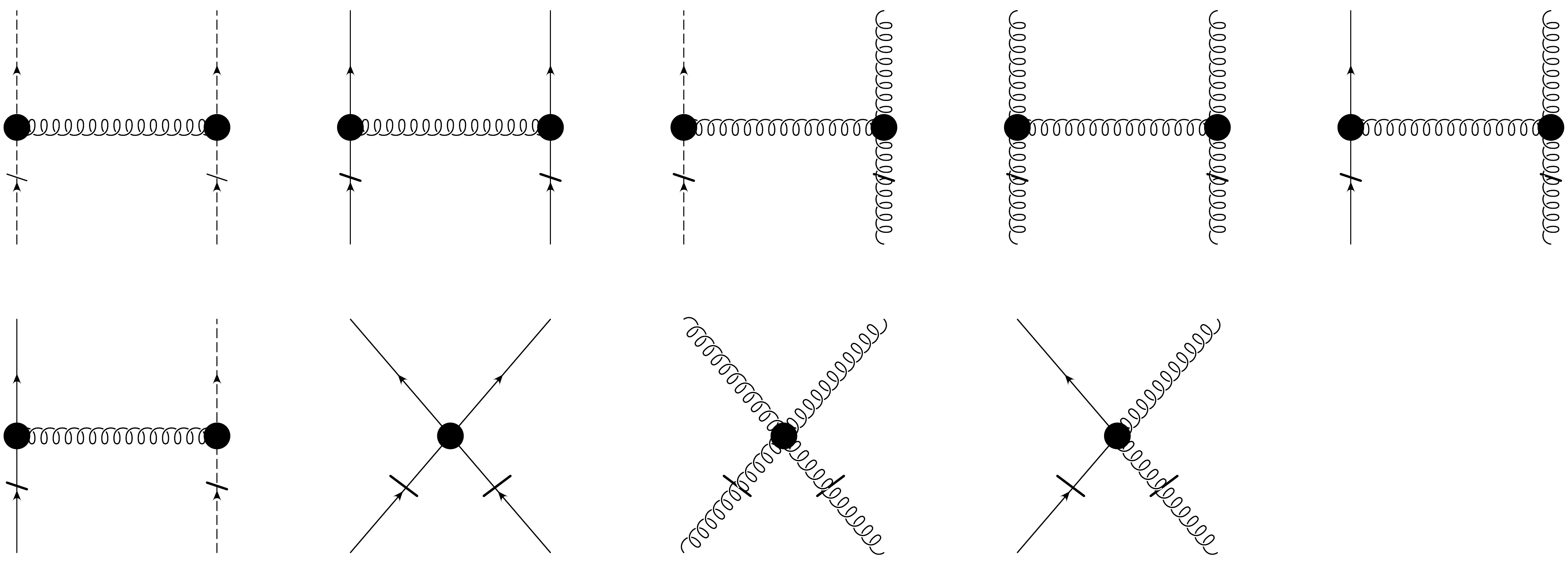}
 \caption{\small{Skeleton expansion elements for given diagrams that raise the loop-order to generate arbitrary graphs.}}
 \label{fig:skel_exp}
\end{figure}

\begin{figure}[!htb]
 \centering
 \includegraphics[width=8cm]{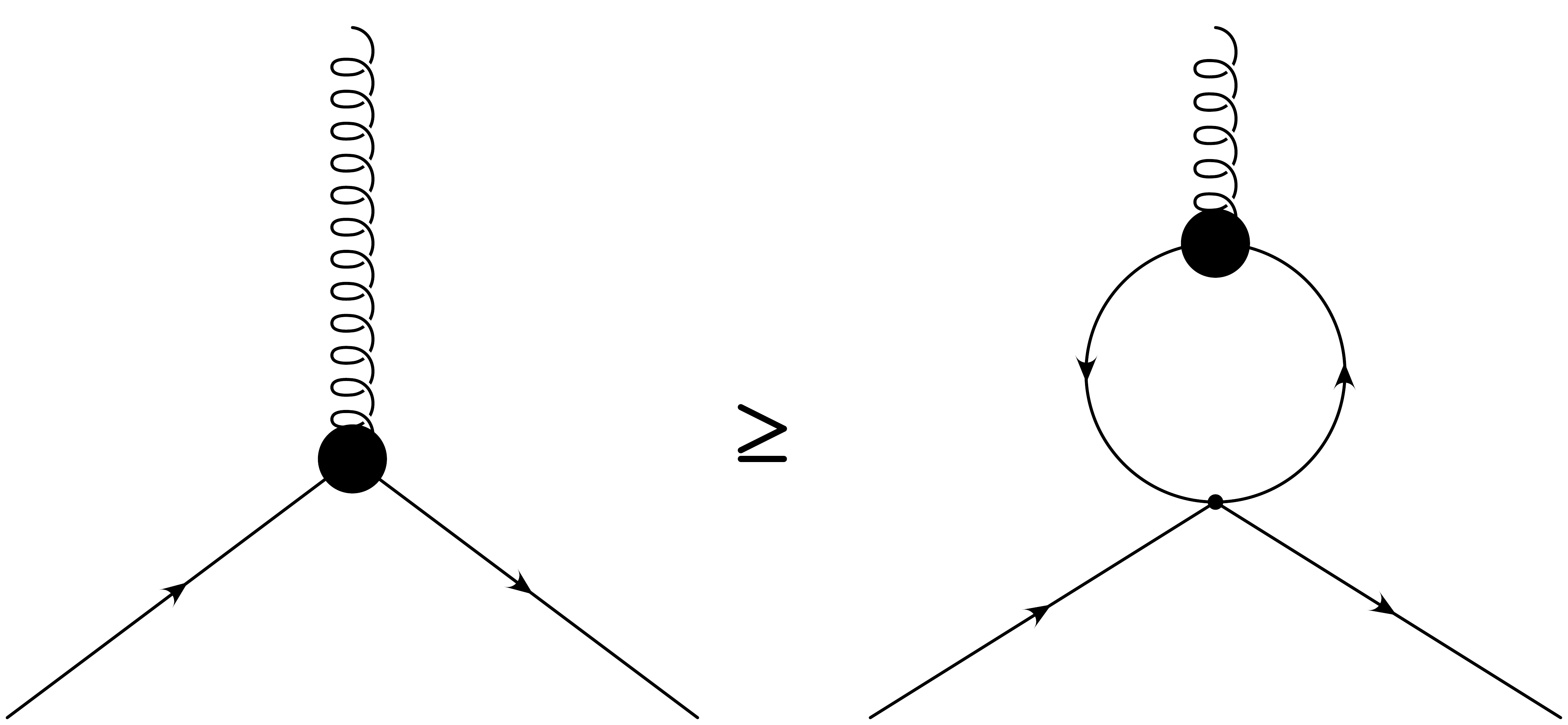}
 \caption{\small{Constraint for the anomalous dimension of the scalar propagator.}}
 \label{fig:dg_det}
\end{figure}

The bare scalar propagator for the massless case is singular, as it scales as $(p^2)^{-1}$
\begin{equation}\label{bare_massless_scalar_prop}
 S^0_{ij}(p^2) = -  \frac{\delta_{ij}}{p^2+m^2} \ \xrightarrow{m^2\rightarrow 0} \ \ - \frac{\delta_{ij}}{p^2} \propto \big( p^2 \big)^{-1}.
\end{equation}
For a finite mass $m^2>0$ there is scaling of the bare scalar propagator $\sim (p^2)^{0}$, as it freezes out at a finite value
\begin{equation}\label{bare_massive_scalar_prop}
 S^0_{ij}(p^2) = -  \frac{\delta_{ij}}{p^2+m^2} \ 
\stackrel{p^2\rightarrow 0}{\longrightarrow}
 \ \ - \frac{\delta_{ij}}{m^2} \propto \big( p^2 \big)^{0}.
\end{equation}
As it will turn out in the calculation it is convenient to define a new quantity $\mu$, that denotes the canonical dimension of the bare scalar propagator in the infrared. The equations (\ref{bare_massless_scalar_prop}) and (\ref{bare_massive_scalar_prop}) suggest the definition of $\mu$ as
\begin{equation}
 \mu=\left\{\begin{array}{cl} 0 & \mbox{for a massless scalar: }m^2=0\\ 1 & \mbox{for a massive scalar: }m^2>0 \end{array}\right. 
\end{equation}
Using the quantity $\mu$, both cases can be written in one equation (\ref{scalar_prop_both})
\begin{eqnarray}\label{scalar_prop_both}
 1-\delta_{s} = & \min \{ 1-\mu, \ 1+\delta_{s}+\delta_{g}+\delta_{sg}, \ 1+\delta_{g}, \ 1+\delta_{s} \} \nonumber \\
 -\delta_{s} = &  \min \{ -\mu, \ \delta_{s}+\delta_{g}+\delta_{sg}, \ \delta_{g}, \ \delta_{s} \}
\end{eqnarray}
I want to emphasise that for a massless scalar particle, and as there are no other scales than the external momentum, in the coupled system of equations for the infrared exponents the canonical dimensions can be subtracted and cancel exactly. Thus only anomalous dimensions appear in the solutions for the infrared exponents. This is no longer true for the massive case, where the canonical dimensions remain in the equations and are involved non-trivially in the solutions, which will become obvious in the subsections about the different cases.\\
Note that due to the parametrization chosen above in eq. (\ref{props}), all the other equations (that do not contain the bare scalar propagator) contain the same terms\footnote{In the whole thesis the notations for the infrared exponents for the cases of a massless and a massive scalar are equal. Due to the sectional separation it will be clear in the following which case is dealed with.}. Thus the system of equations for the infrared exponents of scalar Yang-Mills theory can be set up by counting the contributing graphs in the truncated DSEs\footnote{I want to emphasise at this point that the validity of the obtained solutions was also checked for the full system of Green functions involving also two-loop diagrams.}, which are given in App. \ref{DSEs}. The system is given in App. \ref{App_full_eqn}.\\
An important point in one-loop order is to consider the ghost contribution correctly. This means, that in the DSEs for the scalar-gluon and the two-scalar-two-gluon vertex the diagrams have to be taken into account, that contain a two-scalar-two-ghost vertex. As can be seen in the Lagrangian there is no direct scalar-ghost interaction, but naturally they can interact by interchanging gluons. Thus this vertex has to be expanded in primitively divergent vertices. \\
Since at one-loop order this vertex does not exist, the leading contributions of ghosts have to be diagrams, that interchange gluons between ghosts and scalar particles. These two-loop diagrams stem from a skeleton expansion of the scalar-gluon scattering kernel. It is important to consider a possible ghost-dominance, as it is observed in Yang-Mills theory and quenched QCD. These two-loop diagrams must be considered in the truncation scheme. In fig. \ref{gh_box} and fig. \ref{fig:gh_2s2g} the graphs to be included are given explicitly.
\\
As can be seen in App. \ref{App_full_eqn} the system for the infrared exponents is quite extensive. To determine the infrared exponents of the system of equations in App. \ref{App_full_eqn}, a simplification of the various equations is possible, that is based on the assumption of a stable skeleton expansion. The skeleton expansion is an instruction how to generate higher loop-order diagrams from first order graphs by several replacement rules. Figure \ref{fig:sg_skel_exp} shows possible graphs in the expansions of the 3-gluon vertex DSE, that can be produced by insertions given in figure \ref{fig:skel_exp}. 
\\
According to this the idea that is used to reduce the system is that higher orders in the skeleton expansion should not become more infrared divergent than the first order\footnote{Note that this holds for the assumption, that the prefactors of the diagrams do not vanish nor that they combine such that different diagrams cancel exactly.}, cf. \cite{vonSmekal:1997is}. Otherwise continuous insertions would raise the order of the singularity to an arbitrary high value. Hence the possible insertions into first order and consequently higher order graphs yield several constraints, because the sum of all contributing exponents must be greater or equal zero. \\
Another subtlety is that although higher order diagrams are of the same order of divergence as the first order diagrams, the series involving all terms may diverge. Thus a convergent series of all contributing diagrams is assumed.\\
The insertions for a skeleton expansion are shown in figure \ref{fig:skel_exp}.
The constraints for the infrared exponents are obtained by counting the IR exponents of the extensions in figure \ref{fig:skel_exp}. The crossed out propagators are already part of the initial diagram, \textit{i.e.} they must not be counted in the constraints. An insertion raises the order of loops, therefore another exponent representing the momentum-integral has to be taken into account in the constraints. 
 \begin{figure}[!htb]
 \centering
 \includegraphics[width=3cm]{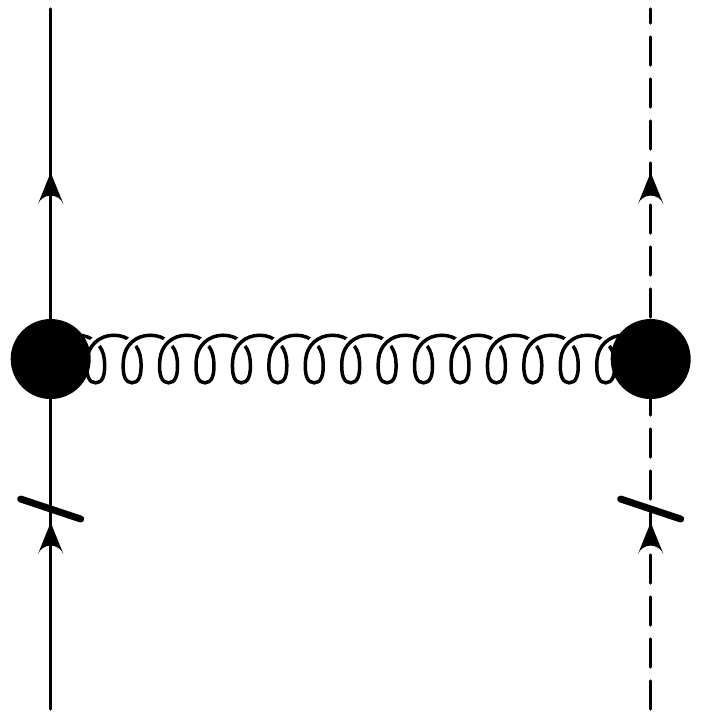}
 \caption{\small{Exemplary graph from the skeleton expansion extensions.}}
 \label{fig:skel_exp_s_g_gh}
\end{figure}
\\
A power counting of the extension in fig. \ref{fig:skel_exp_s_g_gh} yields
\begin{equation*}
\big(p^2\big)^{2} 
\big(p^2\big)^{-1+\delta_{s}} 
\big(p^2\big)^{-1+\delta_{g}}
\big(p^2\big)^{-1+\delta_{gh}}
\big(p^2\big)^{\frac{1}{2}+\delta_{ggh}}
\big(p^2\big)^{\frac{1}{2}+\delta_{sg}}.
\end{equation*}
Thus the exemplary insertion in fig. \ref{fig:skel_exp_s_g_gh} yields an inequality, which serves as a constraint for the infrared exponents:
\begin{eqnarray}\label{eq_skel_exp_s_g_gh}
 \underbrace{2}_{\parbox{1.5cm}{\tiny additional loop-integral}}
\underbrace{-1+\delta_{s}}_{\parbox{1.5cm}{\tiny ghost propagator}}
\underbrace{-1+\delta_{g}}_{\parbox{1.5cm}{\tiny gluon propagator}} 
\underbrace{-1+\delta_{gh}}_{\parbox{1.5cm}{\tiny ghost propagator}}+
\underbrace{\frac{1}{2}+\delta_{sg}}_{\parbox{1.5cm}{\tiny scalar-gluon vertex}}+
\underbrace{\frac{1}{2}+\delta_{ggh}}_{\parbox{1.5cm}{\tiny ghost-gluon vertex}}=
\nonumber \\
=  \delta_{s}+\delta_{g}+\delta_{gh}+\delta_{sg}+\delta_{ggh} \geq 0
\end{eqnarray}
\\ 
Counting also the remaining extensions yields nine constraints:
\begin{align}
&\delta_{g}+2\delta_{gh}+2\delta_{ggh}  \geq 0 \label{firstconstr}\\
&2\delta_{s}+\delta_{g}+2\delta_{sg} \geq 0 \label{constr_2s_g_2sg} \\
&2\delta_{g}+\delta_{gh}+\delta_{3g}+\delta_{ggh} \geq 0 \\
&3\delta_{g}+2\delta_{3g} \geq 0 \\
&\delta_{s}+2\delta_{g}+\delta_{sg}+\delta_{3g} \geq 0 \\
&\delta_{s}+\delta_{g}+\delta_{gh}+\delta_{sg}+\delta_{ggh} \geq 0 \\
&2\delta_{s}+\delta_{4s} \geq 0 \label{constr_4s} \\
&2\delta_{g}+\delta_{4g} \geq 0 \label{constr_4g}\\
&\delta_{s}+\delta_{g}+\delta_{ssgg} \geq 0 \label{lastconstr}
\end{align}
There are two further constraints for the value of the exponent for the propagator of the gluon and the scalar particle, that follow from the bare 4-point functions from the theory. $\delta_{g} \geq 0$ and also $\delta_{s} \geq 0$ due to a simple consideration, which is explained by means of fig. \ref{fig:dg_det} (note that in all illustrations of this kind in this thesis the values that are compared are to be seen as the absolute values of the diagrams, because on this level of investigation no statement about the prefactors is possible.).
The infrared exponent of the dressed scalar-gluon vertex has to be smaller or equal as the overall infrared exponent of the diagram on the r.h.s. in fig. \ref{fig:dg_det}, which involves a scalar loop, \textit{i.e.} it has a higher or equal order of singularity. Considering the power counting of these two diagrams one obtains the inequality $\delta_{sg} \leq 2 \delta_{s}+\delta_{sg}$. Subtracting the dimension of the scalar-gluon vertex leaves the constraint
\begin{equation}\label{delta_s_constr}
\delta_{s} \geq 0,
\end{equation}
A similar relation holds for $\delta_{g} \geq 0$, due to the same arguments, based on the fact, that there is a bare four-gluon vertex in the Lagrangian. The infrared suppression of the gluon propagator is one necessary condition for the Gribov-Zwanziger  \cite{Zwanziger:1991gz, Zwanziger:1992qr} and the Kugo-Ojima \cite{Kugo:1979gm} confinement scenario. In both scenarios the gluon propagator is supposed to be less singular than a simple pole\footnote{There is another solution for the system consistent with $\delta_g \geq 0$, the so-called decoupling solution, see \cite{Aguilar:2008xm, Boucaud:2008ji, Dudal:2007cw, Dudal:2008sp, vonSmekal:2008ws}, which is observed in lattice calculations \cite{Cucchieri:2007md, Bogolubsky:2007ud,Sternbeck:2007ug} and  also the refinement of the Gribov-Zwanziger scenario according to \cite{Dudal:2007cw,Dudal:2008sp}. In this case $\delta_g = 1$, and all other vertices stay trivial. There is a lively discussion, whether this solution is directly comparable to Dyson-Schwinger results, due to complications in the definition of the Landau gauge in lattice QCD and in the continuum \cite{Maas:2008ri}, see  \cite{vonSmekal:2008ws} for a summary of this problem.}. This is fulfilled by the preceding constraint for all $\delta_{g} >0$, because the full infrared exponent of the gluon is $-1+\delta_{g}$.\\
Exploiting the preceding constraints in the equations in App. \ref{App_full_eqn}, several terms drop out, because they are subleading compared to the tree-level term.\\

In the tadpole terms, that occur in the equations of the scalar and gluon propagator the external momentum never enters the inner loop. The tadpole contributions are thus momentum-independent (infinite) constants that are removed in the renormalisation process and cannot be the leading terms compared to possible singularities. Therefore, in the actual calculations these terms can be dropped in the equations for the infrared exponents of the propagators $\delta_{s}$ and $\delta_{g}$. One obtains:
\begin{eqnarray*}
 && -\delta_{s} =  \min \{ -\mu, \ \delta_{s}+\delta_{g}+\delta_{sg} \} \\ && -\delta_{g}  =  \min \{ 0, \ 2\delta_{s}+\delta_{sg},   \ 2 \delta_{gh}+\delta_{ggh} \} 
\end{eqnarray*}

The equation of the ghost is simplified by renormalisation, cf. \cite{vonSmekal:1997vx, vonSmekal:1997is,Lerche:2002ep, Fischer:2002hna, Fischer:2002eq}, which can be done such, that the bare ghost-propagator term drops out. Thus the equation for the ghost exponent is uniquely determined as:
\begin{equation} \label{eq_gh}
\delta_{gh} = -\frac{1}{2}  (\delta_{g}+\delta_{ggh})
\end{equation}
\\
The simplified system is given in eqs (\ref{s_prop}-\ref{ssgg}). 
\begin{eqnarray}
\label{s_prop} -\delta_{s} & = & \min ( - \mu, \delta_{s}+\delta_{g}+\delta_{sg}) 
\\[0.1cm]
\label{g_prop}
-\delta_{g} & = & \min ( 0, 2\delta_{s}+\delta_{sg}, 2 \delta_{gh}+\delta_{ggh} )
\\[0.1cm]
-\delta_{gh} & = &\frac{1}{2}  (\delta_{g}+\delta_{ggh})
\\[0.1cm]
\delta_{3g} & =& \min (  0, 3 \delta_{gh}+2 \delta_{ggh},3 \delta_{s}+2 \delta_{sg},2 \delta_{s}+\delta_{ssgg}, 2 \delta_{s}+\delta_{sg} )\\[0.1cm]
\delta_{ggh} & = & \min ( 0,2 \delta_{g}+\delta_{gh}+ 2 \delta_{ggh} ) \\[0.1cm]
\delta_{sg} & = & \min (  0,\delta_{s}+2 \delta_{g}+2\delta_{sg}, 2 \delta_{g}+\delta_{ssgg}, \delta_{s}+2\delta_{g}+3\delta_{gh}+2\delta_{sg}+2\delta_{ggh}) \nonumber \\ \label{sg_eq_full}
\\[0.1cm]
\delta_{4g} &= & \min (  0, 2 \delta_{s}+\delta_{ssgg}, 3 \delta_{g}+\delta_{3g}+\delta_{4g},3 \delta_{s}+\delta_{sg}+\delta_{ssgg},
 3 \delta_{s}+2 \delta_{sg}, \nonumber \\
&& 4\delta_{g}+3 \delta_{3g}, 4 \delta_{s}+ 3 \delta_{sg},  4 \delta_{gh}+3 \delta_{ggh}
)\\[0.1cm]
\delta_{4s} & = & \min (  0, 2\delta_{g}+\delta_{ssgg}, \delta_{s}+2 \delta_{g}+\delta_{sg}+\delta_{ssgg}, \delta_{s}+2 \delta_{g}+2\delta_{sg},\nonumber \\
&&, 2\delta_{s}+\delta_{g}+\delta_{sg}+\delta_{4s}, 2\delta_{s}+2\delta_{g}+3\delta_{sg}) \label{4s_full_eq} \\[0.1cm]
\label{ssgg} \delta_{ssgg} &= & \min (  0, 2 \delta_{g}+ \delta_{ssgg}, 3 \delta_{g}+\delta_{3g}+\delta_{ssgg},\delta_{s}+2\delta_{g}+\delta_{3g}+\delta_{ssgg}, \nonumber \\
&&, 2 \delta_{s}+\delta_{g}+\delta_{sg}+\delta_{ssgg}, 3 \delta_{s}+ \delta_{sg}+\delta_{4s},\delta_{s}+2\delta_{g}+2\delta_{sg}, \nonumber \\ 
&&,\delta_{s}+2\delta_{g}+\delta_{sg}+\delta_{3g}, \delta_{s}+3 \delta_{g}+ 2 \delta_{sg}+ \delta_{3g}, 2 \delta_{s} + 2 \delta_{g}+ 3 \delta_{sg}, \nonumber \\
&&, 2 \delta_{s}+2 \delta_{g}+2 \delta_{sg}+ \delta_{3g},  3 \delta_{s}+\delta_{g}+3 \delta_{sg}, \nonumber \\
&& , 2 \delta_{s}+ 2 \delta_{g} + 3 \delta_{gh} + 3 \delta_{sg} + 2 \delta_{ggh}, \nonumber \\
&&, \delta_{s}+3\delta_{g}+3\delta_{gh}+2\delta_{sg}+\delta_{3g}+2\delta_{ggh}, \nonumber \\
&&, \delta_{s}+ 2 \delta_{g}+ 4 \delta_{gh}+2 \delta_{sg}+3 \delta_{ggh},2 \delta_{g}+ 4 \delta_{gh}+3\delta_{ggh}+\delta_{ssgg}). \nonumber \\ \label{eq_2s2g_full}
\end{eqnarray}

This system can be solved by hand\footnote{The usual algorithm of computer algebra systems tries all possible combinations. Due to the size of this system the computation is very extensive. The system of equations was implemented on a desktop computer, but did not yield a result, due to too little memory.} when tackled in an effective way, starting with the easier parts, and consecutively plugging the determined values into the remaining equations. 
\\
For the determination of the infrared exponents I start with the calculation of the coupled equations for the ghost-gluon vertex and the ghost propagator. From the two non-zero terms in the ghost-gluon vertex equation one drops out due to the constraints of the skeleton expansion. Thus 
\begin{eqnarray*}
-\delta_{gh} & = &\frac{1}{2}  (\delta_{g}+\delta_{ggh}) \label{ghprop}\\
 \delta_{ggh} & = \min ( &0,2 \delta_{g}+\delta_{gh}+ 2 \delta_{ggh} )
\end{eqnarray*}
Assuming $\delta_{ggh}= 2\delta_{g}+\delta_{gh}+2\delta_{ggh} <0 $ to be the IR-leading term shows, that this choice is not self-consistent with the ghost propagator equation. If this term was the dominating one, the scaling of the ghost-gluon vertex would be 
$$-\delta_{ggh} = 2 \delta_{g}+\delta_{gh}.$$
Setting this relation (l.h.s.) equal $ -\delta_{ggh}=\delta_{g}+2\delta_{gh}$ from the ghost propagator equation (r.h.s) yields
$$-2 \delta_{g}+\delta_{gh} \stackrel{!}{=}-\delta_{g}-2 \delta_{gh} \ \Rightarrow \ \delta_{gh}=\delta_{g}.$$
With this relation each constraint of the skeleton expansion had to be fulfilled. Considering (\ref{firstconstr}) it is obvious that this constellation is no mathematically allowed solution, because
$$
2\delta_{gh}+\delta_{g}+2\delta_{ggh} = 2 \delta_{g}+\delta_{g}+2(-2\delta_{g}-\delta_{gh}) = -3\delta_{g} \leq 0
,$$that can only be fulfilled by $\delta_{g}=0$. This is in contradiction to the initial assumption, which was that $2\delta_{g}+\delta_{gh}+2\delta_{ggh}$ was smaller than zero (and hence the dominating part). As a result, the only possible infrared scaling behaviour of the ghost-gluon vertex in Landau gauge is the trivial one, \textit{i.e.} the vertex stays bare and scales only with its canonical dimension. This non-enhancement of the ghost-gluon vertex, as predicted by \cite{Taylor:1971ff,Marciano:1977su}, can be observed in Yang-Mills theory and quenched QCD as well, which was shown both by means of DSEs \cite{vonSmekal:1997vx,vonSmekal:1997is,Schleifenbaum:2004id,Fischer:2006vf,Alkofer:2008jy} and lattice calculations \cite{Ilgenfritz:2006he,Cucchieri:2008qm,Cucchieri:2004sq}. Therefore it is not relevant up to this point, whether scalars or quarks are coupled to Yang
-Mills theory, but both particles do not change the scaling behaviour of the ghost-gluon vertex, given by 
\begin{equation}
 \delta_{ggh}=0.
\end{equation}
As an immediate result the ghost exponent is proportional to the gluon exponent. Applying this relation in the gluon equation leaves
\begin{equation}
 -\delta_{g} = \min ( 0, \  2\delta_{s}+\delta_{sg}, \ - \delta_{g} ).
\end{equation}
At this point it is obvious that the gluon equation becomes trivial, \textit{i.e.} the term for $\delta_g$ is involved in the equation for $\delta_g$ itself. Thus this terms dominates, and no further statement is possible so far, but it leaves a free parameter to describe the scaling behaviour of the various Green functions, as it is the case in Yang-Mills theory and quenched QCD. In the literature this parameter is usually denoted by $\kappa$ and defined by
\begin{equation}\label{rel_g_gh}
\frac{1}{2} \delta_{g} = \kappa = -\delta_{gh},
\end{equation}
which is a general relation for linear covariant gauges in Yang-Mills theory \cite{Alkofer:2003jr}.\\
Hitherto the decisive point of this calculation is that the scalar contributions in the equations for the ghost and gluon propagator as well as in the ghost-gluon vertex are subleading. Therefore the parameter $\kappa$ has the same value as in Yang-Mills theory or quenched QCD, where the authors of \cite{Watson:2001yv} derived from the ghost propagator DSE, that this parameter has to be in the interval $0\leq \kappa <1$, which is in perfect agreement with numerical calculations that yield $\kappa \approx 0.595$ \cite{Lerche:2002ep, Zwanziger:2001kw, Pawlowski:2003hq}. Thereby the gluon propagator is suppressed in the infrared, whereas the ghost propagator is infrared enhanced.\\

The next step is the determination of the infrared exponent of the scalar propagator 
\begin{equation}\label{s_p}
-\delta_{s} = \min  (-\mu, \ \delta_{s}+\delta_{g}+\delta_{sg}).
\end{equation}
Up to this point there are two possible solutions for $\delta_{s}$, but not both of them ensure stability under the skeleton expansion. By a similar consideration as it was done for the ghost-gluon vertex the non-trivial value of the scalar propagator drops out.\\
Assume that the second term in eq. (\ref{s_p}) dominates. Then the comparison of the constraint (\ref{constr_2s_g_2sg})
$$2\delta_{s}+\delta_{g}+2\delta_{sg} \geq 0$$
and the relation for the scalar propagator
$$-2\delta_{s} = \delta_{g}+\delta_{sg} < 0$$
would yield an inequality
$$-2 \delta_{s} \geq \delta_{g} ,$$
which could only be true iff $\delta_{g}=0$. But this would be inconsistent with the initial assumption $-\delta_{s}<0$. As a result the scalar propagator must be dominated by its bare value
\begin{equation}
\delta_{s}=\mu. 
\end{equation}

These results already suffice to determine the infrared exponent of the three-gluon and the four-gluon vertex. Inserting the values of the known infrared exponents combined with $\mu \leq 0$ and $\delta_g \geq 0$ yields \\
\begin{eqnarray}
\delta_{3g} & =\min ( &  -3\kappa,\ 3\mu+2 \delta_{sg}, \ 2 \mu+\delta_{ssgg}) \label{eq_3g}\\ [0.2cm]
 \delta_{4g} &= \min ( &  -4\kappa, \  4\mu+ 3 \delta_{sg}, \ 2 \mu+\delta_{ssgg},  \ 3 \mu+\delta_{sg}+\delta_{ssgg} ). \label{eq_4g}
\end{eqnarray}
Due to the constraint (\ref{constr_2s_g_2sg}) there is a lower bound for the terms
\begin{eqnarray}
 3 \mu+2 \delta_{sg} & \geq & \mu -2\kappa \ > \ -3\kappa\\
 4\mu+ 3 \delta_{sg} & \geq & \mu -3\kappa \ > \ -4\kappa
\end{eqnarray}
Thus the terms drop out in the appropriate equation. Due to constraint (\ref{lastconstr}) the similar relation
\begin{equation}
 2\mu+\delta_{ssgg} \geq \mu -2\kappa >-3 \kappa \ ( \ >-4\kappa \ )
\end{equation}
holds, so this term cannot be the leading one in both equations (\ref{eq_3g}) and (\ref{eq_4g}).\\
At last the constraints (\ref{constr_2s_g_2sg}) and (\ref{lastconstr}) ensure that
\begin{equation}
\left.\begin{aligned}
 2\mu+\delta_{ssgg} & \geq \mu-2\kappa\\
  \mu+\delta_{sg}& \geq -\kappa,
\end{aligned}
\right\}
+ \quad \Rightarrow 3\mu+\delta_{sg}+\delta_{ssgg} \geq \mu-3\kappa \ > \ -4\kappa.
\end{equation}
I want to point out that this is a very crucial result. At this point all leading IR exponents of the vertices from Yang-Mills theory are determined uniquely (and independent of the mass of a scalar). Thus coupling scalar particles to the Yang-Mills sector does not change the scaling of the pure gauge theory vertices, because the scalar contributions do not alter
the leading parts in the system of equations for the infrared exponents. This is a non-trivial result, which could not be expected before this calculation. The persistence of the ghost-dominance even with scalar contributions in the vertex functions from Yang-Mills theory yields that the result for the leading tensor structure can still be obtained by the general result for an arbitrary vertex function, as given in eq. (\ref{arb_ver}),
\begin{equation}\label{arb_ver}
\Gamma^{n,m} (p^2) \sim ( p^2 )^{(n-m) \kappa}
\end{equation}
wherein $n$ in the number of ghost-antighost pairs and $m$ is the number of gluons in the vertex. Analog results have been found in the case of quarks \cite{Alkofer:2008tt, Schwenzer:2008vt}.\\

\begin{figure}
 \centering
 \includegraphics[width=10cm]{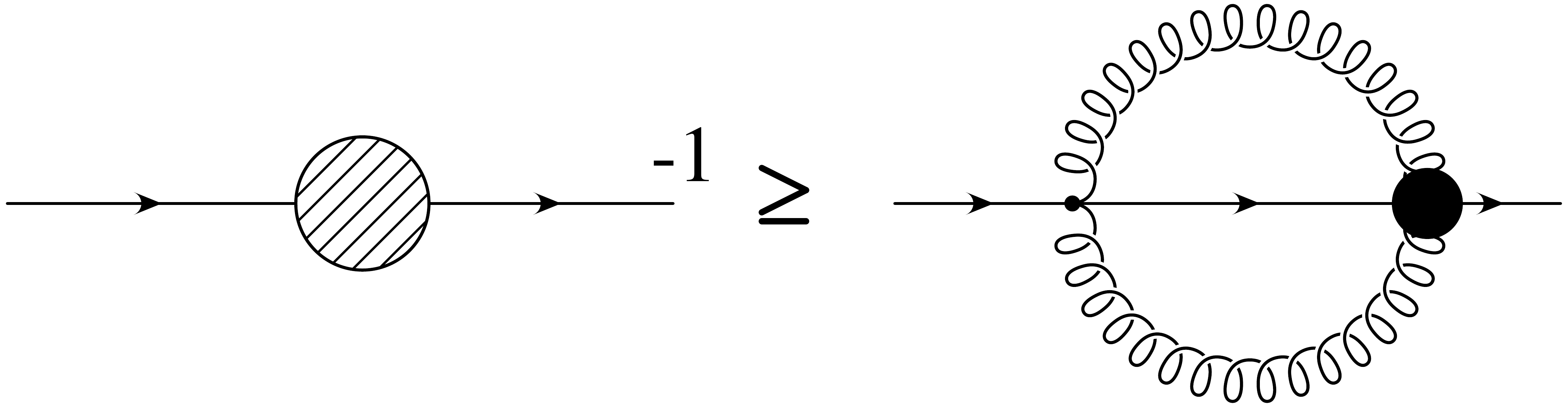}
 \caption{\small{(Two-loop) sunset diagram in the scalar propagator equation.}}
 \label{fig:incons_scal_prop}
\end{figure}

Already at this point it is possible to determine the general value of $\delta_{sg}$ in dependence of $\mu$. Inserting all known infrared exponents into eq. (\ref{sg_eq_full}) leaves
\begin{equation}
\delta_{sg} = \min (0, \ 4\kappa + \delta_{ssgg}, \ \mu+\kappa+2\delta_{sg}).
\end{equation}
With regard to the truncation from above it is not possible to exclude any further terms. But considering that the solutions obtained in this truncation must also hold for the full system of equations, a sunset diagram (thus two-loop), given in fig. \ref{fig:incons_scal_prop}, in the scalar propagator equation gives a further constraint. 
Suppose that $4\kappa+\delta_{ssgg}<0$ is the dominating term in the scalar-gluon equation. Making a power counting for this diagram, the exponent of the r.h.s. of the equation must necessarily be larger or equal to the exponent of the l.h.s. to give a consistent solution. Thus 
\begin{eqnarray}
 1+\mu & \leq & 4+2(-1+\delta_g) -1+\delta_s+\delta_{ssgg} \nonumber \\
0 & \leq & 4\kappa+ \delta_{ssgg}.\label{sun}
\end{eqnarray}
The constraint (\ref{sun}) is in contradiction to the initial assumption, thus the term $4\kappa+\delta_{ssgg}$ cannot dominate in the infrared. \\
The other two scaling exponents of the scalar-gluon vertex are mathematically possible solutions
\begin{equation}
 \delta_{sg} = \ 0 \ \lor
 \ - \mu - \kappa,
\end{equation}
which have also been found in QCD.\\

Table \ref{same_ir_exp} shows the three mathematically possible solutions of the infrared exponents for both, a massless as well as a massive scalar particle. I want to point out explicitely at this point, that the graphs containing a four-scalar or a two-scalar-two-gluon vertices are subleading in all other Green functions. This is an important result of this thesis. The actual values of $\delta_{4s}$ and $\delta_{ssgg}$ cannot be determined simultaneously for the massless and the massive scalar, because there are qualitative differences, which I will explain in the following sections.
\\
Futhermore I emphasise that this scaling behaviour justifies the truncation presented above. By construction all two-loop graphs contain one bare four-point vertex. As it has been shown above, the four-scalar and the two-scalar-two-gluon vertices cannot be the dominating terms, and therefore it is fair to neglect the two-loop graphs except for the diagrams given in fig. \ref{gh_box} and fig. \ref{fig:gh_2s2g}. Considering all two-loop terms in the calculation does not change the final solutions of the system, which has been checked explicitly. It is important to note that due to this the solutions obtained in the one-loop truncation are not an approximation but rather exact self-consistent solutions for the scaling behaviour of the leading tensor structures of the full DSEs in the infrared region.\\
\\
From this point I will split the calculation of the remaining infrared exponents of the four-scalar and two-scalar-two-gluon vertices in two different sections. In these sections I will also interprete the physics of the results.\\
\begin{table}[!htb]\centering
\begin{tabular}{|l||c|c|c|c|c|c|c|}
\hline
infrared exponents & $\delta_{s}$ & $\delta_{g}$ & $\delta_{gh}$ & $\delta_{sg}$ & $\delta_{ggh}$ & $\delta_{3g}$ & $\delta_{4g}$  \\
\hline
trivial solution &$\mu$&0&0&0&0&0&0\\
partial scaling solution & $\mu$ & $2\kappa$ & $-\kappa$ & $0 $ & 0 & $-3\kappa$ & $-4\kappa$ \\ 
full scaling solution & $\mu$ & $2\kappa$ & $-\kappa$ & $- \mu-\kappa$ & 0 & $-3\kappa$ & $-4\kappa$ \\
\hline
\end{tabular}\caption{\small{Uniform infrared fixed points of fundamentally charged scalars.}} \label{same_ir_exp}
\end{table}

\subsection{Massless Scalar Particle}
In the case of a massless scalar particle the values of $\delta_{4s}$ and $\delta_{ssgg}$ simply follow from the lower $n$-point functions. The four-scalar vertex stays bare up to all orders. This can be seen from constraint (\ref{constr_4s}). For a vanishing mass of the scalar $m^2=0$ the value of $\delta_{s}=0$ (as $\mu=0$). Thus the constraint yields
\begin{equation}
 \delta_{4s}\geq 0,
\end{equation}
but the value 0 appears in the \emph{min}-function of equation (\ref{4s_full_eq}), thus the only permitted solution is
\begin{equation}
 \delta_{4s} = 0.
\end{equation}
The exponent of $\delta_{ssgg}$ is determined from equation (\ref{eq_2s2g_full}), which is reduced to 
\begin{equation}
\delta_{ssgg} = \min ( 2 \delta_{sg}, \ \delta_{ssgg}),
\end{equation}
which yields in combination with the constraint (\ref{lastconstr}), that the exponent $\delta_{ssgg}$ depends on the value of $\delta_{sg}$.\\
In table \ref{ir_exp_massless} I complete the solutions for the infrared exponents a massless scalar coupled to Yang-Mills theory.
 \begin{table}[!htb]\centering
\begin{tabular}{|l||c|c|c|c|c|c|c|c|c|}
\hline
infrared exponents & $\delta_{s}$ & $\delta_{g}$ & $\delta_{gh}$ & $\delta_{sg}$ & $\delta_{ggh}$ & $\delta_{3g}$ & $\delta_{4g}$ & $\delta_{4s}$ & $\delta_{ssgg}$ \\
\hline
trivial solution &0&0&0&0&0&0&0&0&0\\
partial scaling solution & 0 & $2\kappa$ & $-\kappa$ & $0 $ & 0 & $-3\kappa$ & $-4\kappa$ & 0 & 0 \\ 
full scaling solution & 0 & $2\kappa$ & $-\kappa$ & $-\kappa$ & 0 & $-3\kappa$ & $-4\kappa$ & 0 & $-2\kappa$\\
\hline
\end{tabular}\caption{\small{Uniform infrared fixed points of massless fundamentally charged scalars.} \label{ir_exp_massless}}
\end{table}

Mathematically there are three possible solutions for this system, that were named according to their scaling properties. I want to emphasise again that all four-point functions do not dominate in the equations of the lower $n$-point functions, so the scaling behaviour of the leading tensor structures in the Green functions is qualitatively the same as for the Green functions in quenched QCD, for which the scaling behaviour has been determined in \cite{Alkofer:2008tt, Schwenzer:2008vt}, which gives rise to the possibility of a comparison of the two systems on a qualitative level. 
Furthermore it is an interesting result that the scalar propagator is dominated by its tree-level term in the infrared. \\
The trivial solution stems from the fact that the ghost equation can be fulfilled trivially, if the gluon propagator is dominated by its tree-level term. Consequently $\kappa = 0$ and this solution cannot be excluded, and hence all exponents vanish as the parameter vanishes in the infrared limit. It is not sure whether this solution is only a mathematical solution, or if it also has physical content for the infrared sector. \\
The partial scaling solution describes the Yang-Mills sector as usual according to its ghost dominance, wherein the scalars decouple in the IR limit. This can be seen in the vanishing exponents of the scalar-gluon and the two-scalar-two-gluon vertex, which are the link from the matter to the Yang-Mills sector, whereas the gluon propagator is suppressed. \\
The full scaling solution yields a scaling scalar-gluon and consistently also a scaling two-scalar-two-gluon vertex. The scalar sector couples to the Yang-Mills vertices, but this low order of singularity is not high enough to yield a linear rising potential. Therefore also in this solution confinement cannot be generated. The authors of \cite{Alkofer:2008tt} found that a similar situation is realised in QCD.

\subsection{Massive Scalar Particle}
The other important case to be presented here is the one of a massive scalar particle, but without a scalar condensate, which corresponds to the case of unbroken symmetry above.
\\
This unbroken symmetry ensures, that the Dyson-Schwinger equations as derived above are not changed by other vertices, that would cause changes in the topology of the equations. 
\\
For a scalar particle with a certain mass $m^2(p^2\rightarrow 0 ) = M^2$ its propagator freezes out at a finite value in the infrared, in contrast to the divergent bare propagator of a massless scalar:
$$S^0_{ij}(p^2) = -  \frac{\delta_{ij}}{p^2+M^2} \ 
\stackrel{p^2\rightarrow 0}{\longrightarrow}
 \ \ - \frac{\delta_{ij}}{M^2} \propto \big( p^2 \big)^{0}.$$
Thus in this case $\mu = 1$, and therefore $\delta_{s} = 1$ and $\delta_{sg}= 0 \lor -1-\kappa$.
In the following the same nomenclature for all anomalous dimensions as in the massless case is used for convenience, cf. Table \ref{dimensions}. \\
With the previous results the exponents for the four-scalar and two-scalar-two-gluon vertex can be analysed. The equations (\ref{4s_full_eq}) and (\ref{eq_2s2g_full}) simplify to
\begin{eqnarray}
  \delta_{4s} & = & \min ( 0, 4 \kappa+\delta_{ssgg}, \ 1+4 \kappa+(0\lor -1-\kappa)+\delta_{ssgg}, \nonumber \\
&&, 1+4 \kappa+2(0\lor-1-\kappa), 2+2\kappa+(0\lor-1-\kappa)+\delta_{4s}\nonumber \\
&&, 2+4\kappa+3(0\lor-1-\kappa)), \label{4s_2} \\   
\delta_{ssgg} &= &\min (  0, \ 3 \kappa+\delta_{ssgg}, \ 2+2\kappa+(0\lor-1-\kappa)+\delta_{ssgg}, \nonumber \\
&& ,3 +(0\lor-1-\kappa)+\delta_{4s}, \ 1+4\kappa+2(0\lor-1-\kappa), \nonumber \\ &&,1+\kappa+(0\lor-1-\kappa), 2 +\kappa+2 (0\lor-1-\kappa), \nonumber \\ 
&& ,   2 + \kappa+ 3 (0\lor-1-\kappa),\ 1+2(0\lor-1-\kappa), \ \delta_{ssgg}). \label{ssgg_2}
\end{eqnarray}
For the rest of the calculation it is convenient to investigate the three different solutions separately.

\subsubsection{Trivial Solution}
All other vertices are dominated by their tree-level terms. Inserting the values into equations (\ref{4s_2}) and (\ref{ssgg_2}) yields
\begin{eqnarray}
 \delta_{4s} & = & 0 \\  
 \delta_{ssgg} & = & 0.
\end{eqnarray}
as the scalar-gluon vertex does not scale. So also the four-gluon and the two-scalar-two-gluon vertex are dominated by their tree-level parts. Also for a massive scalar particle it is not clear whether this solution is physically relevant.

\subsubsection{Partial Scaling Solution}
For a scaling Yang-Mills sector, $\delta_{s}=1$ and $\delta_{sg}=0$ the equations (\ref{4s_2}) and (\ref{ssgg_2}) become trivial and the exponents are equal to the trivial solution
\begin{eqnarray}
 \delta_{4s} & = & 0 \\  
 \delta_{ssgg} & = & 0.
\end{eqnarray}
Therefore this solution describes a decoupling scalar from the scaling Yang-Mills sector, as in the massless case.

\subsubsection{Full Scaling Solution}
A naive calculation of the infrared exponents from the equations (\ref{4s_2}) and (\ref{ssgg_2}) is possible, but the solution for the exponent $\delta_{ssgg}$ bears an ambiguity, as only a range of possible values can be given
\begin{eqnarray}
& \delta_{4s}  =  -1+\kappa \\[0.1cm]
 & -1-2\kappa\leq \delta_{ssgg} <0.
\end{eqnarray}

\begin{figure}[!htb]
 \centering
 \includegraphics[width=7cm]{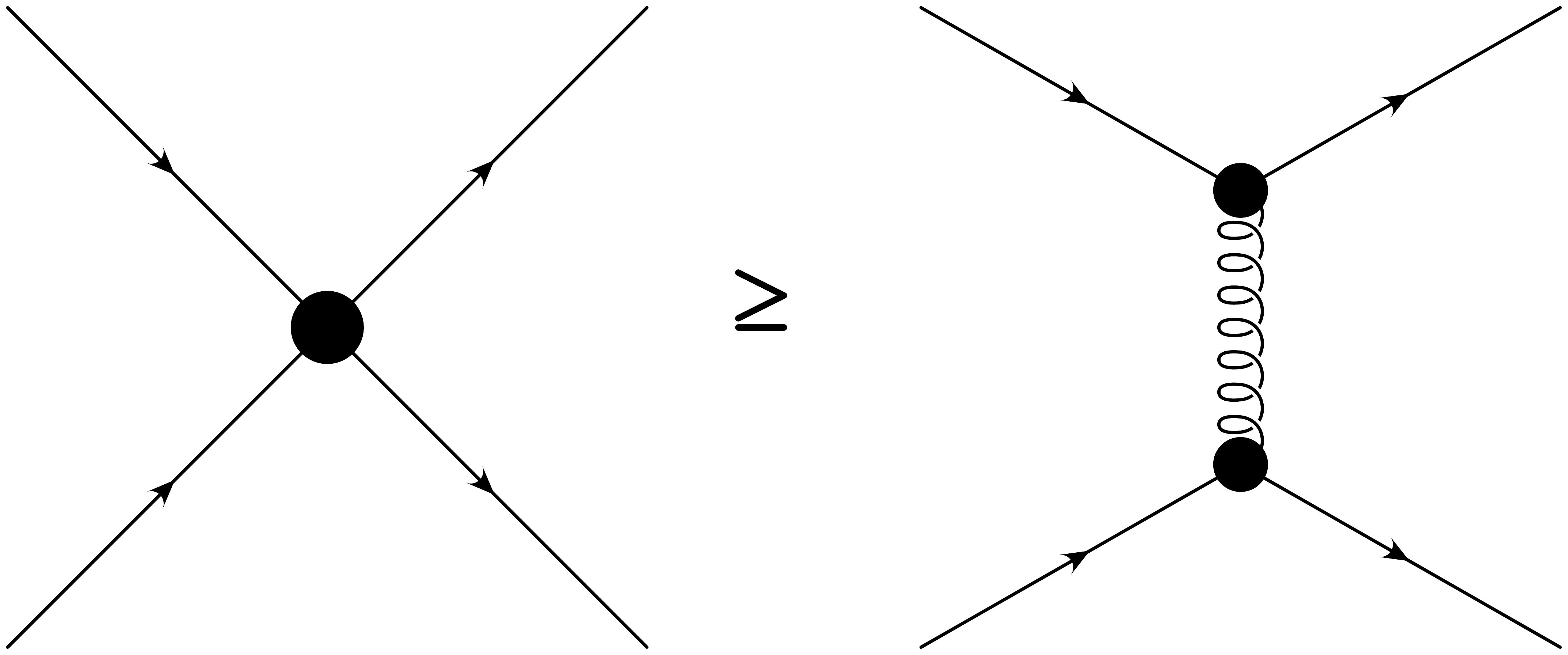}
 \caption{\small{Exemplary diagram in dressed four-scalar vertex.}}
 \label{fig:4s_nonPI}
\end{figure}

Up to this point three solutions remain, the trivial solution, the partial scaling solution and the full scaling solution. The third solution as the only solution with a scaling scalar sector suffers from a problem for the four-scalar and the two-scalar-two-gluon vertex. The degrees of divergence are small, and considering a diagram given in fig. \ref{fig:4s_nonPI}, which is part of the fully dressed vertex, gives evidence, that the obtained infrared exponent may not be the proper one. The simple counting of the uniform exponents of fig. \ref{fig:4s_nonPI} shows, that the order of singularity is larger compared to the dressed vertex
$$\delta_{4s} = -1+\kappa \ \stackrel{?}{\leq} \ -1+2 \kappa + 2 (-1-\kappa) = -2.$$ 
\\

Apparently this approximation considering only the uniform limit\footnote{The calculation was also done with all two-loop diagrams, but did not change the results obtained in the one-loop truncation. This is due to the fact that by construction all two-loop diagrams contain at least one 4-point interaction, which was shown to be subleading in the IR. } fails in giving the correct values for the infrared exponents $\delta_{4s}$ and $\delta_{ssgg}$ for a massive scalar particle. This suggests that soft-singularities may be crucial for the scaling behaviour, as it will be confirmed below.\\
A similar situation is realised in quenched QCD as the authors of \cite{Alkofer:2008tt} showed recently. Diagrams including both soft and hard momenta are the leading diagrams in the IR and kinematic divergencies are related to the mechanism for confinement in quenched QCD. The question is whether this mechanism for confinement also holds for fundamentally charged scalars in Yang-Mills theory. Thus the effect of kinematic divergencies on the scaling behaviour must be taken into account\footnote{Note that the authors of \cite{Braun:2007bx} find a different solution based on the formalism of the functional renormalisation group.}.

\section{Kinematic Divergencies} \label{chapkindiv}
Hitherto only uniform scaling with its divergences was investigated. Furthermore there is also another type of singularities that can emerge, referred to as soft singularities, as presented in recent publications \cite{Alkofer:2008jy, Alkofer:2008tt}. The previously discussed conformal case is based on the assumption that Green functions can be described by a power law in one single scaling variable, which vanishes if and only if all external momenta that enter the vertex scale to zero uniformly. Thus in the uniform case there is only one relevant momentum scale.
\\
Soft singularities can emerge, if there are two different momentum scales involved, but only one of them scales to zero whereas the other one stays finite. In the literature vanishing momenta are referred to as soft momenta, whereas finite momenta are usually called hard momenta.\\
The momentum integral naturally covers all different momentum regions, \textit{i.e.} also finite momenta. Although the momenta stay finite in several kinematic regions, it is not guaranteed, that the leading contribution to the loop integral arises from soft momenta. In special kinematic cases, where only a subset of external momenta vanishes, Green functions may have a different scaling behaviour than in the uniform limit. Note that internal loops can have contributions from both scales, therefore considering hard scales requires a modification of the former equations for the uniform limit too.
\\
As an example I will shortly discuss the three-gluon vertex with its soft singularity as being the clearest one to grasp the idea of soft singularities, due to the complete Bose symmetry of the three-gluon vertex. In the uniform case all three external gluon momenta vanish, but still the internal momenta can be either hard or soft. Due to the hard scales involved, also the equation for the leading infrared exponent for the uniform case receives additional terms originating from hard-loop contributions. As can be seen in Appendix \ref{app_bare} eq. (\ref{bare3g}), the bare three-gluon vertex reads
\begin{equation*}
 \Gamma_{0,\mu\nu\rho}^{abc}(p,q,r) =  i g f^{abc} \left( \delta_{\mu\nu} (p-q)_{\rho} + \delta_{\nu\rho} (q-r)_{\nu}+ \delta_{\rho\mu} (r-p)_{\nu} \right) \delta (p+q+r).
\end{equation*}
Supposing that only one gluon-leg becomes soft yields that the bare vertex is dominated by the remaining hard momenta, other possible tensor structures that depend only on the soft gluon are subleading in the infrared. Therefore for only one vanishing gluon momentum this vertex does not scale canonically as in the uniform limit but only with its anomalous dimension.
\\
A dressed three-gluon vertex with only one soft external momentum possibly scales with a different infrared exponent as the uniform exponent $\delta_{3g}$ discussed in the conformal case, thus there is another equation for the exponent $\delta_{3g}^{g}$ in the soft-gluon limit, that expresses this anomalous dimension for the special kinematic case. Due to the symmetry of the three-gluon vertex there is only one additional equation, because the scaling behaviour does not depend on which gluon is soft. 
\\

The situation becomes more sophisticated for other $n$-point functions. In the case of the ghost-gluon vertex two additional combinations of soft and hard external momenta may reveal different scaling exponents than the uniform vertex. With either a vanishing gluon- or ghost-momentum there are altogether three equations for $\delta_{ggh}$, $\delta_{ggh}^{g}$ and $\delta_{ggh}^{gh}$ to be studied. 
\\
The number of different kinematics rises further for higher $n$-point functions. In general each $n$-point function can have combinations with $n$ or $n-i, \ i \in \lbrace 2,\ldots,n-1 \rbrace $ soft momenta in the infrared. One single hard external momentum is forbidden due to overall momentum conservation in each diagram, but every higher number is allowed. Each kinematic case yields a separate infrared scaling exponent for the vertex function.
\\
For fundamentally charged scalars especially the 4-point functions render a huge variety of possible soft singularities, which yield new equations for the various IR exponents for each kinematic case.
\\
Including soft singularities for fundamentally charged scalars yields a system of equations, which is far too involved to be handled in this thesis. Fortunately this analysis has already been performed in the case of QCD and the similar structure of the equations allows to employ these results in the present case. 
\\

As has been shown above scalar contributions in the DSEs of vertex functions of the pure gauge sector are subleading\footnote{Note that the scalar-loop diagram in the gluon equation is dropped for other reasons. Considering this diagram would eliminate any possibility of observing confinement. This aspect will be discussed further in the next section.}, \textit{i.e.} the same structures as for quenched QCD, as there are no quark loops, dominate the deep infrared behaviour of the pure gauge vertices. The infrared exponents for the kinematic limits of Yang-Mills theory are thus unchanged, their values are listed in table \ref{table_exp_puregauge}, see \cite{Alkofer:2008tt} for their derivation.
\begin{table}[htb]
\centering
\begin{tabular}{|c|c|c|c|c|c|c|c|c|}
\hline
$\delta_{gh}$ & $\delta_g $ & $\delta_{ggh}^{u}$ & $\delta_{3g}^{u}$ & $\delta_{4g}^{u}$ & $\delta_{ggh}^{gh}$ &   $\delta_{ggh}^{g}$ & $\delta_{3g}^{g}$ &  $\delta_{4g}^{g}$ \\ \hline
 $-\kappa$ & $2\kappa$ &  $0$ & $-3\kappa$ & $-4\kappa$ & $0$ & $ 0$ & $1-2\kappa$ & $1-2\kappa$ \\
\hline
\end{tabular}
\caption{\small{Anomalous dimensions for kinematic cases for pure gauge primitively divergent $n$-point functions.}} \label{table_exp_puregauge}
\end{table}

\begin{figure}[!htb]
 \centering
 \includegraphics[width=11cm]{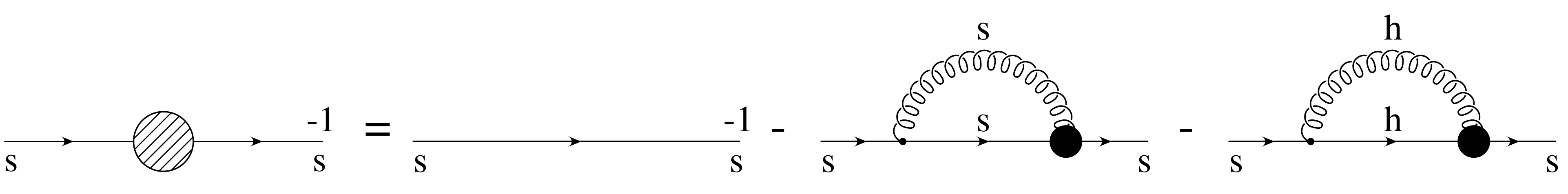}
\caption{\small DSE for the scalar propagator with a diagram for a possible hard momentum routed through the loop. $s$ and $h$ stand for soft and hard momenta.} \label{fig:scal_prop_kin}
\end{figure}

With respect to the different kinematic cases the remaining system consists of the equations for the scalar propagator and the scalar-gluon vertex. In the IR limit the contributions to loop graphs in the DSEs from soft loop momenta of the order of the small external momentum and hard loop momenta that are much larger than these can be separated. The Dyson-Schwinger equation for the scalar propagator is given in fig. \ref{fig:scal_prop_kin}, where different possible momentum routings in one diagram are given as separate diagrams. The caption $s$ denotes a soft momentum, whereas $h$ stands for a finite momentum. 
\\The Dyson-Schwinger equations for the uniform, the soft-gluon and soft-scalar limit are given in fig. \ref{fig:kin_uniform}, \ref{fig:kin_soft_gluon} and \ref{fig:kin_soft_scalar} in Appendix \ref{app_D}.
\\

The determination of the infrared exponents of the vertex in the kinematic limits can be done by comparison with the QCD system based on the following ideas:
\begin{itemize}
 \item{The equations of the scalar propagator and the scalar-gluon vertex turn out to be topologically equal to the corresponding equations for the quark propagator and the quark-gluon vertex in QCD, except for the terms including the bare two-scalar-two-gluon vertex. This also holds for the four-scalar vertex and the quark-antiquark scattering kernel.}
 \item{The differences in the systems of equations for scalars in Yang-Mills theory and QCD are based on the canonical dimensions of the vertices. As it is explicitly shown in App. \ref{app_D} the canonical dimensions in the equations for the scalars cancel in such a way that the leading structure of the equations in QCD is reproduced. In comparison with the calculation in the appendix of \cite{Alkofer:2008tt} the equations for the scalar Yang-Mills theory also yield the same solutions and the values of the exponents $\delta_{s}^{\prime}$\ \footnote{Note the different parametrization than in the calculation in the uniform limit $$ S \sim \big(p^2\big)^{\delta_{s}^{\prime}}$$ for the anomalous dimension of the propagator of a massive scalar.}, $\alpha_{sg}^{u}$ (the anomalous dimension of the scalar-gluon vertex in the uniform limit\footnote{Due to the difference in the canonical dimensions of the scalar-gluon and the quark-gluon vertices the reparametrization $\alpha_{sg}^{u} = \delta_{sg}^{u} + \frac{1}{2}$ for the scalar-gluon vertex is necessary to map quenched QCD onto scalar Yang-Mills theory. The explicit calculation is given in App. \ref{app_D}.}), $\delta_{sg}^{s}$ and $\delta_{sg}^{g}$ are the same as of the analogous quantities in QCD. The infrared exponents are listed in table \ref{expkindiv}.}
 \item{Note that the arguments above hold only because the diagrams containing bare two-scalar-two-gluon vertices are subleading, see App. \ref{app_D} for the explicit discussion.}
\end{itemize}
\begin{table}[!htb]
\centering
 \begin{tabular}{| c | c | c | c |}
\hline
$\delta_{s}^{\prime}$  &  $\alpha_{sg}^{u}$ & $\delta_{sg}^{g}$ & $ \delta_{sg}^{s}$ \\
\hline
$0$ & $ -\frac{1}{2}-\kappa \  \lor \ 0$ & $-\frac{1}{2}-\kappa \ \lor \ 0$ & $0 $ \\
\hline
 \end{tabular}\caption{\small{Infrared exponents for the scalar propagator and the scalar-gluon vertex in different kinematic limits.}}\label{expkindiv}
\end{table}

As a result the statement that soft singularities play a crucial role for the infrared scaling behaviour of higher vertices, as it is proposed in QCD, stays true for fundamentally charged scalars coupled to Yang-Mills theory. 

\begin{figure}[!htb]
 \centering
 \includegraphics[width=10cm]{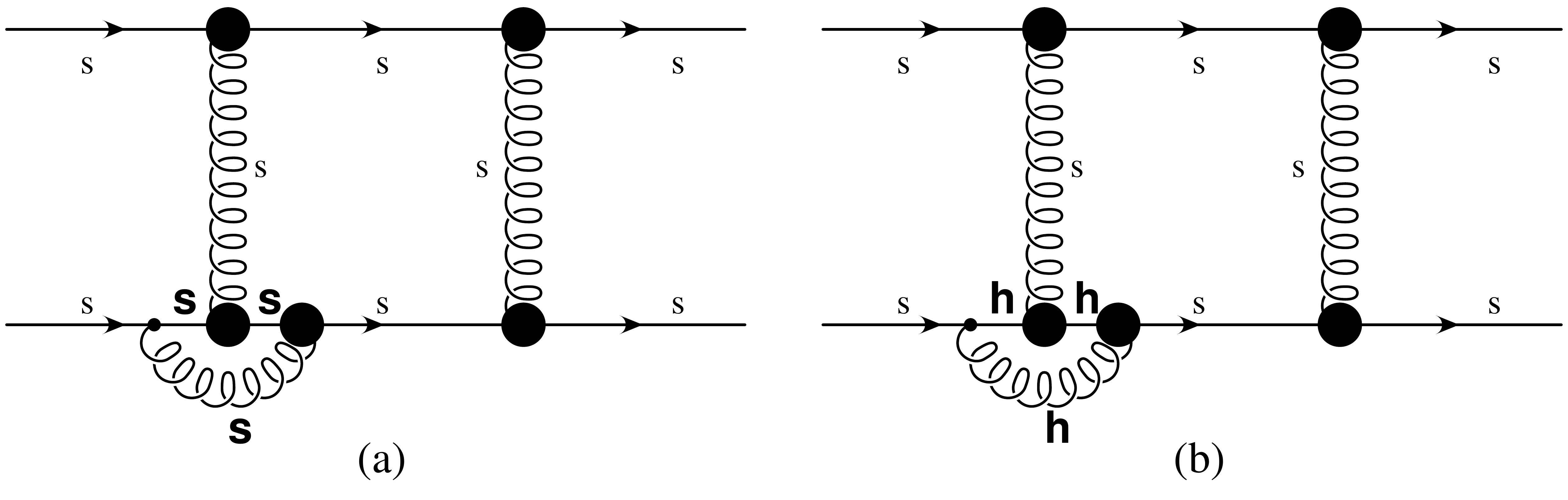}
 \caption{\small{Contribution of the four-scalar-gluon vertex to the four-scalar DSE with inclusion of kinematic divergences.}}
 \label{scat_kernel}
\end{figure}
\begin{figure}[!htb]
 \centering
 \includegraphics[width=10cm]{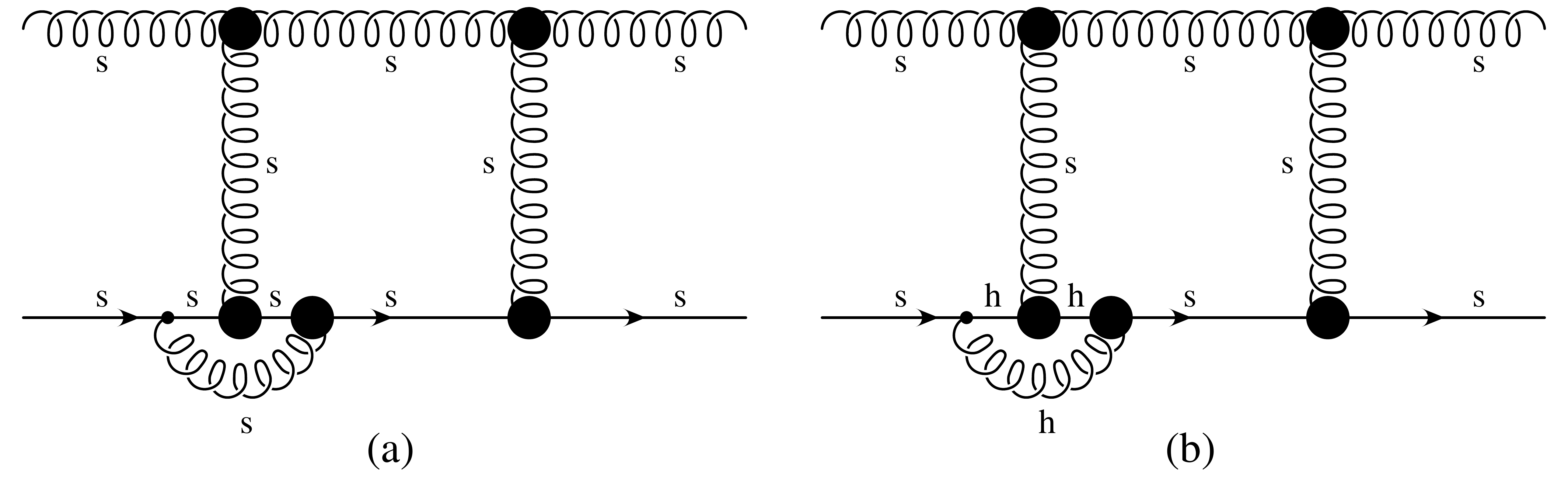}
 \caption{\small{Contribution of the two-scalar-three-gluon vertex to the two-scalar-two-gluon DSE with inclusion of kinematic divergences.}}
 \label{fig:mass_leading2s2g}
\end{figure}

With the strong kinematic divergence in table \ref{expkindiv} the diagram that dominates the four-scalar vertex in the uniform limit\footnote{Note that only diagram \emph{(a)} of fig. \ref{scat_kernel} has been taken into account in the uniform limit.} is the diagram \emph{(b)} in fig. \ref{scat_kernel}. A power counting\footnote{Note that the integration over the hard loop does not give a contribution $(p^2)^{2}$, because the integral is dominated by momenta that stay finite.} yields
$$
2+2(-1+2\kappa)+(-\frac{1}{2}-\kappa) +3(-\frac{1}{2}-\kappa)= -2.
$$
Thus the actual infrared exponent of the four-scalar vertex is much stronger than the value obtained in the uniform limit, and gives the stable solution of
\begin{equation}
 \delta_{4s}=-2.
\end{equation}
Also the value of $\delta_{ssgg}$ is changed by the inclusion of soft-singularities. This becomes obvious by the power counting of the dominating diagram, given in fig. \ref{fig:mass_leading2s2g}.\\
\begin{eqnarray}
& 2+3(-1+2\kappa)+2(\frac{1}{2}-3\kappa)+2(-\frac{1}{2}-\kappa) = -1-2\kappa \nonumber \\ [0.2cm]
& \Rightarrow \ \delta_{ssgg} = -1-2\kappa
\end{eqnarray}
The physical meaning of these exponents is explained by comparison with QCD, which will be discussed in the next chapter.

\chapter{Comparison with Quenched QCD}
In this chapter I will compare the system of fundamentally charged scalars with QCD. This is done on a qualitative level, as the infrared scaling behaviour is determined by means of a power counting analysis. It will be distinguished between the chirally symmetric and the chirally broken phase, as there are severe differences in the origin of the scaling behaviour.
\\

\section{Chirally Symmetric Phase of Quenched QCD}
It is a suprising and important result of this thesis that the fundamentally charged massless scalars feature the same infrared scaling behaviour as chiral quarks in quenched QCD. This could not be guaranteed before the calculation, as there are fundamental differences in the two systems. As a result of the power counting analysis the leading infrared exponents for the primitively divergent vertex functions in scalar Yang-Mills theory stem from diagrams that have analoga in the DSEs in quenched QCD. The solutions that are obtained for massless scalar Yang-Mills theory and for quenched QCD in the chirally symmetric phase, are summarised in table \ref{tab:comp_chir} for the convenience of the reader. This table shows that the anomalous dimensions of the primitively divergent vertex functions are equal, and thus the qualitative infrared scaling behaviour of chirally symmetric quenched QCD and the system of massless fundamentally charged scalars is essentially the same. In this sense quenched QCD can be mapped onto fundamentally charged scalars. \\

The equality of the systems can be explained by a few arguments. At first sight the different nature of interactions seems to be an important difference between the theory containing fundamentally charged scalars and QCD. But as it turned out in chapter \ref{results} this is only a minor difference in the infrared scaling analysis. As already stressed above, scalar particles in the fundamental representation are bosons, whereas quarks are naturally fermions. These different kinds of particles feature different primitively divergent vertices, and as a consequence the system of DSEs for the primitively divergent vertices of fundamentally charged scalars coupled to Yang-Mills theory are far more complicated as the corresponding QCD equations, due to the additional four-scalar and two-scalar-two-gluon vertices. It was therefore a non-trivial question, if these systems had the same infrared power law behaviour. But due to the fact, that these additional four-point vertices are subleading, as determined above, the same diagrams give the leading contributions, in QCD as well as in the scalar Yang-Mills theory. In both systems the pure gauge sector remains unchanged by the coupling of further particles. Considering only the anomalous dimensions, the exponents of the scalar and quark propagator, respectively the scalar-gluon and quark-gluon vertices are equal. \\
Remember that due to the four-scalar vertex there is an additional constraint $\delta_s \geq 0$ for the anomalous dimension of the scalar propagator, see fig. \ref{fig:dg_det}. It is a crucial fact that there is no primitively divergent four-quark vertex in QCD, so that there is no analogous constraint for the quark propagator exponent. This has an impact on the solutions of QCD in the chirally symmetric case, as the renormalisation procedure can be done in different ways. Surprisingly, despite the restrictive dynamics there are more solutions than for scalar Yang-Mills theory. 
\begin{description}
 \item[trivial solution: ]{In this solution the leading anomalous dimensions of all vertices vanish. This solution is also found for scalar Yang-Mills theory. Although the solution for the scaling exponents seems trivial, it is not necessarily a physically irrelevant case. }
\item[scaling solution: ]{Actually this solution contains two different cases, which were both found in scalar Yang-Mills theory. In both cases the Yang-Mills vertices scale according to the general formula (\ref{arb_ver}), but in one case the quark sector is decoupled, \textit{i.e.} the quark propagator and the quark-gluon vertex stay bare up to all orders, whereas in the other case the quark-gluon vertex is divergent in the infrared, as it scales with $(p^2)^{-\kappa}$. So these two solutions have the same scaling behaviour as the scalar propagator and the scalar-gluon vertex in the partial/full scaling solution for scalar Yang-Mills theory.}
\item[other solutions: ]{Due to the complete symmetry of the equations for the quark and ghost propagators, the renormalisation can be done in different ways, which is the origin for other solutions, where \textit{e.g.} the quark propagator is enhanced instead of the ghost propagator, or both of them. I want to emphasise again, that these solutions do not occur in the case of scalar Yang-Mills theory.}
\end{description}
Another surprising fact is that there are less solutions in the case of fundamentally charged massless scalars coupled to the Yang-Mills sector than in chiral QCD, although the system \textquotedblleft looks\textquotedblright easier for the latter case. I want to point out again, that the reason is the additional four-point vertices being subleading in the infrared, and that there is another constraint, which has its origin in the four-scalar vertex.

\begin{table}[!htb]
 \begin{tabular}{|c|c|c|c|c|c|c|c|}
\hline
\multirow{3}{4.5cm}{\centering chirally symmetric quenched QCD}
& $\delta_{g}$& $\delta_{gh}$ & $\delta_{3g}$ & $\delta_{4g}$ & $\delta_{ggh}$ & $\delta_{q}$ & $\delta_{qg}$ \\ \cline{2-8}
&0&0&0&0&0&0&0 \\ 
& $2\kappa$ & $-\kappa$ & $-3\kappa$ & $-4\kappa$ & 0 & 0 & $-\kappa \vee 0$ \\
\hline \hline
\multirow{3}{4.5cm}{\centering massless scalar Yang-Mills theory}
& $\delta_{g}$& $\delta_{gh}$ & $\delta_{3g}$ & $\delta_{4g}$ & $\delta_{ggh}$ & $\delta_{s}$ & $\delta_{sg}$\\ \cline{2-8}
&0&0&0&0&0&0&0 \\ 
& $2\kappa$ & $-\kappa$ & $-3\kappa$ & $-4\kappa$ & 0 & 0 & $-\kappa \vee 0$\\ 
\hline
 \end{tabular}\caption{Infrared exponents for chirally symmetric quenched QCD and massless scalar Yang-Mills theory in the uniform limit.}\label{tab:comp_chir}
\end{table}

\section{Chirally Broken Phase of Quenched QCD}
\begin{figure}[!htb]
 \centering
 \includegraphics[width=10cm]{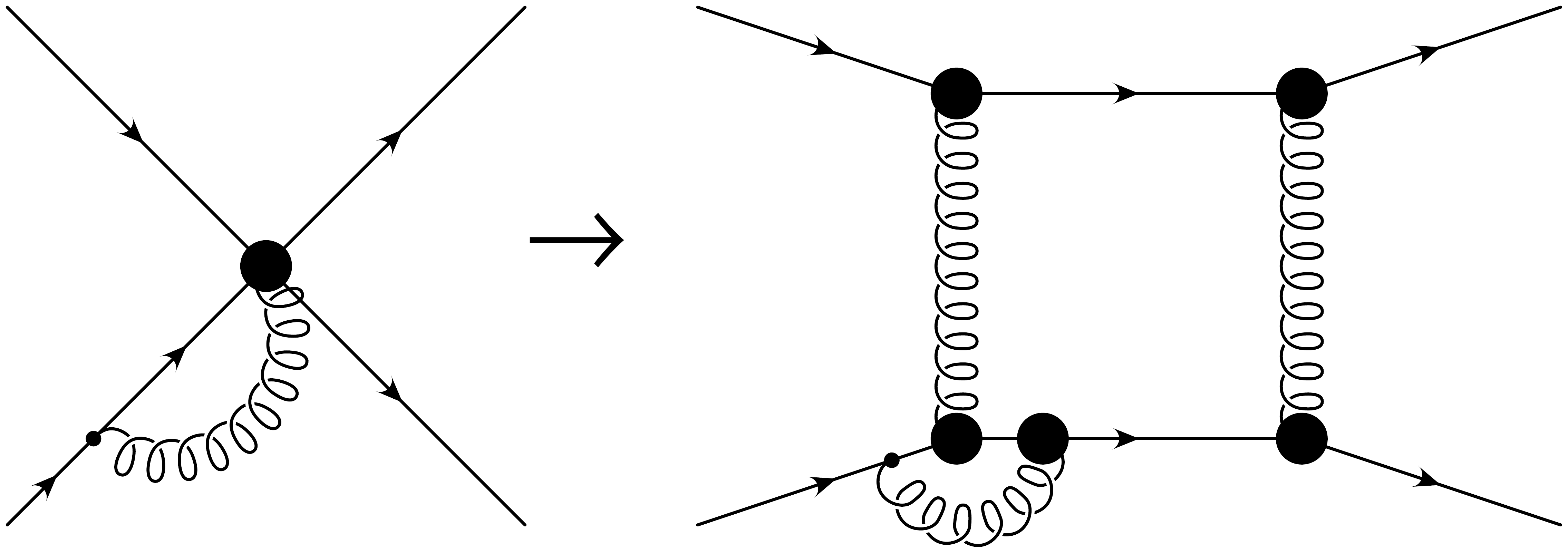}
 \caption{\small{Exemplary diagram in the skeleton expansion of the 5-point function in the quark-antiquark Dyson-Schwinger equation, which gives the dominating infrared exponent.}}
 \label{fig:4s_exp}
\end{figure}

\begin{figure}
 \centering
 \includegraphics[width=13cm]{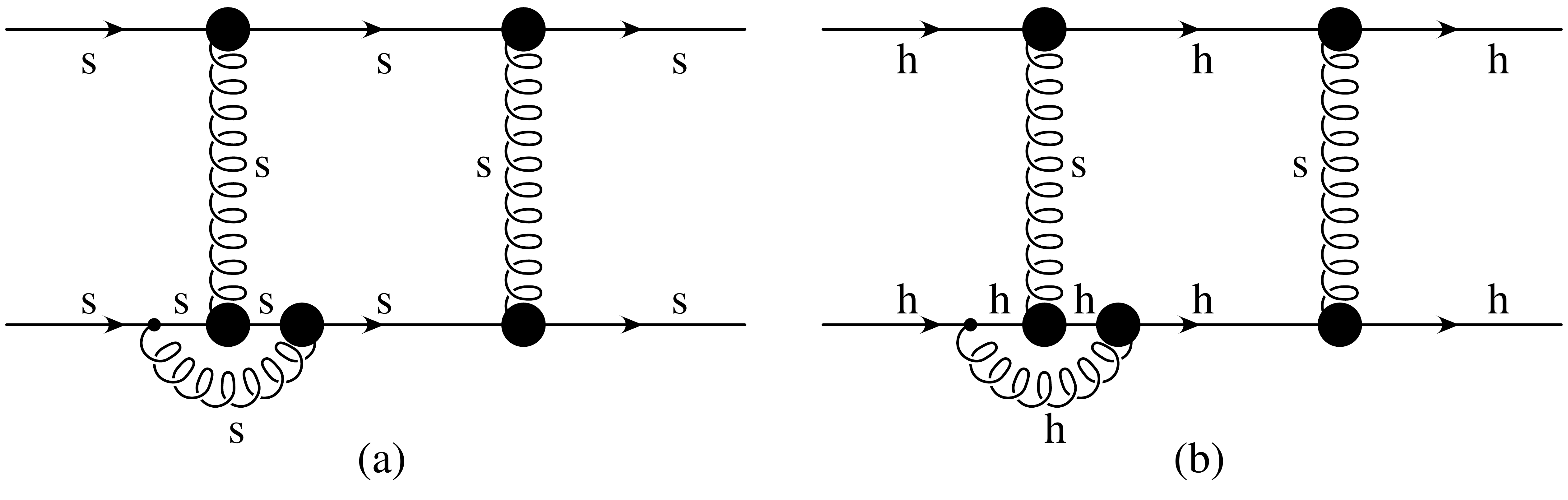}
 \caption{\small{This diagram emerges from a skeleton expansion of the four-quark-gluon vertex in the four-quark DSE. Different limits yield different scaling exponents, \textquotedblleft h\textquotedblright denote hard, \textquotedblleft s\textquotedblright soft momenta. Diagram \emph{(a)} shows the diagram in the uniform limit, whereas diagram \emph{(b)} is the leading diagram in the heavy quark limit with a soft-gluon exchange, which gives the correct infrared scaling exponent to yield a confining potential.}}
 \label{fig:heavy_quark_limits}
\end{figure}

The feature of confinement can be seen in the potential between two static quarks. In the Dyson-Schwinger formalism the relevant quantity is therefore the quark-antiquark scattering kernel\footnote{Note that there is no primitively divergent four-quark vertex in QCD, whereas there is a primitively divergent four-scalar vertex for fundamentally charged scalars.}. The DSE for the four-quark function can be derived according to the algorithm in chapter \ref{chapter2}, and it turns out, that this equation involves similar vertices as the four-scalar vertex in scalar Yang-Mills theory\footnote{It is necessary to mention that there are more diagrams in the four-scalar DSE in leading order, but these graphs involve either one bare four-scalar vertex or one bare two-scalar-two-gluon vertex, thus they are subleading in the infrared and they can be neglected in the power counting analysis. This justifies the determination of scaling exponents for fundamentally charged scalars including kinematic divergencies in Appendix \ref{app_D}, as the system can be mapped to quenched QCD.}
, see fig. \ref{fig:4s_vertex}. The four-quark DSE is given in \cite{Alkofer:2008tt}, where additionally a four-quark-gluon vertex is plotted, which is explicitly given in fig. \ref{fig:4s_exp} here. As this vertex is not a primitively divergent vertex, it has to be expanded by means of a skeleton expansion. In the uniform limit the emerging vertices yield the same order of divergence for the four-quark vertex as it has been obtained from one-loop diagrams. But it turns out that considering kinematic divergencies the infrared scaling behaviour differs from the uniform limit, because the scaling exponent for the four-quark vertex is raised, as I will sketch shortly.\\
In \cite{Alkofer:2008tt} the infrared exponents for the quark propagator and the quark-gluon vertex are calculated\footnote{Here the Dirac scalar parts have the same infrared exponents as the Dirac vector parts of the various $n$-point functions. The mechanism for confinement as it is described in \cite{Alkofer:2008tt} is therefore not crucially dependent on the Dirac structure of the quarks. This served as a motivation for the investigation of fundamentally charged scalars.}. These anomalous dimensions are summarised in table \ref{tab:exp_QCD} in the various kinematic limits, the infrared scaling exponents of the Yang-Mills vertices were given in table \ref{table_exp_puregauge}.\\
\begin{table}[!htb]
 \centering
 \begin{tabular}{|c|c|c|c|}
\hline
$\delta_{q}=0$ & $\delta_{qg}^{u}$ & $\delta_{qg}^{q}$ & $\delta_{qg}^{g}$ \\
\hline
0 & $-\frac{1}{2}-\kappa \ (\lor 0)$ & $ 0$ & $-\frac{1}{2}-\kappa \ (\lor 0)$\\
\hline
\end{tabular}
\caption{\small{Anomalous dimensions for the quark propagator and the quark-gluon vertices in the different kinematic limits. Note that the $\alpha_{sg}^{u}$ for the scalar-gluon vertex in the uniform limit was chosen such, that it has the same parametrization as $\delta_{qg}^{u}$ here.}}
\label{tab:exp_QCD}
\end{table}
With these exponents the skeleton expansion of the four-quark-gluon vertex in the limit of hard external momenta, as can be seen in diagram \emph{(b)} of fig. \ref{fig:heavy_quark_limits}\footnote{In the uniform limit as described above all external as well as internal momenta are soft.}, must be analysed. 

The different limits are illustrated in fig. \ref{fig:heavy_quark_limits} to clarify the various cases and the constituents in the power counting\footnote{Note that the hard momentum integral does not contribute with a $2$ in the exponent.}. \\
In the uniform limit the counting of diagram \emph{(a)} in fig. \ref{fig:heavy_quark_limits} gives 
\begin{equation}
 (p^2)^4 \Bigl((p^2)^{-1+2 \kappa} \Bigr)^{3} \Bigl( (p^2)^{-\frac{1}{2}- \kappa} \Bigr)^{5} = (p^2)^{-\frac{3}{2}+\kappa}.
\end{equation}
\\
If the momenta of the external quarks are hard, whereas the exchanged gluons becomes soft, which means that the quarks are sufficiently spatially separated, the power counting gives (for the solution with a scaling quark-gluon vertex)
\begin{equation}
 (p^2)^2 \Bigl((p^2)^{-\frac{1}{2}-\kappa} \Bigr)^{4} \Bigl( (p^2)^{-1+2\kappa} \Bigr)^{2} = (p^2)^{-2}.
\end{equation}
This is a crucial behaviour, because the scaling exponent of the quark-antiquark scattering kernel is the relevant quantity for the potential between two heavy static quarks. Heuristically confinement means, that only color-singlet states can be detected. By separating two quarks of finite mass, the energy between the particles will become large enough at some point, such that new mesons will be created from the vacuum. In the heavy quark limit this possibility is banned.  But this implies that the energy between the quarks must be infinite in the limit of infinitely separated quarks. Thus a necessary condition for quark confinement is a rising static quark potential. For the value of the infrared exponent of the quark-antiquark scattering kernel this potential is realised, which is shown in eq. (\ref{conf_pot}). From the scaling behaviour of the quark-antiquark scattering kernel in the heavy quark limit one can find the quark potential \emph{V(r)} by the relation
\begin{equation}\label{conf_pot}
  V(r) \sim \int d^3 p \frac{e^{ipr}}{p^4} \sim |r|,
\end{equation}

thus one finds a linearly rising potential for two static quarks.\\

As an important result of this thesis, the same mechanism for confinement also holds for scalars in the fundamental representation. By means of a power counting analysis the infrared exponents of the vertex functions in scalar Yang-Mills theory can be determined, see App. \ref{app_D}, where the system of quenched QCD is mapped onto scalar Yang-Mills theory. As a result the dominating infrared scaling exponents of the vertex functions in scalar Yang-Mills theory are the same as for the analogous functions in quenched QCD. Due to this fact one can argue that the mechanism for quark confinement also holds for heavy scalars. As the DSE for the four-scalar vertex in the heavy scalar limit also involves diagram \emph{(b)} in fig. \ref{fig:heavy_quark_limits}, the dominating scaling exponent for the four-scalar vertex is also $-2$. As there is the same scaling exponent there is qualitatively the same relation, which was given in eq. (\ref{conf_pot}), for the static scalar potential as for the heavy quark potential. Through this linearly rising potential heavy scalars can be confined by the same mechanism as it is proposed for heavy quarks in \cite{Alkofer:2008tt}. 
\\

These results show, that the qualitative scaling behaviour of the infrared dominating vertex functions of fundamentally charged scalars coupled to Yang-Mills theory is the same as for quenched QCD, at least for the cases of massless (massive) scalar particles without a scalar condensate, which can be compared to the chirally symmetric (broken) phase of quenched QCD, where only in the chirally broken phase a confining potential can be seen. Thus this behaviour is reproduced for fundamentally charged scalars. 

\chapter{Conclusion}

In this thesis the DSEs for Yang-Mills theory with a fundamentally charged scalar field were derived. Due to the quartic scalar- and the two-scalar-two-gluon interaction the diagrammatic form of the system of equations is more involved than the system obtained for the primitively divergent Green functions in QCD.\\

\noindent
For the coupled system of nine primitively divergent vertex functions the power counting analysis \cite{Alkofer:2004it} has been performed. It yields the scaling behaviour of the $n$-point functions in the deep infrared which is qualitatively equivalent to that of quenched QCD. \\
Considering a massless scalar particle one does not observe confinement. Characteristic of all mathematical solutions of the system is that \textit{(i)} the scaling of the Yang-Mills vertices in eq. (\ref{arb_ver}) is not altered by the coupling of scalar particles, \textit{(ii)} the scalar propagator stays bare and \textit{(iii)} the order of singularity of the scalar-gluon vertex is not high enough to trigger a linearly rising potential between scalars, thus there is no solution that shows confinement. An interesting and important reason for this is that the four-point vertices are not the dominating parts in the lower $n$-point functions. \\
For a massive particle there are two different cases, which can be distinguished by an either vanishing or finite expectation value of a non-local order parameter $Q$, given in eq. (\ref{order_parameter}). In this work the symmetric case, \textit{i.e.} $\langle Q \rangle = 0$ was investigated, where it was shown, that taking into account only the uniform infrared divergences of the vertices is not a sufficient approximation to fully describe the infrared behaviour of the system. Assuming that four-point functions do not dominate the lower $n$-point functions as shown in the uniform limit, one finds that considering soft singularities yields a solution with a static confinement of massive scalars in the quenched limit. Herein the scalar propagator stays bare, and the Yang-Mills vertices scale according to eq. (\ref{arb_ver}).\\

\noindent
The infrared scaling behaviour of scalars coupled to Yang-Mills theory shows that the confinement mechanism as it was proposed by \cite{Alkofer:2008tt, Schwenzer:2008vt,Schwenzer:2008prep} also holds for bosonic particles. This feature suggests that it is not the Dirac structure of the quark which is crucial for confinement. It is rather an indication, that confinement is exclusively due to non-Abelian gauge dynamics. Thereby, this result supports the confinement mechanism in \cite{Alkofer:2008tt, Schwenzer:2008vt,Schwenzer:2008prep}.\\

\noindent
The comparison of the different systems yields that the same diagrams dominate in the DSEs for the scalar particle, as in the analogous equations in quenched QCD. Thus it is reasonable to take fundamentally charged particles coupled to Yang-Mills theory as a model for quenched QCD. The advantage of this model is that it is easier to find a proper truncation for the DSEs, because there are less tensor structures for bosonic particles as for fermions. \\
In lattice gauge theories dynamical bosons are easier to implement as dynamical fermions. Thus an investigation of the presented model brings about a severe simplification also in lattice studies, which will be done in the near future \cite{Maasprep}. A comparison of the results of both methods will give deeper insight in the validity of the mechanism of confinement, as proposed in \cite{Alkofer:2008tt}.\\

\appendix
\chapter{\normalsize Bare Vertex-Functions} \label{app_bare}
The expressions for the bare propagators and vertices can be calculated from the quantised Euclidean action of Yang-Mills theory with a coupled fundamentally charged scalar field by functional derivations with respect to the in- and out-going particles, that contribute to the bare Green functions. This appendix gives the bare $n$-point functions in momentum space, that are obtained by a Fourier transformation according to the transformation law for an arbitary operator $\hat{O}$.
\begin{equation*}
\hat{O}(x) = \int \frac{d^d p}{(2 \pi^{4})} \ \hat{O} (p) \ e^{-ip\cdot x}, \ \hat{O}(p) = \int d^d x \ \hat{O} (x) \ e^{ip\cdot x}.
\end{equation*}

\begin{table}[p]
 \begin{tabular}{|m{2.4cm}| m{10.5cm}|}
\hline
  \footnotesize bare scalar propagator & \vspace{-0.1cm} \begin{footnotesize} \begin{equation} 
 S_{0,ij}(p^2)  = - \delta_{ij} \frac{1}{p^2+m^2} \label{scprop}
\end{equation}  \end{footnotesize} \vspace{-0.3cm}
\\ 
\hline
  \footnotesize bare gluon propagator & \vspace{-0.1cm} \begin{footnotesize}\begin{equation}  D_{0,\mu \nu}(p^2) =  \left( \delta_{\mu \nu} -\frac{p_{\mu} p_{\nu}}{p^2}  \right)  \frac{1}{p^2} \end{equation} \end{footnotesize} \vspace{-0.3cm}
\\
\hline	
  \footnotesize bare ghost propagator & \vspace{-0.1cm} \begin{footnotesize}\begin{equation}D^{G}_{0}(p^2) = -\frac{1}{p^2} \end{equation} \end{footnotesize} \vspace{-0.3cm}
\\ \hline
  \footnotesize bare scalar-gluon vertex & \vspace{-0.5cm} \begin{footnotesize} \begin{eqnarray}  \Gamma_{0,ij\mu}^{a}(p,q,r) = -\frac{1}{2} g  \left( t^{a} \right)_{ij} r_{\mu} \delta ( p-q+r )
 \end{eqnarray} \end{footnotesize} \vspace{-0.4cm}
\\ 
\hline
  \footnotesize bare ghost-gluon vertex & \vspace{-0.2cm} \begin{footnotesize} \begin{equation}  \Gamma_{0,\mu}^{abc}(p,q,r) = - i g f^{abc} q_{\mu} \delta (p-q+r) \end{equation} \end{footnotesize} \vspace{-0.4cm}
\\ 
\hline
  \footnotesize bare three-gluon vertex & \vspace{-0.6cm} \begin{footnotesize} \begin{eqnarray} \label{bare3g} & \Gamma_{0,\mu\nu\rho}^{abc}(p,q,r) &=  i g f^{abc} \delta  (p+q+r) \nonumber \\ 
&& \Big[ \delta_{\mu\nu} (p-q)_{\rho}  + \delta_{\nu\rho} (q-r)_{\nu}+ \delta_{\rho \mu} (r-p)_{\nu} \Big]  \end{eqnarray} \end{footnotesize} \vspace{-0.6cm}
\\ 
\hline 
  \footnotesize bare four-gluon vertex & \vspace{-0.6cm} \begin{footnotesize} \begin{eqnarray} &\Gamma_{0,\mu\nu\rho\sigma}^{abcd} (p,q,r,s) = & g^2 \delta (p+q+r+s) \times \nonumber \\ 
	&& \times \Big[ f^{jab}f^{jcd} (\delta_{\mu\rho} \delta_{\nu\sigma}-\delta_{\mu\sigma}\delta_{\nu\rho}) \nonumber  \\
	&& \ \ \ + f^{jac} f^{jbd} ( \delta_{\mu\nu} \delta_{\rho\sigma}-\delta_{\mu\sigma}\delta_{\nu\rho} ) \nonumber  \\ 
	&& \ \ \ + f^{jad} f^{jbc} ( \delta_{\mu\nu} \delta_{\rho\sigma}-\delta_{\mu\rho}\delta_{\nu\sigma}) \Big]   \end{eqnarray} \end{footnotesize} \vspace{-0.6cm}\\
\hline
   \footnotesize bare four-scalar vertex & \vspace{-0.6cm} \begin{footnotesize} \begin{eqnarray}  \Gamma_{0,ijkl}(p,q,r,s) = \frac{\lambda}{3!} \delta_{ij} \delta_{kl} \delta (p-q+r-s)  \end{eqnarray} \end{footnotesize}\vspace{-0.6cm} 
\\
\hline
  \footnotesize bare two-scalar-two-gluon vertex & \vspace{-0.6cm} \begin{footnotesize} \begin{eqnarray}\Gamma_{0,ij\mu\nu}^{ab} (p,q,r,s) = g^2 \left( \left(t^{a}\right)_{kj}\left(t^{b}\right)_{ji} +\left(t^{b}\right)_{kj}\left(t^{a}\right)_{ji} \right) \times \nonumber\\ \times \ \delta_{\mu\nu} \delta (p-q+r-s) 
 \end{eqnarray} \end{footnotesize}\vspace{-0.6cm} \\ 
\hline
 \end{tabular}
\end{table}

\chapter{\normalsize Dyson-Schwinger Equations}\label{DSEs}
\begin{figure}[!htb]
 \centering
 \includegraphics[width=13.8cm]{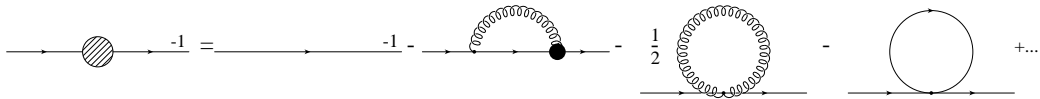}
 \caption{\small{Dyson-Schwinger equation for the scalar propagator.}}
 \label{fig:scalar_prop}
\end{figure}

\begin{figure}[!htb]
 \centering
 \includegraphics[width=13.8cm]{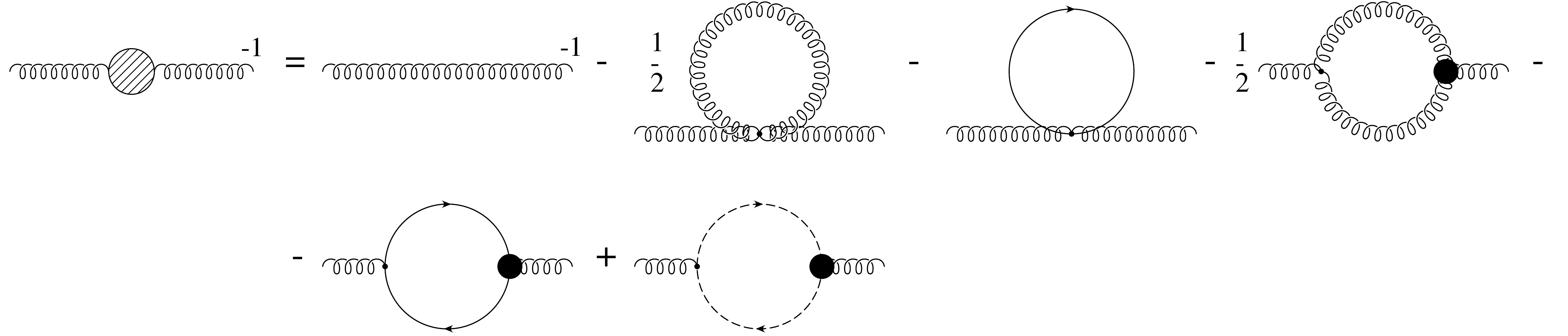}
 \caption{\small{Dyson-Schwinger equation  for the gluon propagator.}}
 \label{fig:gluon_prop}
\end{figure}

\begin{figure}[!htb]
\centering
\includegraphics[width=10cm]{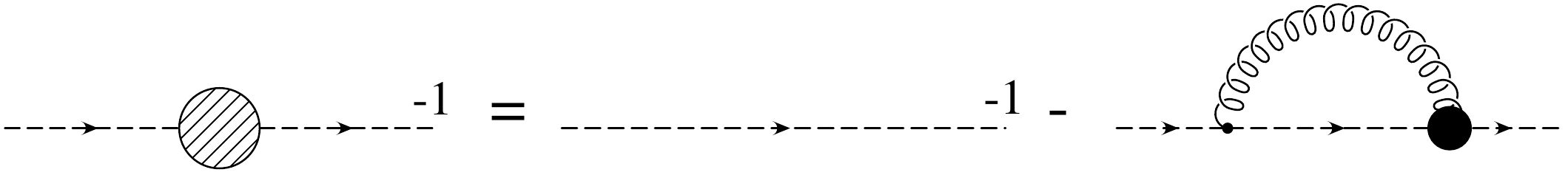}
\caption{\small{Dyson-Schwinger equation  for the ghost propagator.}}
\label{fig:ghost_prop}
\end{figure}

\begin{figure}[!htb]
\centering
\includegraphics[width=13.8cm]{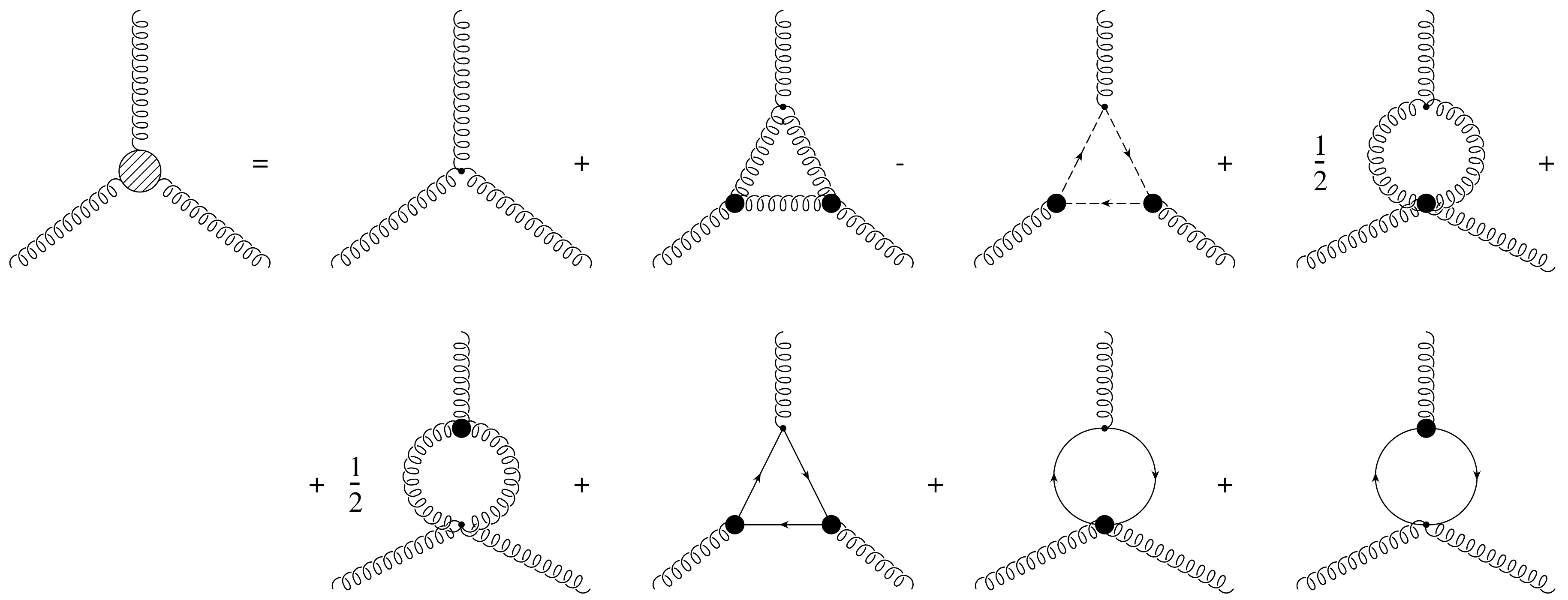}
\caption{\small{Dyson-Schwinger equation  for the 3-gluon vertex.}}
\label{fig:3g_vertex}
\end{figure}

\begin{figure}[!htb]
\centering
\includegraphics[width=13.8cm]{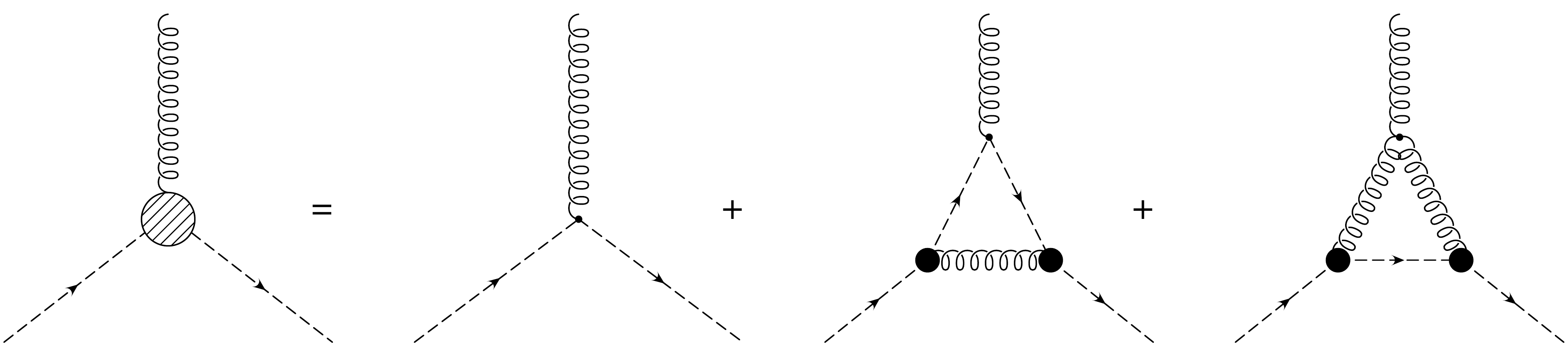}
\caption{\small{Dyson-Schwinger equation  for the ghost-gluon vertex.}}
\label{fig:ggh_vertex}
\end{figure}

\begin{figure}[!htb]
\centering
\includegraphics[width=13.8cm]{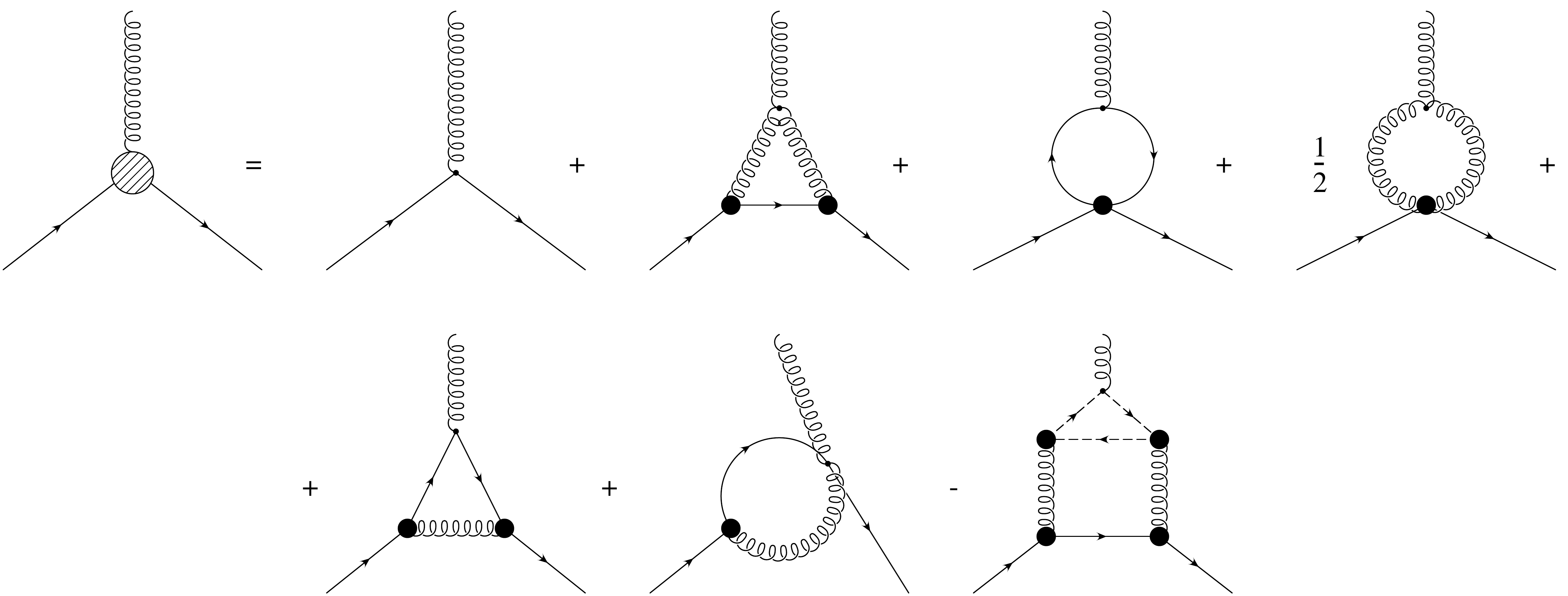}
\caption{\small{Dyson-Schwinger equation  for the scalar-gluon vertex.}}
\label{fig:sg_vertex}
\end{figure}

\begin{figure}[!htb]
\centering
\includegraphics[width=13.8cm]{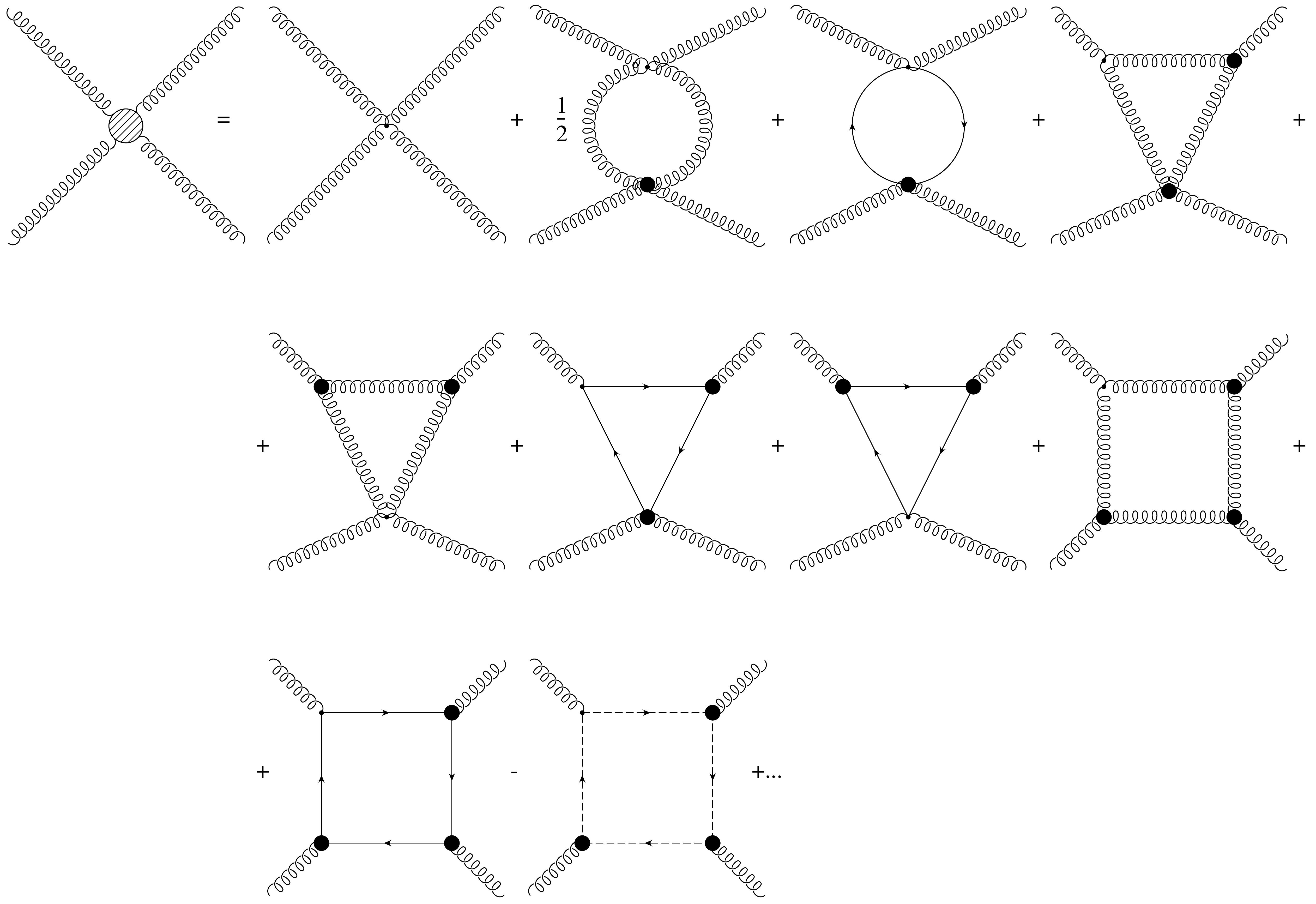}
\caption{\small{Dyson-Schwinger equation  for the 4-gluon vertex.}}
\label{fig:4g_vertex}
\end{figure}

\begin{figure}[!htb]
\centering
\includegraphics[width=13.8cm]{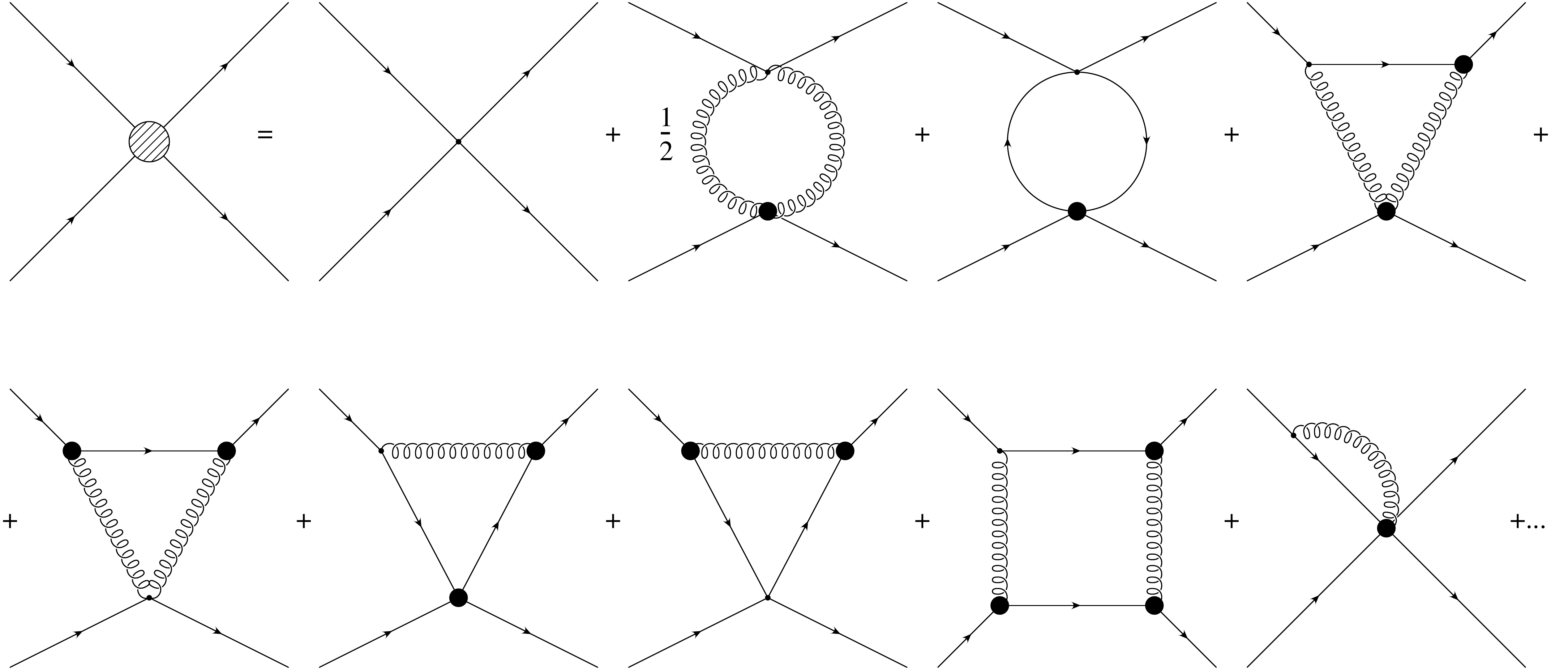}
\caption{\small{Dyson-Schwinger equation  for the 4-scalar vertex.}}
\label{fig:4s_vertex}
\end{figure}

\begin{figure}[!htb]
\centering
\includegraphics[width=13.8cm]{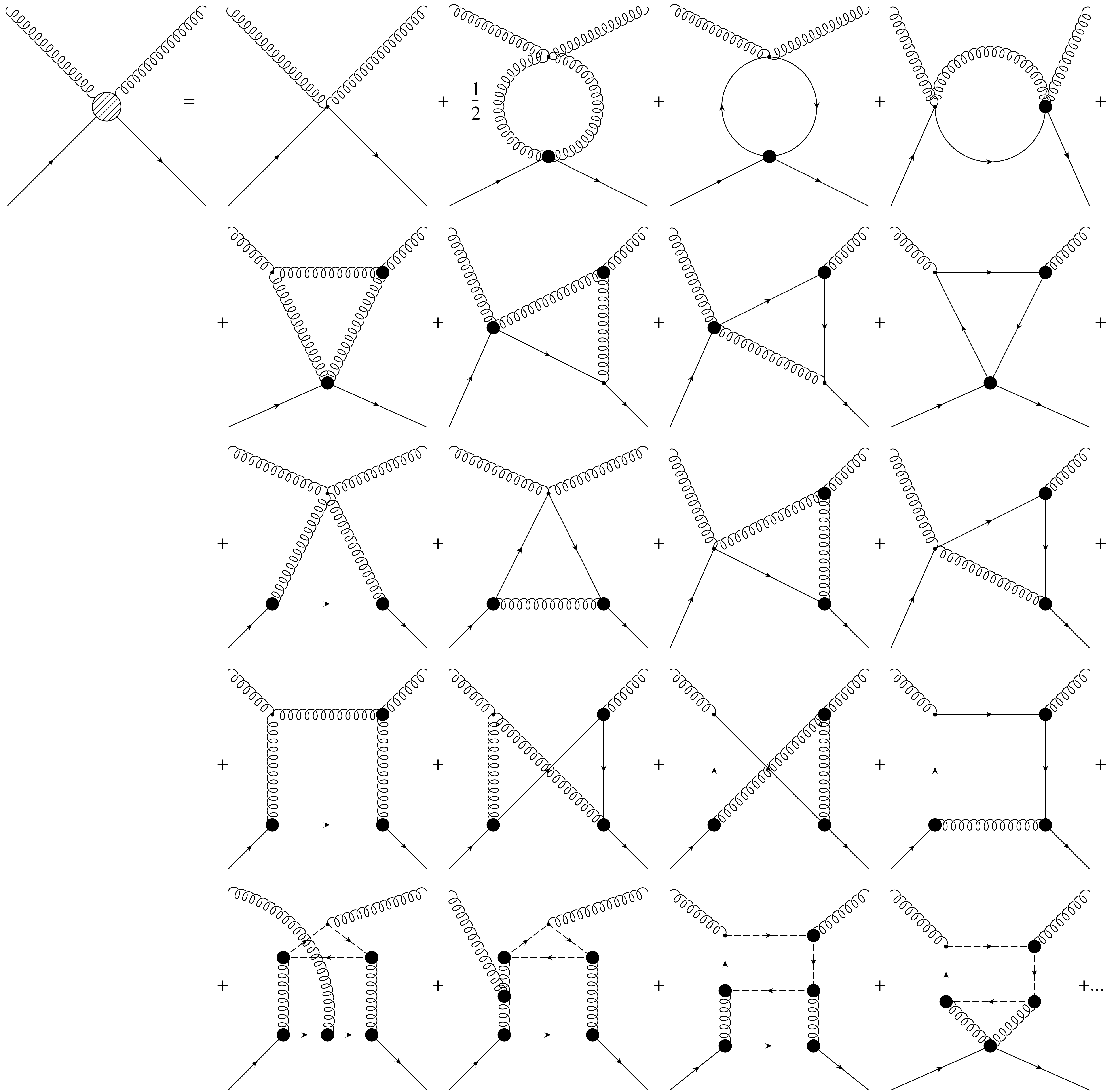}
\caption{\small{Dyson-Schwinger equation  for the 2-scalar-2-gluon vertex.}}
\label{fig:2s2g_vertex}
\end{figure}

The figures  \ref{fig:scalar_prop}, \ref{fig:gluon_prop}, \ref{fig:ghost_prop}, \ref{fig:3g_vertex}, \ref{fig:ggh_vertex}, \ref{fig:sg_vertex}, \ref{fig:4g_vertex}, \ref{fig:4s_vertex}, \ref{fig:2s2g_vertex} show only the truncated DSEs according to the truncation presented in the main text. Only topologically different diagrams have to be taken into account with respect to the power counting analysis, because topologically equal ones contribute with the same exponents. The \textquotedblleft \emph{+...}\textquotedblright in the diagrams denote various permutations, which I omit for brevity. Internal propagators are dressed.\\
The full DSEs have been checked with \emph{DoDSE}, a \emph{Mathematica}-package for the algorithmic derivation of DSES, cf. \cite{Alkofer:2008nt}.
\clearpage

\chapter{\normalsize Full System of Equations for the Infrared Exponents of Scalar Yang-Mills theory}\label{App_full_eqn}
The power counting gives a coupled system of equations for the infrared exponents of the primitively divergent $n$-point functions. In the uniform limit this system is given by:
\begin{eqnarray*}
 1 -\delta_{s} & = & \min  ( 1-\mu, 1+\delta_{s}+\delta_{g}+\delta_{sg}, 1+\delta_{g}, 1+\delta_{s} ) 
\\
1-\delta_{g} & = & \min ( 1, 1+ \delta_{g}, 1+ \delta_{s}, 1+2 \delta_{g}+ \delta_{3g}, 1+2\delta_{s}+\delta_{sg}, \\ 
&& ,1 + 2 \delta_{gh}+\delta_{ggh} )
\\
1-\delta_{gh} & = & \min ( 1, 1+\delta_{g}+\delta_{gh}+\delta_{ggh})
\\
\nicefrac{1}{2}+\delta_{3g} & = & \min ( \nicefrac{1}{2} , \nicefrac{1}{2}+ 3 \delta_{g}+2 \delta_{3g}, \nicefrac{1}{2}+3 \delta_{gh}+2 \delta_{ggh}, \nicefrac{1}{2}+ 2\delta_{g}+\delta_{4g}, \\
&&, \nicefrac{1}{2}+2 \delta_{g}+\delta_{3g},\nicefrac{1}{2}+3 \delta_{s}+2 \delta_{sg},  \nicefrac{1}{2}+ 2 \delta_{s}+\delta_{ssgg}, \\ && ,\nicefrac{1}{2}+2 \delta_{s}+\delta_{sg} )\\
\nicefrac{1}{2}+ \delta_{ggh} & = & \min  \nicefrac{1}{2}, \nicefrac{1}{2}+\delta_{g}+2 \delta_{gh}+2\delta_{ggh}, \nicefrac{1}{2}+2 \delta_{g}+\delta_{gh}+ 2 \delta_{ggh} ) \\
\nicefrac{1}{2}+\delta_{sg} & = & \min ( \nicefrac{1}{2}, \nicefrac{1}{2}+\delta_{s}+2 \delta_{g}+2\delta_{sg}, \nicefrac{1}{2}+2 \delta_{s}+\delta_{4s}, \nicefrac{1}{2}+2 \delta_{g}+\delta_{ssgg}, \\
&&, \nicefrac{1}{2}+2\delta_{s}+\delta_{g}+2\delta_{sg}, \nicefrac{1}{2}+\delta_{s}+\delta_{g}+\delta_{sg}, \\ &&, \nicefrac{1}{2}+ \delta_{s}+2\delta_{g}+3\delta_{gh}+2\delta_{sg}+2\delta_{ggh})
\\
\delta_{4g} &= & \min ( 0, 2 \delta_{g}+ \delta_{4g}, 2 \delta_{s}+\delta_{ssgg}, 3 \delta_{g}+\delta_{3g}+\delta_{4g},3 \delta_{g}+2 \delta_{3g},\\
&&,3 \delta_{s}+\delta_{sg}+\delta_{ssgg} 3 \delta_{s}+2 \delta_{sg}, 4\delta_{g}+3 \delta_{3g}, 4 \delta_{s}+ 3 \delta_{sg}, \\ && ,4 \delta_{gh}+3 \delta_{ggh} 
)\\
\delta_{4s} & = & \min ( 0, 2\delta_{g}+\delta_{ssgg},  2\delta_{s}+\delta_{4s}, \delta_{s}+2 \delta_{g}+\delta_{sg}+\delta_{ssgg},\\
&&, \delta_{s}+2 \delta_{g}+2\delta_{sg}, 2\delta_{s}+\delta_{g}+\delta_{sg}+\delta_{4s},2\delta_{s}+\delta_{g}+2\delta_{sg}, \\&& ,2\delta_{s}+2\delta_{g}+3\delta_{sg}) \\
\delta_{ssgg} &= & \min ( 0, 2 \delta_{g}+ \delta_{ssgg}, 2\delta_{s}+\delta_{4s}, \delta_{s}+\delta_{g}+\delta_{ssgg},3 \delta_{g}+\delta_{3g}+\delta_{ssgg},\\
&&,\delta_{s}+2\delta_{g}+\delta_{3g}+\delta_{ssgg}, 2 \delta_{s}+\delta_{g}+\delta_{sg}+\delta_{ssgg}, 3 \delta_{s}+ \delta_{sg}+\delta_{4s},\\ 
&&,\delta_{s}+2\delta_{g}+2\delta_{sg}, \delta_{s}+2\delta_{g}+\delta_{sg}+\delta_{3g}, 2\delta_{s}+\delta_{g}+2\delta_{sg}, \\
&&, \delta_{s}+3 \delta_{g}+ 2 \delta_{sg}+ \delta_{3g}, 2 \delta_{s} + 2 \delta_{g}+ 3 \delta_{sg},  2 \delta_{s}+2 \delta_{g}+2 \delta_{sg}+ \delta_{3g}, \\
&& ,  3 \delta_{s}+\delta_{g}+3 \delta_{sg}, 2 \delta_{s}+ 2 \delta_{g} + 3 \delta_{gh} + 3 \delta_{sg} + 2 \delta_{ggh}, \\
&&,\delta_{s}+3\delta_{g}+3\delta_{gh}+2\delta_{sg}+\delta_{3g}+2\delta_{ggh}, \\ && ,\delta_{s}+ 2 \delta_{g}+ 4 \delta_{gh}+2 \delta_{sg}+3 \delta_{ggh},2 \delta_{g}+ 4 \delta_{gh}+3\delta_{ggh}+\delta_{ssgg})
\end{eqnarray*}

\chapter[Scalar-Gluon Vertex in Different Kinematic Regions]{\normalsize DSEs and Power Counting for the Scalar-Gluon Vertex for Different Kinematic Regions}\label{app_D}
\begin{figure}[!htb]
 \centering
 \includegraphics[width=13.8cm]{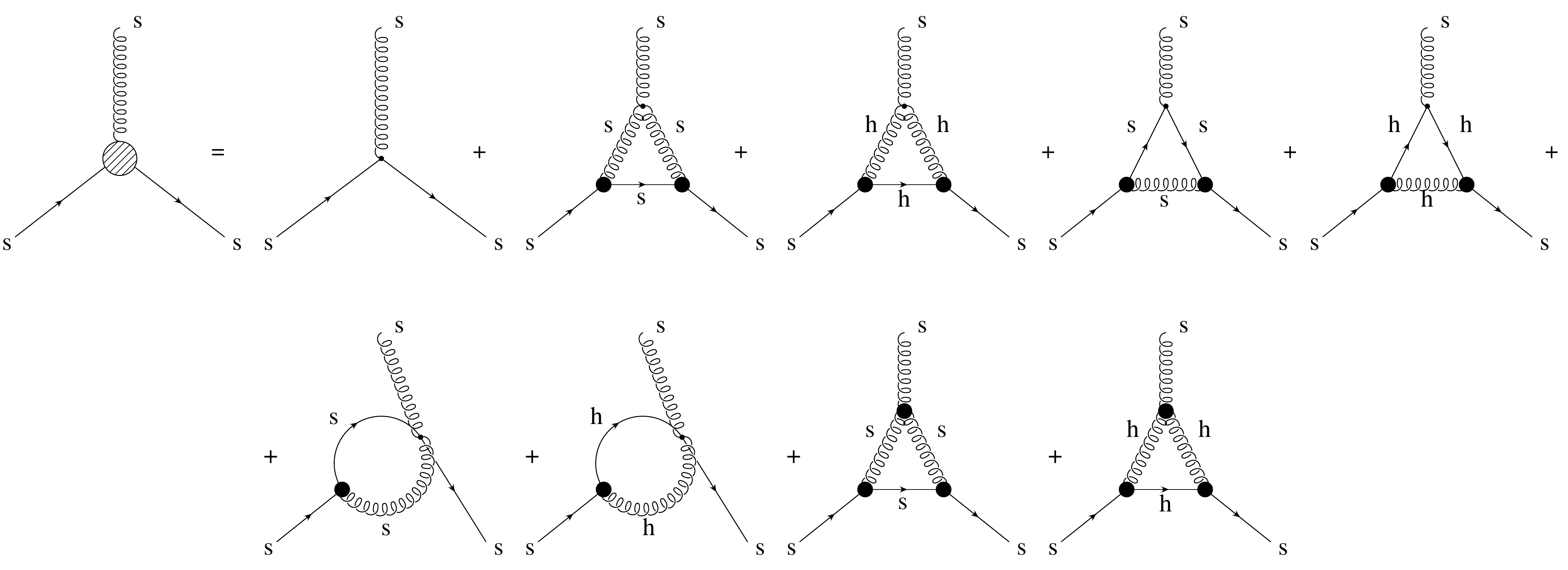}
\caption{\small{DSE for the scalar-gluon vertex in the uniform limit.}}
\label{fig:kin_uniform}
\end{figure}

\begin{figure}[!htb]
 \centering
 \includegraphics[width=13.8cm]{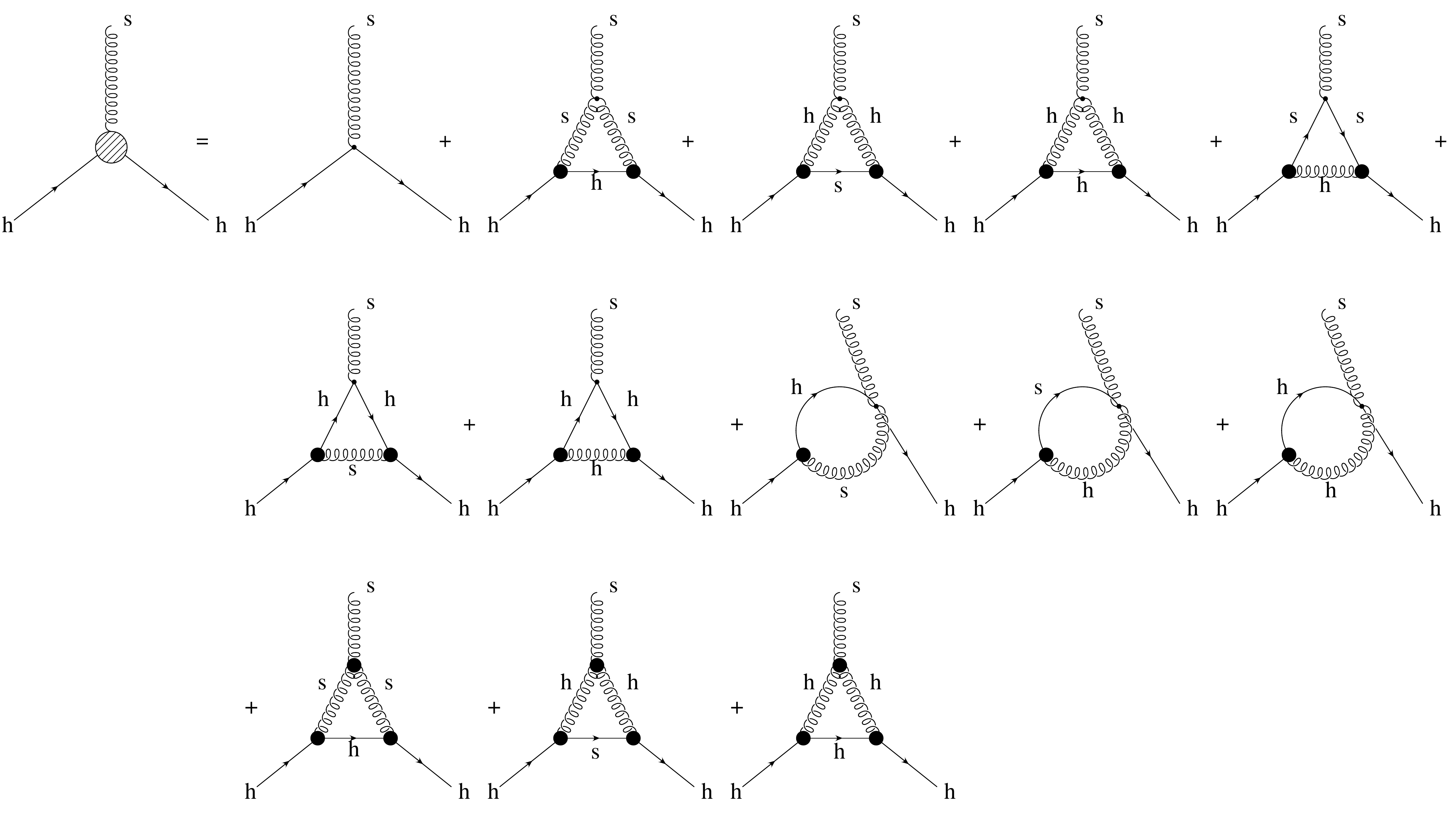}
 \caption{\small{DSE for the scalar-gluon vertex in the soft gluon limit.}}
 \label{fig:kin_soft_gluon}
\end{figure}

\begin{figure}[!htb]
 \centering
 \includegraphics[width=13.8cm]{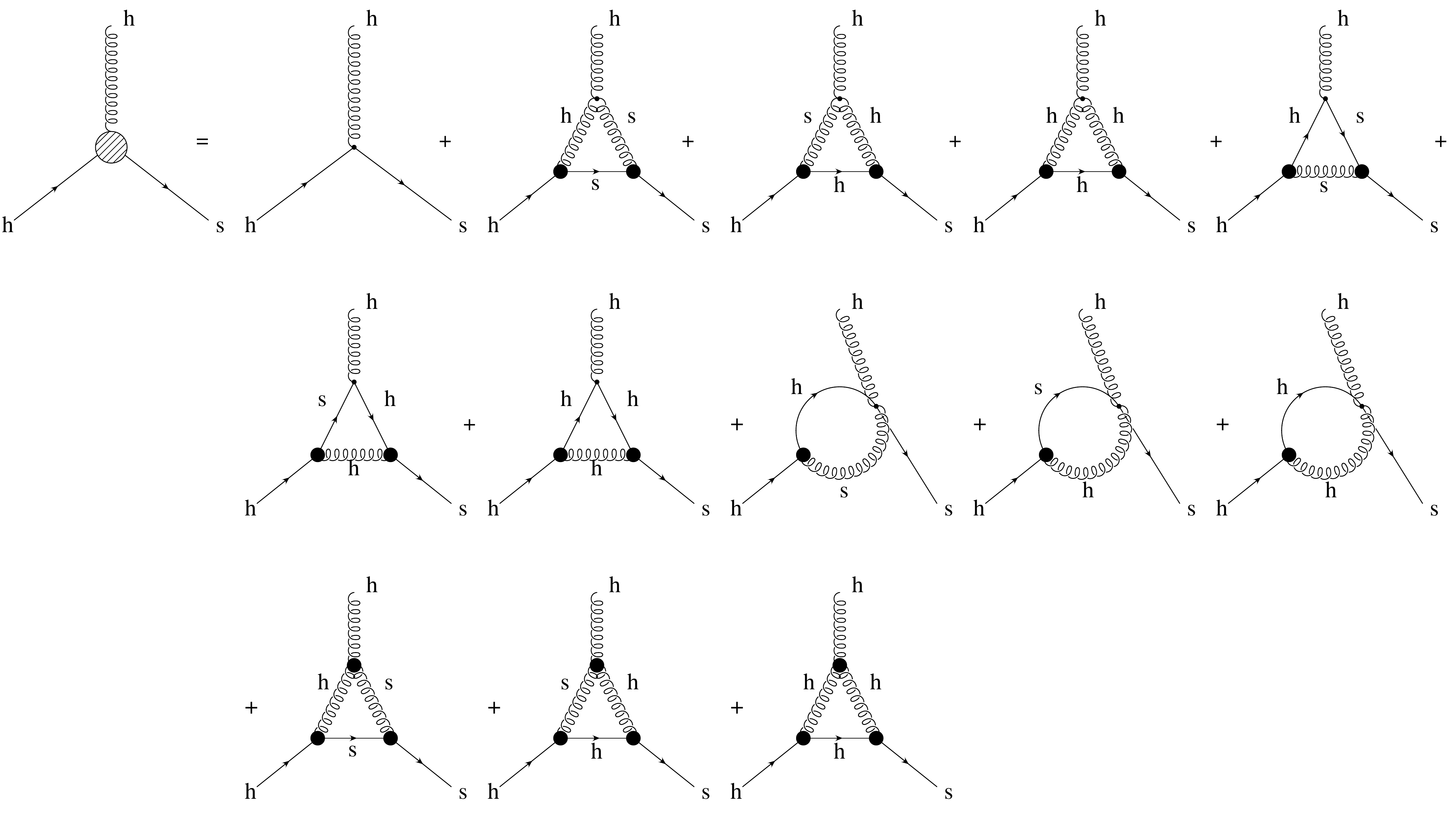}
 \caption{\small{DSE for the scalar-gluon vertex in the soft scalar limit.}}
 \label{fig:kin_soft_scalar}
\end{figure}

\begin{figure}[!htb]
 \centering
 \includegraphics[width=4cm]{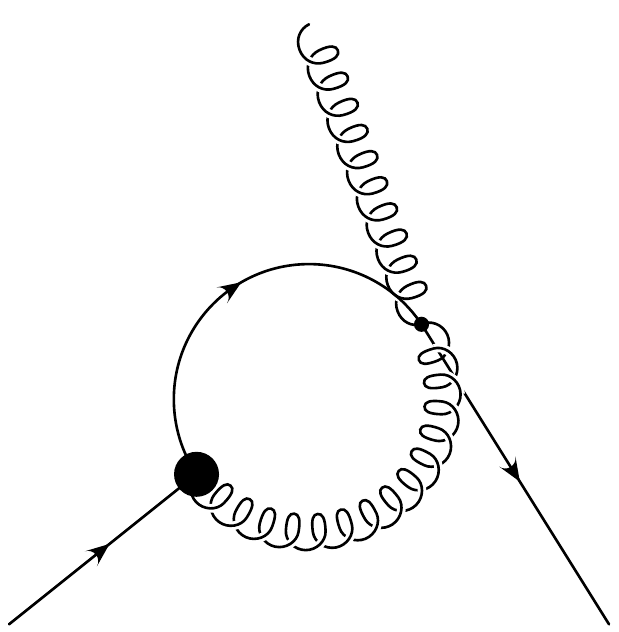}
 \caption{\small{Additional diagram for the scalar-gluon DSE of fundamentally charged scalars in Yang-Mills theory, that is not involved in the quark-gluon vertex DSE.}}
 \label{fig:add_diag_for_kin}
\end{figure}

In this appendix the possibility of kinematic divergencies is considered. As shown in chap. \ref{chapkindiv} several kinematic limits can exist in a 3-point function, which may yield a different scaling behaviour. The exponents for the Yang-Mills sector are known and remain unchanged by inclusion of scalars. For the scalar-gluon vertex the scaling exponents have to be determined simultaneously with the exponent for the scalar propagator. This is done as in the uniform case by means of a power counting analysis. \\
The Dyson-Schwinger equation for the scalar propagator in the uniform limit, which is the only possible limit for a two-point function, is given in fig. \ref{fig:scal_prop_kin}. The equations for the scalar-gluon vertex in the different kinematic limits are given in the figures \ref{fig:kin_uniform}, \ref{fig:kin_soft_gluon} and \ref{fig:kin_soft_scalar}. Contributions from different momentum regions in an integral are expressed in separate diagrams. Captions $s$ and $h$ denote soft and hard momenta.
\\
\\
\\
The power counting analysis is performed with respect to the parametrization for the scalar propagator and the scalar-gluon vertex as given in the equations \ref{eq_para_scal_prop}, \ref{eq_para_uniform}, \ref{eq_para_soft_gluon} and \ref{eq_para_soft_scalar}, which is chosen with regard to a comparison to quenched QCD. Note that the scalar propagator $\delta_{s}^{\prime}$ in this case is differently parametrised as in the calculation in the uniform limit. As it approaches a finite value for vanishing momenta this parametrization is reasonable.
\begin{eqnarray}
S               & \sim & \big(p^2\big)^{\delta_{s}^{\prime}} \label{eq_para_scal_prop} \\
\Gamma_{sg}^{u} & \sim & \big(p^2\big)^{\nicefrac{1}{2}+\delta_{sg}^{u}} \label{eq_para_uniform}\\
\Gamma_{sg}^{g} & \sim & \big(p^2\big)^{\delta_{sg}^{g} } \label{eq_para_soft_gluon}\\
\Gamma_{sg}^{s} & \sim & \big(p^2\big)^{\delta_{sg}^{s}} \label{eq_para_soft_scalar}
\end{eqnarray}
The usual power counting is combined with the values for the Yang-Mills vertex scaling exponents, given in table \ref{table_exp_puregauge}. Further a re-definition of the exponent of the scalar-gluon vertex in the uniform limit $\alpha_{sg}^{u}=\delta_{sg}^{u}+\nicefrac{1}{2} $ is done, because of the different canonical dimension of the scalar-gluon and quark-gluon vertices. Thus this definition is in agreement with the parametrization of \cite{Alkofer:2008tt}. Altogether this leaves a system of equations, given in eqs. (\ref{kin_scal})-(\ref{kin_sg_s}). This system is very similar to the one for quenched QCD, except for the terms that contain a bare 4-point function. These terms will be shown to be subleading.
\\
\begin{small}
\begin{eqnarray}
\label{kin_scal}- \delta_{s}^{\prime} & = & \min \Big(  0, \ 2+\delta_{s}^{\prime}+\delta_{g}+\delta_{sg}^{u}, \ 1+\delta_{sg}^{s} \ \Big) \\
\alpha_{sg}^{u} & = & \min \Big(  \nicefrac{1}{2}; \ \nicefrac{1}{2}+4\kappa+\delta_{s}^{\prime}+2\alpha_{sg}^{u}, \ 2\delta_{sg}^{s}+1; \ 
\nicefrac{3}{2}+2\delta_{s}^{\prime}+2\kappa+2\alpha_{sg}^{u} , \nonumber \\ && , 2\delta_{sg}^{s}+1; \ ,1+\delta_{s}^{\prime}+2\kappa+ \alpha_{sg}^{u}, \  \delta_{sg}^{s}+\nicefrac{1}{2}; \ \nicefrac{1}{2}+\kappa+\delta_{s}^{\prime}+2\alpha_{sg}^{u}, \nonumber \\ && ,2-2\kappa+2\delta_{sg}^{s} \Big) \label{alpha_sg}
\\
\delta_{sg}^{g} & = & \min \Big(  0; \ \nicefrac{1}{2}+4\kappa+2\delta_{sg}^{g}, \ 3+\delta_{s}^{\prime}+2\delta_{sg}^{s}, \ 0; \ \nicefrac{7}{2}+2\delta_{s}^{\prime}+2\delta_{sg}^{s}, \nonumber \\ && ,1+2\kappa+2\delta_{sg}^{g}, \ 0; \ 1+2\kappa+\delta_{sg}^{g}, \nicefrac{5}{2}+\delta_{s}^{\prime}+\delta_{sg}^{s}, \ 0; \ \nicefrac{1}{2}+\kappa+2\delta_{sg}^{g}, \nonumber \\ &&, 3+\delta_{s}^{\prime}+1-2\kappa+2\delta_{sg}^{s}, \ 1-2\kappa \Big) \label{kin_sg_g}
\\
 \delta_{sg}^{s} & = & \min \Big( 0, \ \nicefrac{3}{2}+2\kappa+\delta_{s}^{\prime}+\alpha_{sg}^{u}+\delta_{sg}^{s}, \ 
\nicefrac{3}{2}+2\kappa+\delta_{sg}^{g}+\delta_{sg}^{s}, \ \delta_{sg}^{s}+\nicefrac{1}{2}; \nonumber \\ 
&&  ;1+2\kappa+\delta_{s}^{\prime}+\delta_{sg}^{g}+\alpha_{sg}^{u},3+\delta_{s}^{\prime}+2\delta_{sg}^{s}, \  \delta_{sg}^{s}+\nicefrac{1}{2}; \ 1+2\kappa+\delta_{sg}^{g}, \nonumber \\ && ,\nicefrac{5}{2}+\delta_{s}^{\prime}+\delta_{sg}^{s}, \ 0; \ \nicefrac{3}{2}+\delta_{s}^{\prime}+1+\delta_{sg}^{s}+\alpha_{sg}^{u}
,\nicefrac{5}{2}+\delta_{sg}^{g}+\delta_{sg}^{s}, \ \delta_{sg}^{s}+\nicefrac{1}{2} \Big) \nonumber \\ && \label{kin_sg_s}
\end{eqnarray}
\end{small}
As the DSEs in App. \ref{DSEs} show there are additional diagrams which do not occur in QCD. These diagrams stem from the \textquotedblleft sheep\textquotedblright -diagram, which is not part of the QCD equations, because of a lack of bare two-quark-two-gluon vertex. The graph of interest is given again explicitly in fig. \ref{fig:add_diag_for_kin} for illustrational reasons.

The consequence for the scalar-gluon DSE in the uniform limit is the occurence of the following terms
\begin{eqnarray}
& 1+\delta_{s}^{\prime} + 2 \kappa + \alpha_{sg}^{u}, \label{u_con1}\\
& \delta_{sg}^{s}+\nicefrac{1}{2},\label{u_con2}
\end{eqnarray}
which must be shown to be subleading. The first eq. (\ref{u_con1}) cannot dominate, because all terms in the sum of $\alpha_{sg}^{u}$ are constrained to be $\geq 0$. Thus this term cannot be leading.\\
If the second term were the leading one it would have to be $<0$. But for any number $a<0$: $2a<a$. Thus the term in (\ref{u_con2}) cannot dominate, because there is the term
\begin{equation}
 2  \delta_{sg}^{s} + 1 \ \biggl( =  2 \times \big( \delta_{sg}^{s}+\nicefrac{1}{2} \big) \leq \delta_{sg}^{s}+\nicefrac{1}{2} \  \forall \ \delta_{sg}^{s} \leq -\nicefrac{1}{2} \biggr),
\end{equation}
which has its analogon in the QCD equation.\\
In the soft-gluon limit the terms, that must be shown to be subleading, are
\begin{eqnarray}
 & 1+2\kappa+\delta_{sg}^{g}, \label{g_con1}\\
& \nicefrac{5}{2}+\delta_{s}^{\prime}+ \delta_{sg}^{s}. \label{u_con2_a}
\end{eqnarray}
The first term (\ref{g_con1}) cannot dominate, because $\delta_{sg}^{g}$ appears in the equation for itself in a sum with a positive definite term. Thus it has to be subleading.\\
For the second term (\ref{u_con2_a}) one can use a constraint from the linear term in eq. (\ref{kin_sg_s}):
\begin{equation}
 \nicefrac{5}{2}+\delta_{s}^{\prime}+ \delta_{sg}^{s} \geq 0.
\end{equation}
The two terms under investigation in the soft-scalar limit are 
\begin{eqnarray}
 & \delta_{sg}^{s}+\nicefrac{1}{2}, \label{s_con1}\\
 & 1+2\kappa+\delta_{sg}^{g} .\label{s_con2}
\end{eqnarray}
For (\ref{s_con1}) the idea is equal as in the soft-gluon limit. $\delta_{sg}^{s}$ occurs in the equation for itself in a sum with a positive definite term. So it is necessarily subleading. \\
For the term (\ref{s_con2}) there is a constraint from the linear term in eq. (\ref{kin_sg_g}). Thus
\begin{equation}
 1+2\kappa+\delta_{sg}^{g} \geq 0.
\end{equation}
This analysis implies that the terms in the equations for $\alpha_{sg}^{u}$, $\delta_{sg}^{g}$ and $\delta_{sg}^{s}$, that have no comparable terms in the QCD equations, must be subleading. As a result the system of equations has the same solutions as for quenched QCD, namely the infrared exponents of the scalar propagator $\delta_{s}^{\prime}$ and the scalar-gluon vertex in the possible kinematic limits $\alpha_{sg}^{u}$, $\delta_{sg}^{g}$ and $\delta_{sg}^{s}$. The solutions are listed in the main text in table \ref{expkindiv}.

\newpage
\thispagestyle{plain}
\section*{Acknowledgments}
First of all I especially thank my supervisor, Prof. Dr. Reinhard Alkofer (Univ. Graz) for offering me the chance to write this thesis and to immerge myself into an interesting and hot topic of modern physics. I am very grateful for the guidance and support I received.\\
Another big \textquotedblleft thank you\textquotedblright \ I want to say to my second supervisor Dr. Kai Schwenzer (Univ. Graz, Univ. St. Louis), for listening to my noob's questions and for his extraordinary patience in answering them.\\
I am also very grateful to the people who co-initiated my thesis, Prof. Dr. Jeff Greensite (Univ. San Francisco) and Prof. Dr. \v{S}tefan Olejn\'{\i}k (Univ. Bratislava), with whom I also enjoyed discussions about the results of my thesis.\\
Furthermore I am grateful for having enjoyed many fruitful discussions with Prof. Dr. Christian S. Fischer (Tech. Univ. Darmstadt), Dr. Axel Maas (Univ. Graz) and Prof. Dr. Jan M. Pawlowski (Univ. Heidelberg).\\
I also want to thank Markus \textit{Q.} Huber (Univ. Graz) for plenty discussions, helping me in the early beginning of derivation of the DSEs and for help on checking the DSEs with his package DoDSE.\\
I also appreciate the critical reading of the draft of my thesis by Dr. Axel Maas, Markus Q. Huber and Veronika Macher (Univ. Graz).
\newpage
\addcontentsline{toc}{chapter}{Bibliography}
\bibliography{./diploma_thesis_ref.bib}
\bibliographystyle{./bibstyle}
\end{document}